\documentclass[journal]{IEEEtran}
\usepackage[english]{babel}
\usepackage{hyperref}
\usepackage[utf8]{inputenc}
\usepackage{balance}
\usepackage{xspace}
\usepackage{bbm}
\usepackage{rotating}
\usepackage{url}
\usepackage{csquotes}
\usepackage{paralist}
\usepackage{multirow}
\usepackage{amsmath,amssymb,mathtools,amsthm}
\usepackage{graphicx}
\ifCLASSOPTIONcompsoc
    \usepackage[caption=false, font=normalsize, labelfont=sf, textfont=sf]{subfig}
\else
\usepackage[caption=false, font=footnotesize]{subfig}
\fi
\usepackage{xfrac}
\usepackage{array}
\usepackage{algorithm,algorithmic}
\usepackage{textcomp}
\usepackage{siunitx}
\usepackage{microtype}
\usepackage[usenames,dvipsnames,table]{xcolor}
\usepackage{pgfplots}
\pgfplotsset{compat=newest,unit code/.code={\si{#1}},plot coordinates/math parser=false,grid style={lightgray}}
\usepgfplotslibrary{units,external,groupplots}
\usetikzlibrary{positioning,angles,quotes,patterns,shapes}
\tikzsetexternalprefix{tikz/}
\tikzstyle{block} = [draw, rectangle, minimum height=2em, minimum width=5em]
\tikzstyle{addon} = [draw, rectangle, rounded corners]
\tikzstyle{pinstyle} = [pin edge={<-,thin,black}]
\tikzstyle{pinstyle2} = [pin edge={->,thin,black}]
\tikzstyle{mult} = [draw, isosceles triangle]
\tikzstyle{circ} = [draw, circle]
\tikzstyle{coord} = [coordinate]
\tikzstyle{circ2} = [draw, circle,minimum width=3pt, inner sep=0]
\tikzset{>=latex}
\tikzset{radiation/.style={{decorate,decoration={expanding
waves,angle=90,segment length=4pt}}}}
\usepackage{booktabs}
\usepackage{pifont}
\newcommand{\cmark}{\textcolor{green!50!black}{\ding{51}}}%
\newcommand{\xmark}{\textcolor{red}{\ding{55}}}%

\graphicspath{{./img/}}

\newtheorem{example}{Example}
\newcommand{\fakepar}[1]{\vspace{1mm}\noindent\textbf{#1.}}

\DeclareSIUnit{\belmilliwatt}{Bm}
\DeclareSIUnit{\dBm}{\deci\belmilliwatt}

\DeclareMathOperator*{\E}{\mathbb{E}}

\DeclareMathOperator*{\R}{\mathbb{R}}

\newcommand{\transp}{\text{T}}

\let\originalleft\left
\let\originalright\right
\renewcommand{\left}{\mathopen{}\mathclose\bgroup\originalleft}
\renewcommand{\right}{\aftergroup\egroup\originalright}

\newcommand\figref[1]{Fig.~\ref{#1}}
\newcommand\tabref[1]{Table~\ref{#1}}
\newcommand\secref[1]{Sec.~\ref{#1}}

\newcommand{\eg}{e.g.,\xspace}
\newcommand{\ie}{i.e.,\xspace}
\newcommand{\etc}{etc.\xspace}
\newcommand{\capt}[1]{\mdseries{\emph{#1}}}
\newcommand{\cf}{cf.\xspace}
\newcommand{\iid}{i.i.d.\xspace}

\newcommand{\cps}{CPS\xspace}

\newcommand{\ap}{AP\xspace}

\newcommand{\cp}{CP\xspace}

\newcommand{\dpp}{DPP\xspace}
\newcommand{\bolt}{Bolt\xspace}

\newcommand{\Tupdate}{\ensuremath{T_U}\xspace}
\newcommand{\Tdelay}{\ensuremath{T_D}\xspace}

\newcommand{\dBm}{\ensuremath{\,\text{dBm}}\xspace}

\usepackage{ifthen}
\newboolean{authnotes}

\setboolean{authnotes}{true}
\ifthenelse{\boolean{authnotes}}
{
\newcommand{\fm}[1]{\footnote{{\bf\color{blue} Fabian: #1}}}
\newcommand{\mz}[1]{\footnote{{\bf\color{blue} Marco: #1}}}

\newcommand{\db}[1]{\footnote{{\bf\color{green!50!black} Dominik: #1}}}
\newcommand{\st}[1]{\footnote{{\bf\color{purple!90!black} Sebastian: #1}}}
\newcommand{\lt}[1]{\footnote{{\bf\color{orange!50!black} Lothar Thiele: #1}}}
\newcommand{\sttodo}[1]{{\color{purple!90!black} [Seb-todo: #1]}}
}
{
\newcommand{\fm}[1]{}
\newcommand{\mz}[1]{}
\newcommand{\db}[1]{}
\newcommand{\st}[1]{}
\newcommand{\lt}[1]{}
\newcommand{\sttodo}[1]{}
}

\usepackage{fancyhdr}
\newcommand{\mytitle}{\textbf{Accepted final version.}
To appear in \textit{Proceedings of the IEEE}.\\
\copyright 2020 IEEE. Personal use of this material is permitted. Permission
from IEEE must be obtained for all other uses, in any current or future
media, including reprinting/republishing this material for advertising or
promotional purposes, creating new collective works, for resale or
redistribution to servers or lists, or reuse of any copyrighted component of
this work in other works.}
\begin{document}
% \title{Feedback Control Goes Wireless:\\ Reliable Stabilization and Coordination over Low-Power Networks}
%\title{Adaptive and Reliable Wireless Control\\ for Smart Manufacturing}
%\title{Wireless Control for Smart Manufacturing:\\State of the Art and Open Challenges}
\title{Wireless Control for Smart Manufacturing:\\Recent Approaches and Open Challenges}


\author{Dominik~Baumann$^*$,
        Fabian~Mager$^*$,
        Ulf~Wetzker,
        Lothar~Thiele~\IEEEmembership{Member,~IEEE},
        Marco~Zimmerling,
        and Sebastian~Trimpe~\IEEEmembership{Member,~IEEE}% <-this % stops a space
\thanks{This work was supported in part by the German Research Foundation (DFG) within SPP 1914 (grants ZI 1635/1-1 and TR 1433/1-1) and 
the Emmy Noether project NextIoT (grant ZI 1635/2-1), the German Federal Ministry of Education and Research (BMBF) project Veritas (grant 01IS18073A), the Cyber Valley Initiative, and the Max Planck Society.}%
\thanks{$^*$ Equal contribution.}%
\thanks{D.~Baumann is with the Intelligent Control Systems Group, Max Planck Institute for Intelligent Systems, Stuttgart, Germany; e-mail: dbaumann@tuebingen.mpg.de}% <-this % stops a space
\thanks{F.~Mager and M.~Zimmerling are with the Networked Embedded Systems Lab, TU Dresden, Dresden, Germany; e-mail: fabian.mager@tu-dresden.de, marco.zimmerling@tu-dresden.de.}% 
\thanks{U.~Wetzker is with the Industrial Wireless Communication group, Fraunhofer Institute for Integrated Circuits Division Engineering of Adaptive Systems, Dresden, Germany; e-mail: ulf.wetzker@eas.iis.fraunhofer.de}% 
\thanks{L.~Thiele is with the Computer Engineering group, ETH Zurich, Zurich, Switzerland; e-mail: thiele@ethz.ch}%<-this % stops a space
\thanks{S.~Trimpe is with the Institute for Data Science in Mechanical Engineering, RWTH Aachen University, Aachen, Germany and with the Intelligent Control Systems Group, Max Planck Institute for Intelligent Systems, Stuttgart, Germany; e-mail: trimpe@dsme.rwth-aachen.de}
%\thanks{Manuscript received Month XX, 2020; revised Month XX, 2020.}
}

\maketitle

%%%%%%%%%%%%%%%%%%%%%%%%%%%%%%%%%%%%%%%%%%%%%
% Added after final submission
%%%%%%%%%%%%%%%%%%%%%%%%%%%%%%%%%%%%%%%%%%%%%
\thispagestyle{fancy}	% final submitted: empty
\pagestyle{empty}

% !TEX root = ../paper.tex

\begin{abstract}
Smart manufacturing aims to overcome the limitations of today's rigid assembly lines by making the material flow and manufacturing process more flexible, versatile, and scalable.
The main economic drivers are higher resource and cost efficiency as the manufacturers can more quickly adapt to changing market needs and also increase the lifespan of their production sites.
The ability to close feedback loops fast and reliably over long distances among mobile robots, remote sensors, and human operators is a key enabler for smart manufacturing.
Thus, this article provides a perspective on control and coordination over wireless networks.
Based on an analysis of real-world use cases, we identify the main technical challenges that need to be solved to close the large gap between the current state of the art in industry and the vision of smart manufacturing.
We discuss to what extent existing control-over-wireless solutions in the literature address those challenges, including our own approach toward a tight integration of control and wireless communication.
In addition to a theoretical analysis of closed-loop stability, practical experiments on a cyber-physical testbed demonstrate that our approach supports relevant smart manufacturing scenarios.
The article concludes with a discussion of open challenges and future research directions.

%Closing feedback loops fast and over long distances is key to emerging cyber-physical applications
%such as robot motion control, swarm coordination, or smart manufacturing. These applications
%typically require sensor-to-controller and controller-to-actuator delays of at most tens of
%milliseconds. Low-power wireless communication technology is interesting due to its low cost, small
%form factor, and flexibility, especially when devices support multi-hop communication. This article
%illustrates the research challenges in harnessing those benefits and provides an overview of the state
%of the art. We specifically outline a co-design approach that integrates feedback control and
%coordination with reliable real-time communication over wireless multi-hop networks. This approach
%is the first to provably guarantee closed-loop stability for physical processes with linear time-
%invariant dynamics despite dynamic changes between a well-defined set of operating modes.
%Further, the underlying many-to-all communication allows each agent to take decisions based on
%global knowledge, supporting an efficient implementation of tasks such as synchronization or
%consensus. Experiments on a cyber-physical testbed with 30 low-power wireless devices and 10 cart-
%pole systems demonstrate reliable feedback control and coordination with mode changes over
%multi-hop networks for sensor-to-controller and controller-to-actuator delays of 20 to 100
%milliseconds. The article concludes with a discussion of open problems and research directions.
\end{abstract}

% \begin{IEEEkeywords}
% IEEE, IEEEtran, journal, \LaTeX, paper, template.
% \end{IEEEkeywords}
% !TEX root = ../paper.tex

%\mz{Intro must be completely rewritten. Dominik has some good suggestions. Overall, the intro can be quite short as the following sections provide an overview of the vision, use cases, challenges, etc. After an introductory paragraph on the vision and promises of smart manufacturing, we should motivate why control and coordination of wireless is important and hence the focus of this article. Then, we also need to say something about the road map and possibly highlight the 2-3 key messages of this article.}

\IEEEPARstart{M}{anufacturing} and several other industrial sectors are increasingly caught between a rapidly growing demand for individualized, high-quality products, and the constant pressure to maximize profit margins.
To successfully handle this balancing act, the manufacturing industry is
%at the beginning
in the early phases of a revolution: Driven by advances in digitalization and automation, \emph{smart manufacturing} promises more flexible, versatile, and scalable material flows and manufacturing processes through plants that can be reconfigured based on the individual product and overall process requirements~\cite{jeschke2016industrial}.
These plants will consist of \emph{physical systems} (\eg machines, storage systems, supply-chain entities) with sensing, computation, communication, and actuation capabilities.
Reconfigured and automated through edge- or cloud-based services, the physical systems will interact autonomously with one another and human operators.
Smart manufacturing will thus give rise to a new bread of complex, large-scale \emph{cyber-physical systems (\cps)} that are safety-critical and rely on networked and distributed control architectures.

Due to the complex interactions among the different entities, wired communication systems, such as fieldbusses and Industrial Ethernet, will reach their limits.
Instead, \emph{wireless communication} enables much higher flexibility, allowing for on-demand plant reconfigurability while mitigating cable breaks and faulty connections~\cite{mcmillan09}.
Wireless communication is also more robust to certain external influences, including heat, humidity, abrasive substances, and undamped vibrations.
Completely untethered, mobile physical systems are possible through the use of batteries, which can be recharged using energy harvesting or wireless power transfer techniques~\cite{bhatti16}.
Furthermore, to reach into tiny spaces and to cover large distances, physical systems built from \emph{embedded hardware} and capable of \emph{multi-hop communication} will be crucial~\cite{ahlen2019toward,Akerberg2011}.
Taken together, wireless communication offers unprecedented flexibility, cost efficiency, and robustness and is, therefore, a key enabler for smart manufacturing.

Despite many advantages, using wireless instead of wired communication also poses significant challenges, for example, in terms of reliability and security.
Moreover, because the manufacturing tasks are defined by algorithms, the software and algorithmic components are closely linked to the physical manufacturing processes.
Hence, cyber and physical components are in \emph{feedback} with each other and cannot be designed in isolation.
Instead, joint designs are needed to leverage their full potential.
In particular, the physical process will dictate timing and safety requirements, which pose challenges especially on the wireless communication and embedded systems, which drive the algorithms (\eg control, learning, monitoring) that control the overall manufacturing process.
The situation is exacerbated when many feedback loops are closed in a flexible and ad-hoc manner across the same network and possibly over large distances, while requiring fast and reliable updates, as is the case in many envisioned smart manufacturing use cases.

\fakepar{Contributions and road map}
%\db{adding a graphic here that illustrates the structure of the whole article would be great.}
\begin{figure}
\centering
\includegraphics[width=0.85\linewidth]{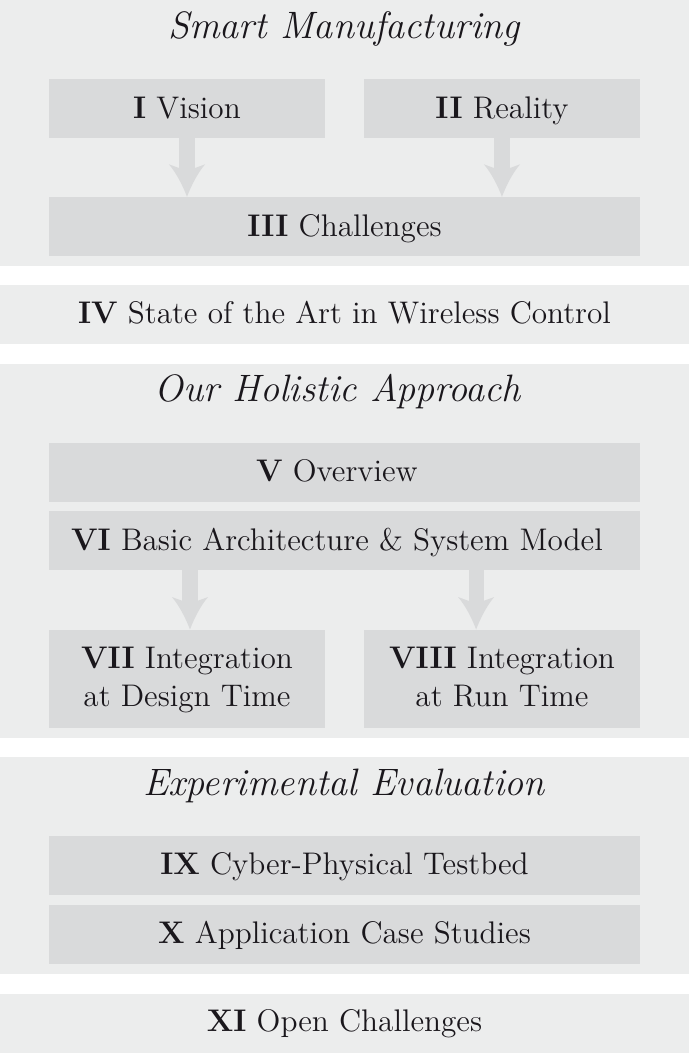}
\caption{Structure of the article.}
\label{fig:roadmap}
\vspace{-5mm}
\end{figure}
This article provides a perspective on control and coordination over wireless networks as a core technology for future smart manufacturing. \figref{fig:roadmap} depicts the structure of the article. To ground the subsequent discussion, we start by presenting commonly discussed future use cases of smart manufacturing and derive a unified vision from these (\secref{sec:vision}).  We then contrast this vision with concrete examples of wireless technology in industrial practice today (\secref{sec:sota_industry}), and analyze the gap between vision and practice.  Based on the found reality gap, we derive requirements and challenges of wireless control for smart manufacturing (\secref{sec:challenges}), and review the state of the art (\secref{sec:sota}).  In particular, we discuss to what extent existing approaches in the literature address core desiderata, such as support for fast feedback, multi-hop communication,
% mode changes,
stability guarantees, and resource efficiency.
We then focus on our own approach from recent work
%\cite{} -> we could cite our papers here, but I kind of decided against it, as this would presumably force us to discuss differences with this article in greater detail.
that jointly addresses these aspects through a tight integration of control and wireless communication, both at design time (\ie prior to operation)~\cite{mager2019feedback,baumann2019fast} and at run time (\ie during operation)~\cite{baumann2019control}. \secref{sec:approach} introduces our co-design approach, followed by the presentation of the basic architecture and unified system model (\secref{sec:architecture}).  Through \emph{integration at design time} (\secref{sec:design_time}), we are able to achieve fast feedback control (on the order of tens of milliseconds) over multi-hop wireless networks with formal guarantees on closed-loop stability.
% that captures all  components of the \cps (embedded real-time system, wireless communication, control, and linear system dynamics).
To further increase adaptability and resource efficiency, we deepen the integration toward  \emph{run time} (\secref{sec:run_time}) by having the control system inform the communication system about its current needs for data transmission, and empowering the communication systems to react accordingly.  The developed approaches are illustrated through several experiments on a cyber-physical testbed consisting of ten inverted pendulum systems as fast physical systems and 20 wireless embedded nodes (\secref{sec:testbed} and \secref{sec:eval}).   The article ends in \secref{sec:outlook} with a discussion of open challenges that are not yet addressed by the state of the art in wireless control (including our own work), but are vital for realizing the full vision of smart manufacturing and thus provide ample opportunities for important future research.

\fakepar{Difference to prior publications} This article makes several novel contributions compared with our prior work~\cite{mager2019feedback,baumann2019fast,baumann2019control,baumann2018evaluating} by: (\emph{i}) articulating a vision of smart manufacturing from the perspective of wireless control and putting this vision in contrast to the current reality in the manufacturing industry; (\emph{ii}) working out the main challenges for smart manufacturing in general and wireless control in particular to close the gap between vision and reality; (\emph{iii}) providing a unified, tutorial-style description of how our state-of-the-art approaches, which have appeared in specialized venues and separate communities, are able to address for the first time some of these challenges; (\emph{iv}) reporting on new experiments on an extended cyber-physical testbed that demonstrate the capabilities of our approach; (\emph{v}) and highlighting the key research questions yet to be answered.

\fakepar{Relation to Industry 4.0}
Enabling wireless control constitutes one important building block toward realizing the vision of smart manufacturing.
This vision is often described in the context of \emph{Industry~4.0}~\cite{adolphs2015reference,lin2017industrial,lu2016current}.
% With reference to the reference architecture proposed in~\cite{adolphs2015reference}, the concepts discussed herein aim toward integrating physical systems with the digital world and, in particular, enabling communication between machines and agents and of machines and agents with remote computing units.
Relating the concepts discussed herein with the reference architecture proposed in~\cite{adolphs2015reference}, we aim at enabling wireless control and coordination
% \mz{Why only communication? Wouldn't it be more appropriate to write ``control'' or ``closing feedback loops'' something similar CPS-ish here and in the next sentence?}
of entities \emph{within} a smart factory.
% That way, they provide access to necessary data.
% Regarding the hierarchy, we here consider connections within a factory.
This can be further extended by also establishing communication between different factories, possibly worldwide, through the \emph{Internet of Things}.
% \db{First draft of a paragraph relating this paper to industry 4.0}

% Previous version:
%This article provides a perspective on smart manufacturing with a particular focus on the wireless control system that will form the backbone of envisioned manufacturing tasks.  Starting with a discussion of real-world use cases, we develop an abstract and unifying view of smart manufacturing from the perspective of communication and control.  Based on this, we will derive requirements and challenges for the wireless control in manufacturing, followed by a discussion of the state of the art.  We then present a recent co-design of wireless embedded and control systems that addresses many of the key challenges.  In particular, we achieve fast feedback control over multi-hop wireless networks by a tight co-design of the wireless embedded and control systems (\emph{integration at design time}).  To further increase the flexibility and resource efficiency, we deepen the integration of communication and control by having the control system inform the communication system about its needs for data transmission at run time (\emph{integration at run time}).  We present theoretical guarantees and experimental results on a cyber-physical testbed consisting of ten inverted pendulums as physical processes and 20 wireless embedded nodes.  The article closes by putting the proposed solution into perspective of the broader vision and providing an outlook of further important research aspects.

% !TEX root = ../paper.tex

\section{Smart Manufacturing: Potential Use Cases and Vision}
\label{sec:vision}

To provide some context and concrete motivation for wireless control in future manufacturing, this section outlines envisioned use cases and the associated benefits (\secref{sec:usecases:conveyorbelts} to \secref{sec:usecase:mining}).  Abstracting from these use cases, we derive in \secref{sec:abstract_view} the main features and a unifying vision of smart manufacturing.
%This section outlines envisioned use cases and the associated benefits of smart manufacturing, followed by a more abstract view on the overall vision and promises.

\subsection{Reconfigurable Conveyor Belts}
\label{sec:usecases:conveyorbelts}

In traditional factories, conveyor belts have two ends: one serves as an input, and the other one serves as an output.
The machines alongside a conveyor belt are arranged such that the product can be manufactured in a step-by-step fashion.
None of the machines in the chain is redundant.
The whole setup is carefully designed in advance and follows the same procedure for its entire lifespan to manufacture the ever same product.

By contrast, conveyor belts in
%\uw{This use case is very common (future might be misleading) and IDs are based on barcodes or RFIDs.}
% -> ST: good point, deemphasized "future" here
smart factories are capable of handling multiple types of products at once~\cite{wang2016implementing}.
Each product has a unique ID, and the conveyor belt is equipped with multiple redundant machines.
The process relies on machines taking information communicated by the product (\ie the unique ID), extracting which processing steps are necessary, and distributing the workload across the machines such that every product is manufactured as fast as possible without sacrificing on product quality.
For this, conveyor belts can be \enquote{closed,} \eg arranged in a circular shape without input and output, enabling various production routes.
Coordination is required between product and manufacturing machines as well as among the machines distributed along the conveyor belt.
While wired or mixed wired/wireless communication solutions may be possible in certain scenarios to realize this coordination, the use of wireless networks arguably provides the highest flexibility.
% which should both happen over wireless communication networks to enhance flexibility.

%\uw{In general there is no feedback channel since it only requires unidirectional communication. Maybe this should no be the first use case since the connection to the content of this article is no that obvious}
%\db{No real feedback control in the classic sense, but kind of distributed control. And it's the example I find everywhere, so I think it would  be nice to have and we can also relate to it at some points. But not entirely sure about that one.}

\subsection{Autonomous Drones}
\label{sec:usecase:drones}

A predominant and widely-cited example of \cps in general is autonomous drones.
Due to their agility, flexibility, and capability to operate in three-dimensional space, drones are also of interest for many use cases in future manufacturing~\cite{maghazei2019drones}.
Those use cases include visual inspection and monitoring of factory automation machinery, sensing tasks, transporting objects, or delivering goods.
In laboratory contexts and as demonstrations, drones have also been used for cooperative manipulation and transportation~\cite{michael2011cooperative} and \mbox{for construction tasks~\cite{augugliaro2014flight}.}

Taking the example of inspection, drones can act in different ways.
A drone can be fully autonomous, regularly monitoring a production plant.
Alternatively, a human operator can define waypoints that a drone should follow while the drone transmits image or video data for visual inspection.
Next to different levels of autonomy, drones also offer great flexibility as they can switch between different tasks.
For instance, a drone monitoring a production plant can interrupt this task momentarily to carry a spare part across the factory hall to a location where it is urgently needed.
While these application examples, performed by a single drone, already constitute useful tasks that can improve the efficiency and quality level of a smart factory, it is swarms or fleets of multiple drones that can leverage the full potential of autonomous flying vehicles.
Fleets of drones can, for example, carry parts or goods that are too heavy for a single drone, jointly monitor larger plants, or directly participate in the manufacturing process.
In all cases, coordination among drones is essential to ensure that they fulfill their tasks and to prevent drones from crashing into each other.
Typically, each drone is equipped with an embedded microcontroller for local control tasks (\eg stabilizing the flight), while distributed control tasks (\eg coordinating the position of every drone in a fleet) require wireless communication among the drones~\cite{Hayat2016}.

\begin{figure*}
\centering
\includegraphics[width=\linewidth]{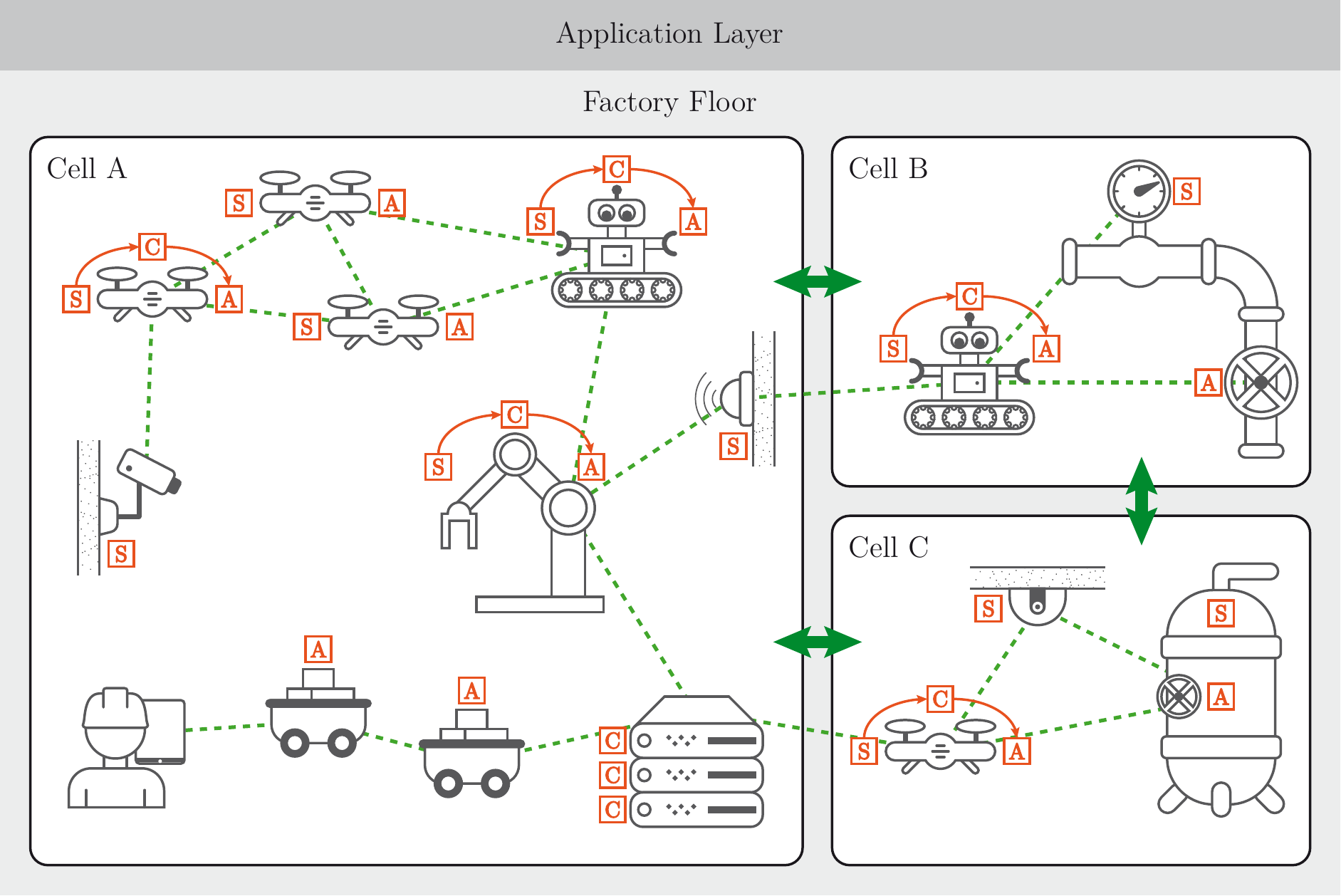}
\vspace{-6mm}
\caption{Abstract view of a future smart manufacturing system. Distributed agents are equipped with sensing (S), computing (C), and actuation (A) capabilities, or a subset thereof.  Depending on the manufacturing task at hand, which can change dynamically at run time, the agents interconnect wirelessly (green dashed lines depict communication links) to form teams within an automation cell, among cells across the factory hall, and with global wired communication infrastructure to access, for example, cloud services.}
\vspace{-2mm}
\label{fig:smartManufacturingAbstraction}
\end{figure*}

\subsection{Remote Control of Mobile Robots}
\label{sec:usecase:remote}
When multiple mobile robots act in the same workspace, be it in the air or on the ground, planning and coordination among them is often essential.
In particular, when planning actions for one robot it can be beneficial to take information from other agents into account for improving safety, minimizing reaction time, and optimizing resource usage.
%Enabling drones or, more generally, mobile robots, to locally take decisions on the next action, taking into account information from other agents, is oftentimes beneficial since it minimizes reaction times between computing and applying actuation commands.
When decision making happens locally on an agent, information from other agents must be wirelessly communicated.  Alternatively, in many scenarios it is desirable to partly or entirely outsource the necessary computations to edge devices or cloud services executing inside large data centers~\cite{sabella2018industrial,abbenseth2017cloud}.
%The same may hold for related tasks such as localization and mapping \cite{abbenseth2017cloud}.
%However, for several applications, it may also be desirable to (partly or entirely) outsource such computations to edge computing devices or cloud services~\cite{sabella2018industrial}.
Outsourcing computations can bring many benefits.
Edge devices and in particular cloud services offer more compute power, memory, and can exploit additional data sources.
This makes the application of more sophisticated control and planning algorithms possible.
Further, stored data can be used to improve the algorithms' decisions over time, for example, via machine learning techniques.
The edge devices or cloud services then act as an \enquote{external brain.}

\subsection{Mining Industry}
\label{sec:usecase:mining}
While not belonging to smart manufacturing in a strict sense, the mining industry shares many design considerations and has an intrinsic need for autonomous and wireless technologies.
%As a final example, we take a look at the mining industry.
%Although not belonging to smart manufacturing in a strict sense, mining shares many design aspects and has an intrinsic need for smart and wireless technologies.
One reason for this is that mining environments are typically extraordinarily hazardous and dangerous to humans;
numerous accidents have been reported over the last decade~\cite{pei2009anchor}.
A first step toward reducing the risks of human accidents in the mine is the deployment of wireless sensor networks that localize the workers in the mine~\cite{pei2009anchor} and monitor the environment~\cite{li2009underground}.
This already comes with severe challenges because large spaces have to be covered by the wireless sensor network.
However, it would be even more desirable to employ autonomous agents or to control tasks in the mine remotely.
On the one hand, if autonomous or remotely controlled agents could fulfill all tasks in the mine, this would entirely eliminate the risk of human accidents in the mine.
On the other hand, it would also enable substantial savings and thus also be economically viable, since pumping fresh air to the workers in the mine is extremely costly.
While companies are already moving into this direction~\cite{morton2018underground}, the vision of fully autonomous agents working independently in the mine has not been realized to date.

\subsection{The Vision of Smart Manufacturing}
\label{sec:abstract_view}
%Abstracting from the above-mentioned use cases,
Multiple heterogeneous sensing-computing-actuation units are at the heart of the above-mentioned use cases.
As shown in \figref{fig:smartManufacturingAbstraction}, a unit may involve one main modality (\eg sensing for a stationary sensor in a mine or computing for an edge device) or all modalities together (\eg a drone that senses its environment, locally computes a path, and actuates its motors to move along that path).  We refer to these sensing-computing-actuation units as \emph{agents}.
%Abstracting from the above-mentioned use cases,
The agents are generally coupled with the physical world:
%\st{I think this is a core point; from this we can derive requirements such as: physics dictate certain time scales/update rates, and requirements such as stability/safety.  We could take this up in following sections.}
sensors measure physical quantities and actuators act on their environment.
The connection of the two happens through algorithms that execute on embedded computers.
%while the sensor information is connected to the actuator commands through algorithms in the `cyber space'.
Using sensors and actuators, software and processing are thus linked to the physical world, which is the defining characteristic of CPS~\cite{rajkumar2010cyber}.
An integral aspect of future smart manufacturing systems are \emph{multiple} agents (or multiple \cps) that, each on an individual level, connect the physical world and the cyber space, but are at the same time also interconnected with each other to jointly perform manufacturing tasks.
At an abstract level, the overall smart manufacturing system
%(as well as any other large-scale CPS)
is therefore given by multiple agents that \emph{interconnect} in various ways, execute different types of \emph{algorithms} (control, optimization, learning, planning, \etc), and can thus flexibly and jointly perform diverse tasks that contribute to the overall manufacturing process.
%types of manufacturing tasks.
% as defined by local or distributed algorithms.
%perform different types of \emph{algorithmic tasks}\mz{What about ```... and perform different types of manufacturing tasks as defined by localized or distributed algorithms.''}.

Interconnections occur in various forms and on different hierarchical levels.  At the core of many smart manufacturing visions are teams of mobile agents that interconnect with each other.  In all the above-mentioned use cases, inter-agent communication is wireless (at least partially) to support the required flexibility, mobility, ease of installation and maintenance, and so on.  Mobile agents may additionally connect with stationary agents (\eg fixed sensors), human operators, and the infrastructure (\eg edge devices), as illustrated in \figref{fig:smartManufacturingAbstraction}.  Depending on the use case, the multi-agent networks differ in the number of agents supported (\eg two for remote control of a mobile robot, or hundreds in the drone scenario) and the physical distance covered (\eg direct peer-to-peer links on the conveyor belt, and long-distance multi-hop communication in the mining scenario).  Further, one may distinguish different levels of communication (according to \cite{ruth2017communication}), also indicated in \figref{fig:smartManufacturingAbstraction}: A \emph{local automation cell} comprises multiple interconnected agents, a \emph{factory hall} interconnects several cells, and factories interconnect through the \emph{application layer communication} to global networks, such as the Internet or cloud services.
The need for mobility typically decreases with increasing communication level.
While higher-level communication between cells or with the cloud can often resort to cabled routing-based networking solutions, wireless communication is critical on the cell and factory hall levels.
%at the agent level\mz{Would be better to be more specific here and use the just-introduced terminology, e.g., by saying ``on the cell and factory hall levels''.}.
Depending on the envisioned scenarios, wireless communication may extend across several cells or the whole factory hall (\eg coordinated inspection with drones).
%Notwithstanding the importance of wireless communication at the agent and cell level, higher-level communication with the cloud can often resort to cable-based networking solutions.

%\st{Should we specifically also talk about communication protocols?  We could consider these are part of the communication system (interconnections), as well as an algorithmic component.}
A key prerequisite for the envisioned flexibility in smart manufacturing is the ability to perform various types of tasks that are enabled and defined through algorithms and software.
% components and that contribute to the manufacturing process.
For instance, the same drone may switch from an inspection task to bringing an urgently needed tool to a worker from one moment to the next, just by downloading a new algorithm from the cloud in a split second.
%the same AGV may transport a heavy box and, from one moment to the next, be equipped with a screw-dringing robot arm to peform, just by downloading a new algorithm.
Algorithms in smart manufacturing can either use only local information and resources available to a single agent (localized) or use information and resources from multiple agents (distributed), which is supported through interconnections and the communication system.
%contributing to the overall manufacturing process.
%Such manufacturing tasks are defined by the logic that is embodied by some algorithm or sthat uses either only local information and resources available to a single agent (localized) or uses information and resources from multiple agents (distributed)

%Such algorithms may be local, or distributed as supported by communication system.
%Algorithmic tasks\mz{Is this established terminology? I don't quite like algorithmic here as the purpose of the tasks is not to execute some algorithm but to contribute to the manufacturing process; rather, the manufacturing tasks are defined by the logic that is embodied by some algorithm that uses either only local information and resources available to a single agent (localized) or uses information and resources from mutliple agents (distributed). At least, this is my view ;-) Not sure if that makes any sense.} are multifaceted in smart manufacturing.

%\mz{Should we put a 'fakepar' here with the work 'control' in it? The whole subsection is quite long and it would be good to give the text a bit more 'visual' structure, e.g., through 'fakepars', if possible.}
%\db{If we do this we should have multiple fakepars and those should also somehow capture the essence of the subsection. At the moment it's not clear to me how to best do this.}
% -> Seb:  No, I would not.  Also, let's not worry, I have doubts whether they will allow us these fakepars in a final version anyway.

Many different types of algorithms are relevant and jointly define the smart manufacturing process.
Nevertheless, ultimately, all algorithms operate on data from physical sensors or received over the network, and decisions lead to actions in the physical world (the manufacturing).
Hence, \emph{control} algorithms are an essential component of any manufacturing system.
Controllers compute actuator commands based on sensor information to achieve a desired control objective, including
%of the closed-loop system
stability, precise positioning, or robust performance in the face of uncertainty.
%stability, robustness, and tracking performance.

For the design of a control algorithm, the question of what information is available as input is of key importance and leads to different control architectures.
% Which control architecture is available is of key importance in the control design and for the achievable performance.
 In a \emph{centralized} architecture, the controller has global information; that is, it has access to all relevant sensor readings based on which it computes all actuator commands.  In a \emph{decentralized} architecture, the control computation is distributed among multiple controllers, each of which has access only to local sensors and computes commands only for local actuators; in a purely decentralized architecture, no information is shared between the controllers.  Finally, \emph{distributed} control resides in between the two extremes: control computations are distributed, but some information is shared between the controllers.  Typically, centralized control yields the best performance as compared with decentralized control~\cite{fowler2003formation} because it has maximal information available.

 In smart manufacturing, the control architecture is not given but can be influenced and adapted, in particular through the communication system.
 %\mz{Is it really only the communication system that can affect the control architecture?}
For example, wireless links can be established between otherwise decoupled controllers, at the cost of using extra resources such as communication bandwidth and energy.
Thus, depending on the communication support, control may be easier or harder, and the co-design of control and communication in smart manufacturing gives significant additional design freedom over traditional fixed, cable-based architectures \cite{lunze1992feedback}.  In general, the smart manufacturing system comprises a multitude and various types of feedback loops and control algorithms across different hierarchies.
% This gives significant additional design freedom over traditional fixed architectures, and multi-agent control may become easy or hard depending on the overall CPS design.
Besides control, typical algorithmic components are \emph{optimization}, \emph{planning},
% and prediction,
\emph{data fusion},
%for monitoring,
and \emph{machine learning}.
%\mz{OK, now I start to see why the term algorithmic may make sense. It just felt a little strange when it occured for the first time, still for the reasons given in my other comment: ultimate purpose is manufacturing, but I see now that not every (software) task the system performs (e.g., machine learning, sensing, etc.) \emph{directly} translates into an advancement of the manufacturing process (but certainly \emph{indirectly}).}
%\st{Your comments on this were very valuable.  The term `algorithmic tasks' I invented ;), but I see your point that it can be misleading.  I thus rewrote this.  Does it make more sense to you now?}
%All algorithms in the end operate on data from sensors or received over the network, and decisions ultimately lead to actions in the physical world.
Here, architectural aspects play a similar role as discussed for control~\cite{manyika1995data,boyd2011distributed,verbraeken2019survey}.

A key characteristic of future smart manufacturing systems is the demand for unprecedented \emph{flexibility} and \emph{adaptability} of the overall system.  This concerns both the interconnections and the algorithms.  Agents need to flexibly interconnect to form heterogeneous teams that constitute an automation cell or may even span across multiple cells.
%\mz{I am noticing that we are not using the above-introduced terminology, e.g., we speak about ``teams'' rather than ``cells''.}
%\st{So, do we want to use Klaus' terminology and should we globally change this?} to solve a given manufacturing task (\eg a team of drones whose size depends on the object to be transported).
Similarly, algorithms have to adapt to the task at hand.  Adaptation may happen locally on a computing unit, through the interaction of many agents, or by downloading a new algorithm from the cloud, for example.

\figref{fig:smartManufacturingAbstraction} abstracts envisioned smart manufacturing use cases and illustrates key system characteristics.
Before developing requirements and challenges for realizing this vision, we discuss in the next section how this vision relates to the current reality in the manufacturing industry.

% Seb: my notes
%
%\sttodo{Sketch:}
%\begin{itemize}
%\item multiple distributed sensing, actuation, processing units; may include single functions (such as a temperature sensor...) or combine all (AGV); these have \emph{interconnections} and perform certain \emph{algorithmic tasks}
%\item important: physical entities, sensing and actuation connects cyber to the physical world; core implications: dictates time scales, requirements such as safety...
%\item various interconnections: inter-agent, with infrastructure (downloading new control algorithm), with cloud, remote controlled
%\item algorithmic tasks: feedback control of multi-agent systems -- can be local, distributed, central; local often worst performance, thus want some interconnection [paper Raff]; in general, multitude of/multiple types of feedback loops on different hierarchies
%\item other algorithmic tasks: planning, data fusion for monitoring/prediction, learning; similar architectural aspects as for contrl
%\item flexibility and adaptability are key: forming connections, updating algorithms, ...
%\item Possible distinction [1, Wehrle]:
%– Local Automation Cells: Need fast and reliable
%communication, but single-hop\newline
%– Factory Hall: Contains multiple cells, need interconnections,
%multi-hop, also fast and reliable\newline
%– Additional Layer Communication: Connection to
%the internet, cloud, etc.
%\end{itemize}

% !TEX root = ../paper.tex

\begin{figure*}[!tb]
% please contact ulf.wetzker@eas.iis.fraunhofer.de regarding the use of these images outside of this article
 \centering
 \subfloat[Factory floor.\label{fig:factory_floor}]{%
  \includegraphics[width=0.49\textwidth]{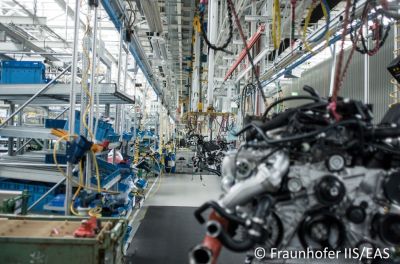}}
 \hfill
 \subfloat[Cordless power tool.\label{fig:cordless_power_tool}]{%
  \includegraphics[width=0.49\textwidth]{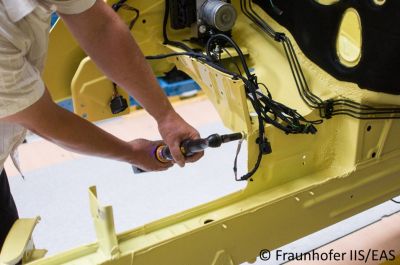}}
 \caption{Motor assembly line at Mercedes-Benz Ludwigsfelde GmbH in Germany. }
 \label{fig:use_case} 
\end{figure*}

\section{Smart Manufacturing: Industry Examples}
\label{sec:sota_industry}
%\db{I would reserve the term' state of the art' for the literature review and would not use it here.}
%\st{Agree -- I doubt that we can fully summarize the state-of-the-art in industry; thus I also prefer `examples'.}
% -> Seb:  Yes, let's NOT use SOTA here.
%
%While the previous section develops an abstract vision of smart manufacturing,
%While above we outlined potential use cases for and a general vision of smart manufacturing, 
%we now turn to examples that can actually be found in the industry today.

We now present three example applications, where wireless technology, as a key enabler for smart production, is actually adopted by the manufacturing industry today.
%In particular, we present three specific applications, where wireless technology, as a key enabler for smart manufacturing, is adopted.
% in the manufacturing process, a key enabler for future smart manufacturing as discussed in the previous section.
The discussion is based on our experience at Fraunhofer IIS in troubleshooting real-world industrial communication systems.
Afterward, we examine the gap between envisioned and existing use cases.

\subsection{Automated Guided Vehicles} 
\label{sec:usecase:agv}

The automotive industry progressively uses automated guided vehicles (AGVs) to drive car bodies to the respective assembly station. 
In the considered example, 50 AGVs are commissioned on the first factory floor and arrive at the assembly line on the second floor by use of an elevator. 
All AGVs are wirelessly connected to a central control system, which plans the routes for each of them according to the car model, the model's individual configuration, and the utilization of the different assembly stations. 
This way, the factory floor is used more effectively without being tied to a specific vehicle model. 

Communication between the control system and the AGVs is based on IEEE~802.11 (\ie Wi-Fi) in the unlicensed \SI{2.4}{\giga\hertz} frequency band. 
After more than a year of productive operation, unpredictable communication faults occurred more and more frequently, resulting in unplanned stops of the AGVs. 
This was caused by an ever-increasing use of the \SI{2.4}{\giga\hertz} frequency band by additional Wi-Fi and Bluetooth-based applications, which led to temporary coexistence problems.
Such coexistence problems 
%(in particular in-network coexistence) 
are among the hardest to troubleshoot in real-world wireless networks~\cite{wetzker2016troubleshooting}.
%On the other hand, moving to a licensed frequency band, for example, by deploying a small LTE cellular base station, is often out of scope as the running costs are too high especially for small and medium-sized enterprises (SMEs), companies may not be able to service the network themselves leading to long delays in case of failures, and network providers gain access to the communicated data, which may be sensitive.
Unfortunately, moving to a licensed frequency band (\eg a 4G/LTE or 5G cellular network) is not a viable option for many applications as running costs are high; companies are unable to operate the network themselves, leading to long service deployment times, demanding reliability, latency, and real-time requirements cannot be met; and existing security mechanisms do not offer sufficient protection for sensitive data~\cite{siemens2019requirements,zvei2019connected}.
So far, technological advances have only been able to contain the coexistence problem, but not to solve it.
A co-design of control and communication is the most promising approach to ensure the availability of the system.

\subsection{Cordless Power Tools} 
\label{sec:usecase:tools}

In machine construction, cordless power tools are frequently used to give workers maximum freedom of movement and to access hard-to-reach places.
To further improve the working conditions and facilitate continuous quality assessment, these tools can be connected to a central control system, which sets the torque for each screw individually and logs the measured values in a database.
In the concrete use case, situated in a motor assembly line at Mercedes-Benz Ludwigsfelde GmbH in Germany (see \figref{fig:use_case}), 18 cordless power tools are connected to a central programmable logic controller (PLC) via Bluetooth across a \SI{100}{\meter} by \SI{500}{\meter} factory floor.
While this wireless solution provided the desired freedom of movement, the reliability was not satisfactory. 
Especially in hard-to-reach places, the shadowing effect of both the workers and the cordless power tools resulted in severe message loss.
Such message loss may be prevented by introducing redundancy in the network, for example, by installing more nodes that form a mesh topology and forward messages on behalf of other nodes (\ie multi-hop communication).
The star topology of Bluetooth, however, proved non-ideal for dealing with shadowing issues.

\subsection{Parametrization of Automotive Electronics} 
\label{sec:usecase:automotive_electronics}

The various electronic control units (ECUs) installed in a car during production are initially not customized for the specific vehicle model and equipment. 
A parametrization of the generic software components running on the ECUs is carried out while going through multiple test stations (\eg chassis dynamometer, advanced driver-assistance systems, 
%(ADAS)     % -> ST: removed, as I don't think we use this again in the paper
sensor calibration, and the like). 
During this calibration procedure, the worker needs to set up the control system by connecting the test station to the ECU through vehicle communication interfaces (VCIs). 
In the considered use case, the VCIs are based on Wi-Fi operating in the unlicensed \SI{5}{\giga\hertz} frequency band. 
Up to 30 of these free-moving wireless VCIs are used on a multi-level all-metal factory floor of \SI{100}{\meter} by \SI{150}{\meter}.
The coverage of this area is provided by 15 Wi-Fi access points. 
% Multiple logistics applications co-exist such as based on WiFi\db{I already saw 'Wi-Fi' and 'Wi-Fi', tried to unify it to 'Wi-Fi'. Needs to be checked again in the end}, Bluetooth, Siemens MOBY U RFID, or Zigbee.
% To avoid interference problems, those are hosted in the \SI{2.4}{\giga\hertz} band.
Since the \SI{5}{\giga\hertz} frequency band is shared with weather radar stations, the access points need to perform \enquote{in service monitoring} for the presence of radar pulses.
If radar pulses are detected, an access point must move to another channel within ten seconds of \enquote{channel move time.} 
In the considered use case, unfounded detections of radar pulses by the access points occurred. 
This led to sudden channel changes and a reconnection of the VCIs, which caused delays or even interruptions of the calibration procedure.

\subsection{The Reality of Smart Manufacturing}

These real-world application examples are in stark contrast to the envisioned use cases of smart manufacturing presented in \secref{sec:vision}.
Especially the use of wireless communication today is restricted to applications that are neither time- nor safety-critical.
Closing sensing-computing-actuation loops continuously over wireless communication networks, as put forward by the smart manufacturing vision, is extremely far from what industry is currently willing to take up, although the wish to wirelessly interconnect robots, rotating parts, and AGVs in automated high-rise warehouses is commonly expressed by many of our industry partners.
However, already in the comparably simple real-world use cases discussed above, wireless communication causes several issues.
These issues are mainly related to the reliability of wireless communication because, for example, shadowing effects and coexistence problems can cause severe, unpredictable message loss or complete disconnections.
Thus, the main reason for the reluctance of industry to adopt wireless technology for more challenging and critical tasks appears to be a lack of trust in the reliability of wireless solutions.

\section{Challenges for Wireless Control in Smart Manufacturing}
\label{sec:challenges}

% \db{@Marco, Fabian: One of the reviewers required a (quick) discussion of communication protocols, their impact on robustness, error rate, network topology, and possible limitations on actual deployment, cybersecurity problems. Could you check whether it makes sense to include such a discussion here? I'm not convinced that this is the right place for this\ldots}

% To close the gap between the current reality and future vision of smart manufacturing, three main challenges must be solved.
The comparison between vision and reality of smart manufacturing revealed that there clearly exists a large gap. 
Our analysis also identified three main challenges that need to be solved to close this gap.
We detail those challenges next.

\subsection{Dependability}

The real-world examples and envisioned use cases clearly demonstrate that smart manufacturing systems are large-scale, complex, and \emph{safety-critical} \cps.
For this reason, \emph{dependability} is of utmost importance, which entails that the systems must be provably secure and function as intended under realistic attacker and fault models.
In addition to broken sensors, failing devices, software bugs, side-channel attacks, and so on, these models must account for message loss and disconnections due to the notorious unreliability of wireless communication.
Based on these attacker and fault models, rigorous proofs of dependability properties along with an end-to-end validation of the systems on real-world smart manufacturing demonstrators are needed to promote industrial \emph{and} societal acceptance of wireless technology in safety-critical applications---\emph{there seems no alternative path to realizing the smart manufacturing vision}.

Concerning wireless control, the \emph{stability} of the feedback loops in a smart manufacturing system is arguably the most basic dependability property that has to be provably satisfied.
To guarantee closed-loop stability and also achieve the application-specific control performance, the ability to provide \emph{fast} feedback over a wireless communication network is essential to keep up with the dynamics of the physical (often mechanical) systems in smart manufacturing.  
For example, controlling the motion of a remote robot and coordinating a fleet of drones, as mentioned in the envisioned use cases from \secref{sec:vision}, require update intervals of tens of milliseconds~\cite{abbenseth2017cloud,Preiss2017}.
Such real-time requirements must also be met when wireless communication occurs over \emph{multiple hops}, which is key, for example, when large distances in mining or across a whole factory floor are to be covered. 

%\db{(pointers for `Challenges' section -- Case Mining Industry) Abstracting challenges: For wireless sensor network saving energy is important (low-power, event-triggered), multi-hop communication, safety-critical.
%For remote control from outside the mine we need fast update intervals over large distances (difficult environments, drones/agents can easily crash if signals come too late)}

\subsection{Adaptability}

Smart manufacturing systems also have to be \emph{highly adaptive} to support on-demand reconfigurability of conveyor belts, dynamic assignment of new tasks to individual drones, or the reformation of entire fleets of AGVs.
This means in particular that the systems must adapt at run time to changes in application requirements, hardware and software components, and dynamics of the environment in which they operate.
As a result of such adaptability, the number, types, and capabilities of agents, the algorithms they execute, and how they are interconnected can change in unforeseen and complex ways throughout the lifespan of a smart manufacturing system.

With regard to wireless control, a first step toward adaptability is to enable the system to switch at run time between a fixed set of \emph{operating modes}, while ensuring closed-loop stability and control performance.
A mode may correspond to a distinct manufacturing task or conveyor belt configuration, and \emph{mode changes} may be triggered by the cloud-based automation logic or the arrival of a product at a conveyor belt.
For an effective collaboration of multiple agents, a \emph{distributed implementation}, where every agent locally uses information received over the network to solve a common control task, is highly beneficial.
\emph{Multi-hop communication} supports unforeseen interconnections; for example, no additional access points must be installed if suddenly the need for a larger wireless coverage arises.

%Similarly, \emph{multi-hop communication} enables more flexible interconnections among agents, workers, and other infrastructure.

%\db{(pointers for `Challenges' section -- Case Autonomous Drones:) Challenges mainly in communication with mobile nodes and distributed control, for inspection tasks where images/video data are sent around also data sizes can be a challenge}
%\db{Here the AGV example with computation power at agent-level might be used.
% We need to be fast, local agents only have limited computational power, we have many agents and large physical spaces (2 floors).
% It's hard to bring the energy aspect in, I think it's communication is not that critical for energy in all the use cases we're mentioning.
% Might talk about additional remote sensors.}

\subsection{Efficiency}

The algorithms (control, optimization, learning, \etc), communication protocols, and other software components need to operate \emph{efficiently} because of limited compute power, memory capacity, and wireless bandwidth.
This is particularly relevant on the agent level, where embedded hardware such as low-cost microcontrollers and wireless radios are desirable for economic, size, or weight considerations.
Energy efficiency is not only a concern for battery-powered drones, remote energy-harvesting sensors, and the like, but also for cloud-based services~\cite{Masanet2020}. 

With regard to wireless control, models and methods are needed that increase control performance (which may translate into shorter manufacturing cycles and higher product quality) with the least resources.
To achieve this goal, controllers should, for example, \emph{save resources} whenever possible and \emph{reallocate} otherwise unused resources to avoid wasting them.

\section{State of the Art in Control over Wireless}
\label{sec:sota}

This section reviews efforts reported in the literature to solve the wireless control challenges outlined in the previous section.

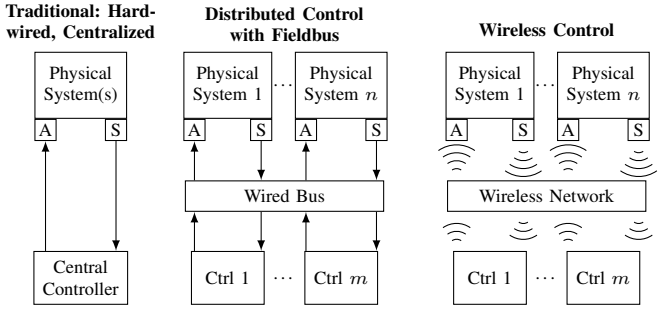
\begin{figure}
\centering
\tikzset{radiation/.style={{decorate,decoration={expanding waves,angle=0,segment length=3pt}}}}
\tikzset{>=latex}
\tikzsetnextfilename{Historical_sep}
\begin{tikzpicture}
\tikzset{font=\scriptsize}
 \node[draw,rectangle, minimum height = 2em,align=center](controller1){Central\\Controller};
% \node[draw,rectangle, minimum width = 4em, minimum height = 2em,right = 1em of controller1](controller2){Controller~\num{2}};
\node[draw,rectangle,minimum height = 2.5em,above = 5em of controller1,align = center](system1){Physical\\System(s)};
\node[draw,rectangle,below=0em of system1.south west, anchor = north west,inner sep=0,minimum width=0.8em, minimum height=0.8em](act1){A};
\node[draw,rectangle,below=0em of system1.south east, anchor = north east,inner sep=0,minimum width=0.8em, minimum height=0.8em](sens1){S};
\node[above=1.3em of system1.center,align=center]{\textbf{Traditional: Hard-}\\ \textbf{wired, Centralized} };
% \node[draw,rectangle,minimum width = 4em, minimum height = 2.5em,above = 5.8em of controller2,align = center](system2){Physical\\System~\num{2}};
% \node[draw,rectangle,below=0em of system2.south west, anchor = north west](act2){A};
% \node[draw,rectangle,below=0em of system2.south east, anchor = north east](sens2){S};
\draw[->] (controller1.north-|act1)--  (act1);
\draw[->] (sens1)--  (controller1.north-|sens1);
% \draw[->](controller2.north-|act2)--  (act2);
% \draw[->] (sens2)--  (controller2.north-|sens2);
\end{tikzpicture}
\tikzsetnextfilename{Historical_bus}
\begin{tikzpicture}
\tikzset{font=\scriptsize}
\node[draw,rectangle, minimum height = 2em, minimum width=2.75em](controller1){Ctrl~\num{1}};
\node[draw,rectangle, minimum height = 2em,minimum width=2.75em,right = 1.5em of controller1](controller2){Ctrl~$m$};
\node at($(controller1)!0.5!(controller2)$){\scriptsize{\ldots}};
\node[draw, rectangle, minimum width = 7.5em,above = 2.5em of $(controller1)!0.5!(controller2)$, minimum height = 1em](network){Wired Bus};
\node[draw,rectangle,minimum height = 2.5em,above = 5em of controller1,align = center](system1){Physical\\System~\num{1}};
\node[draw,rectangle,below=0em of system1.south west, anchor = north west,inner sep=0,minimum width=0.8em, minimum height=0.8em](act1){A};
\node[draw,rectangle,below=0em of system1.south east, anchor = north east,inner sep=0,minimum width=0.8em, minimum height=0.8em](sens1){S};
\node[draw,rectangle,minimum height = 2.5em,above = 5em of controller2,align = center](system2){Physical\\System~$n$};
\node[draw,rectangle,below=0em of system2.south west, anchor = north west,inner sep=0,minimum width=0.8em, minimum height=0.8em](act2){A};
\node[draw,rectangle,below=0em of system2.south east, anchor = north east,inner sep=0,minimum width=0.8em, minimum height=0.8em](sens2){S};
\node at($(system1)!0.5!(system2)$){\scriptsize{\ldots}};
\node[above=1.3em of $(system1)!0.5!(system2)$,align=center]{\textbf{Distributed Control}\\ \textbf{with Fieldbus}};
\draw[->] (controller1.north-|act1)--  (network.south-|act1);
\draw[->] (network.south-|sens1)--  (controller1.north-|sens1);
\draw[->](controller2.north-|act2)--  (network.south-|act2);
\draw[->] (network.south-|sens2)--  (controller2.north-|sens2);
\draw[->] (network.north-|act2)--  (act2.south);
\draw[->](sens2.south)--  (network.north-|sens2);
\draw[->] (network.north-|act1)--  (act1.south);
\draw[->] (sens1.south)--  (network.north-|sens1);
\end{tikzpicture}
\hspace{0.3cm}
\tikzsetnextfilename{Historical_wireless}
\begin{tikzpicture}
\tikzset{font=\scriptsize}
\node[draw,rectangle, minimum height = 2em,minimum width=2.75em](controller1){Ctrl~\num{1}};
\node[draw,rectangle, minimum height = 2em,minimum width=2.75em,right = 1.5em of controller1](controller2){Ctrl~$m$};
\node at($(controller1)!0.5!(controller2)$){\scriptsize{\ldots}};
\node[draw, rectangle, minimum width = 7.5em,above = 2.5em of $(controller1)!0.5!(controller2)$, minimum height = 1em](network){Wireless Network};
\node[draw,rectangle,minimum height = 2.5em,above = 5em of controller1,align = center](system1){Physical\\System~\num{1}};
\node[draw,rectangle,below=0em of system1.south west, anchor = north west,inner sep=0,minimum width=0.8em, minimum height=0.8em](act1){A};
\node[draw,rectangle,below=0em of system1.south east, anchor = north east,inner sep=0,minimum width=0.8em, minimum height=0.8em](sens1){S};
\node[draw,rectangle,minimum height = 2.5em,above = 5em of controller2,align = center](system2){Physical\\System~$n$};
\node[draw,rectangle,below=0em of system2.south west, anchor = north west,inner sep=0,minimum width=0.8em, minimum height=0.8em](act2){A};
\node[draw,rectangle,below=0em of system2.south east, anchor = north east,inner sep=0,minimum width=0.8em, minimum height=0.8em](sens2){S};
\node at($(system1)!0.5!(system2)$){\scriptsize{\ldots}};
\node[above=1.5em of $(system1)!0.5!(system2)$,align=center]{\textbf{Wireless Control}};
\draw[radiation,decoration={angle=35}] ([shift={(0,0.05cm)}]controller1.north-|act1)--  ([shift={(0,-0.05cm)}]network.south-|act1);
\draw[radiation,decoration={angle=35}] ([shift={(0,-0.05cm)}]network.south-|sens1)--  ([shift={(0,0.05cm)}]controller1.north-|sens1);
\draw[radiation,decoration={angle=35}] ([shift={(0,0.05cm)}]controller2.north-|act2)--  ([shift={(0,-0.05cm)}]network.south-|act2);
\draw[radiation,decoration={angle=35}] ([shift={(0,-0.05cm)}]network.south-|sens2)--  ([shift={(0,0.05cm)}]controller2.north-|sens2);
\draw[radiation,decoration={angle=35}] ([shift={(0,0.05cm)}]network.north-|act2)--  ([shift={(0,-0.05cm)}]act2.south);
\draw[radiation,decoration={angle=35}] ([shift={(0,-0.05cm)}]sens2.south)--  ([shift={(0,0.05cm)}]network.north-|sens2);
\draw[radiation,decoration={angle=35}] ([shift={(0,0.05cm)}]network.north-|act1)--  ([shift={(0,-0.05cm)}]act1.south);
\draw[radiation,decoration={angle=35}] ([shift={(0,-0.05cm)}]sens1.south)--  ([shift={(0,0.05cm)}]network.north-|sens1);
\end{tikzpicture}
\vspace{-5mm}
\caption{Evolution of control architectures.
%: from traditional centralized to wireless control.
\capt{Physical systems with sensors (S) and actuators (A) are connected to controllers (Ctrl) over point-to-point wires (left), a wired bus (center), or a wireless network (right).}}
%
% ST: Previous version
%\caption{Different control architectures showing multiple systems with sensors (S) and actuators (A) connected to controllers (Ctrl) over point-to-point wires (left), wired bus (center), or wireless network (right).}
%\vspace{-4mm}
\label{fig:historical}
\end{figure}

\begin{table*}
\caption{Design space of wireless \cps that have been evaluated on physical platforms and real wireless networks.}
\label{tab:rel_work}
\centering
\begin{tabular}{cccccccc}
\toprule
& \multicolumn{2}{c}{\hspace{3em} \textbf{Dependability}} & \multicolumn{3}{c}{\hspace{-2em} \textbf{Adaptability}} & \multicolumn{2}{c}{\textbf{Efficiency}} \\[0.5em]
\multirow{2}{*}{Work} & Stability & Fast update & Multi- & Mode & Distributed & Reallo- & Resource \\
& analysis & intervals & hop & changes & implementation & cation & savings\\
\midrule
\cite{araujo2011self,Araujo2014} & \cmark & \xmark & \xmark & \xmark & \xmark & \cmark & \cmark \\
\cite{Eker2001} & \cmark & \cmark & \xmark & \xmark & \xmark & \xmark & \xmark \\
\cite{Ploplys2004} & \cmark & \cmark & \xmark & \xmark & \xmark & \xmark & \xmark \\
\cite{Bauer2014} & \cmark & \cmark & \xmark & \xmark & \xmark & \xmark & \xmark \\
\cite{heshmati2014self} & \cmark & \cmark & \xmark & \xmark & \xmark & \xmark & \cmark \\
\cite{santos2014adaptive,santos2015aperiodic} & \cmark & \cmark & \xmark & \xmark & \xmark & \xmark & \xmark\\
\cite{stanoev20closedloop} & \xmark & \cmark & \xmark & \xmark & \xmark & \xmark & \xmark \\
\cite{Hernandez2011} & \xmark & \cmark & \xmark & \xmark & \xmark & \xmark & \xmark \\
\cite{Ceriotti2011} & \xmark & \xmark & \cmark & \xmark & \xmark & \xmark & \xmark \\
\cite{Saifullah2014} & \xmark & \xmark & \cmark & \xmark & \xmark & \cmark & \cmark \\
\cite{ma2018efficient} & \xmark & \xmark & \cmark & \xmark & \xmark & \xmark & \cmark\\
\rowcolor{lightgray!50!white}\cite{baumann2019fast} & \cmark & \cmark & \cmark & \cmark & \cmark & \xmark & \xmark \\
\rowcolor{lightgray!50!white}\cite{baumann2019control} & \xmark & \cmark &  \cmark & \xmark & \cmark & \cmark & \cmark\\
\bottomrule
\end{tabular}
\end{table*}

%\subsection{Evolution of Control Architectures}

Such efforts are required as closing feedback loops over a wireless network can make the control design significantly more difficult compared to traditional control architectures~\cite{yang2005networked,LuGr14,HeNaXu07,BaAn07}.
Traditionally, sensors and actuators are connected to a centralized controller through \emph{point-to-point wires} (\figref{fig:historical}, left) \cite{BaAn07}.
Although centralized control is beneficial because the controller has global information, it is often impractical for large-scale systems.
An alternative is decentralized control, where the system is split into a number of subsystems, each connected to a local controller, but without signal transfer between them~\cite{yang2005networked,LuGr14}.
However, decentralized control can exhibit poor performance, and it may not even be possible to achieve closed-loop stability~\cite{yang2005networked}.
Hence, communication networks in the form of \emph{wired busses} were introduced~\cite{BaAn07} (\figref{fig:historical}, center), which are still widely used in automation and control today \cite{Th05,trimpeCSM12}.
Replacing a wired bus with a \emph{wireless network} (\figref{fig:historical}, right), as required for smart manufacturing to close the sensing-computing-actuation loops shown in \figref{fig:smartManufacturingAbstraction}, leads to communication imperfections including longer transmission delays, larger jitter, as well as higher and correlated message loss (see \secref{sec:approach} for a detailed discussion).

Over the past two decades, a lot of research in the control community has considered these communication imperfections in the form of \emph{theoretical} stability analyses for different control architectures, transmission delay distributions, message loss models, \etc
Numerous \emph{simulation} case studies have also been carried out, for instance, based on the WirelessHART industrial standard~\cite{Li2016,Ma2018}.
This line of research has led to a principled understanding of the wireless control problem, and excellent surveys~\cite{Hespanha2007,Zhang2013} review the developed approaches and results.

However, as discussed in \secref{sec:sota_industry}, one of the primary concerns in industry today is a lack of trust in the reliability of wireless solutions.
This trust can only be established by complementing rigorous theoretical analyses with real experiments on realistic cyber-physical testbeds~\cite{Lu2016a}.
Therefore, we focus our discussion in the following on control-over-wireless approaches that have been validated on physical platforms and real wireless networks.

\tabref{tab:rel_work} qualitatively compares such approaches currently found in the literature by assessing whether the wireless control challenges outlined in \secref{sec:challenges} have been addressed.
Although these constitute basic challenges toward realizing the smart manufacturing vision and many challenges are yet to be solved, as discussed in \secref{sec:outlook}, we can see that none of the existing approaches is capable of addressing all basic challenges.

Having a closer look at \tabref{tab:rel_work}, we observe that only~\cite{baumann2019fast,baumann2019control} support a \emph{distributed implementation}\footnote{Another work that demonstrated distributed control over wireless is~\cite{mager2019feedback}, a prior conference version of~\cite{baumann2019fast}. 
In general,~\cite{mager2019feedback} provides the same properties as~\cite{baumann2019fast}, the main difference being that~\cite{mager2019feedback} does not support mode changes.}.
Distributed implementation is understood in the sense that agents are equipped with local computing units and use their own sensor readings \emph{and} information received over the communication network to locally compute new actuation commands.
All other approaches support only a central computing unit that computes the new actuation commands for all agents.
However, support for both architectures is needed to enable the use cases from \secref{sec:vision}.

Further, motion control for mobile robots, for example, needs to happen at \emph{fast update intervals} (\ie tens of milliseconds) to keep up with the physical dynamics of the agents.
At the same time, in large factories or for remote control in mines, large distances are to be covered, requiring \emph{multi-hop communication}.
We see that apart from~\cite{baumann2019fast,baumann2019control} no solution can cater for this crucial requirement.
References~\cite{araujo2011self,Araujo2014} control a double-tank system, where update intervals of around \SI{1}{\second} are sufficient, over a single-hop network.
Several works consider different variants of inverted pendulum systems~\cite{Eker2001,Ploplys2004,Bauer2014,stanoev20closedloop,Hernandez2011} or mobile robots~\cite{heshmati2014self,santos2014adaptive,santos2015aperiodic}, demanding update intervals of at most tens of milliseconds.
These approaches are all restricted to single-hop networks.
Wireless control over multi-hop networks is demonstrated in~\cite{Ceriotti2011}, where the lightning in an operational road tunnel is controlled, in~\cite{Saifullah2014} which considers power capping management in a data center, and in~\cite{ma2018efficient} where Matlab simulations of physical plants are controlled.
The update intervals in those setups are on the order of several seconds.
While the majority of the works that consider control over single-hop networks provide some kind of stability analysis, only~\cite{mager2019feedback,baumann2019fast} can do this for multi-hop networks.
However, since a practical demonstration can never cover all potential situations that may be encountered during a real execution in a smart factory, a theoretical analysis is absolutely indispensable.
Naturally, this becomes more challenging under multi-hop communication.
Also, the demand for (basic) adaptation at run time through \emph{mode changes} (\eg changes between a fixed set of application tasks) are exclusively demonstrated in~\cite{baumann2019fast}.

% A practical end-to-end evaluation is essential to establish trust in wireless control solutions.
% Nevertheless, since a practical demonstration can never cover all potential situations that may be encountered during a real execution in a smart factory, a theoretical analysis is absolutely indispensable.
% Nearly half of the works in \tabref{tab:rel_work} provide a theoretical stability analysis, yet only~\cite{baumann2019fast} does this for a multi-hop communication network.
% A stability analysis for a wireless \cps needs to take the key properties of the wireless system into account.
% Naturally, this becomes more challenging under multi-hop communication.

Another key characteristic of smart factories is that many agents need to transmit information over the wireless channel, such as their current positions, latest sensor values, or actuation commands.
Moreover, at least some of the agents and infrastructure sensors attached to walls or moving equipment in a smart factory may be powered by batteries.
As the bandwidth of a wireless channel is limited, and wireless radios draw considerable power, messages should only be sent if necessary.

Motivated by this, the control community has developed a variety of event-triggered and self-triggered methods~\cite{heemels2012introduction,miskowicz2018event}.
Both methods send information only upon the occurrence of certain events, such as an error growing too large.
Event-triggered control algorithms take this decision instantaneously.
Thus, the communication system cannot reallocate resources that are not needed by the control system.
To enable this, the control system needs to inform the communication system in advance about its communication demands.
This can be realized through self-triggered control, where at each communication instant the control system already decides when it needs to transmit information the next time.
\tabref{tab:rel_work} therefore also lists integrations of self-triggered control with communication systems to enable \emph{resource savings} and \emph{reallocation}.
Next to self-triggered approaches,~\cite{Saifullah2014} considers a network manager which checks the communication demands of all agents and assigns slots accordingly.
However, as the update intervals considered in~\cite{Saifullah2014} are \SI{20}{\second} or longer, this solution is not applicable to the fast feedback loops in future smart manufacturing systems.
In summary,~\cite{baumann2019control} is the only work that can combine resource reallocation and resource savings with fast update intervals, multi-hop communication, and a distributed implementation.

Our analysis of the state of the art shows that~\cite{baumann2019fast,baumann2019control} are superior to other prior work.
Thus, in the following sections, we illustrate the approach presented in~\cite{baumann2019fast,baumann2019control} in more detail.

\begin{figure*}[!tb]
 \centering
 \includegraphics[width=\linewidth]{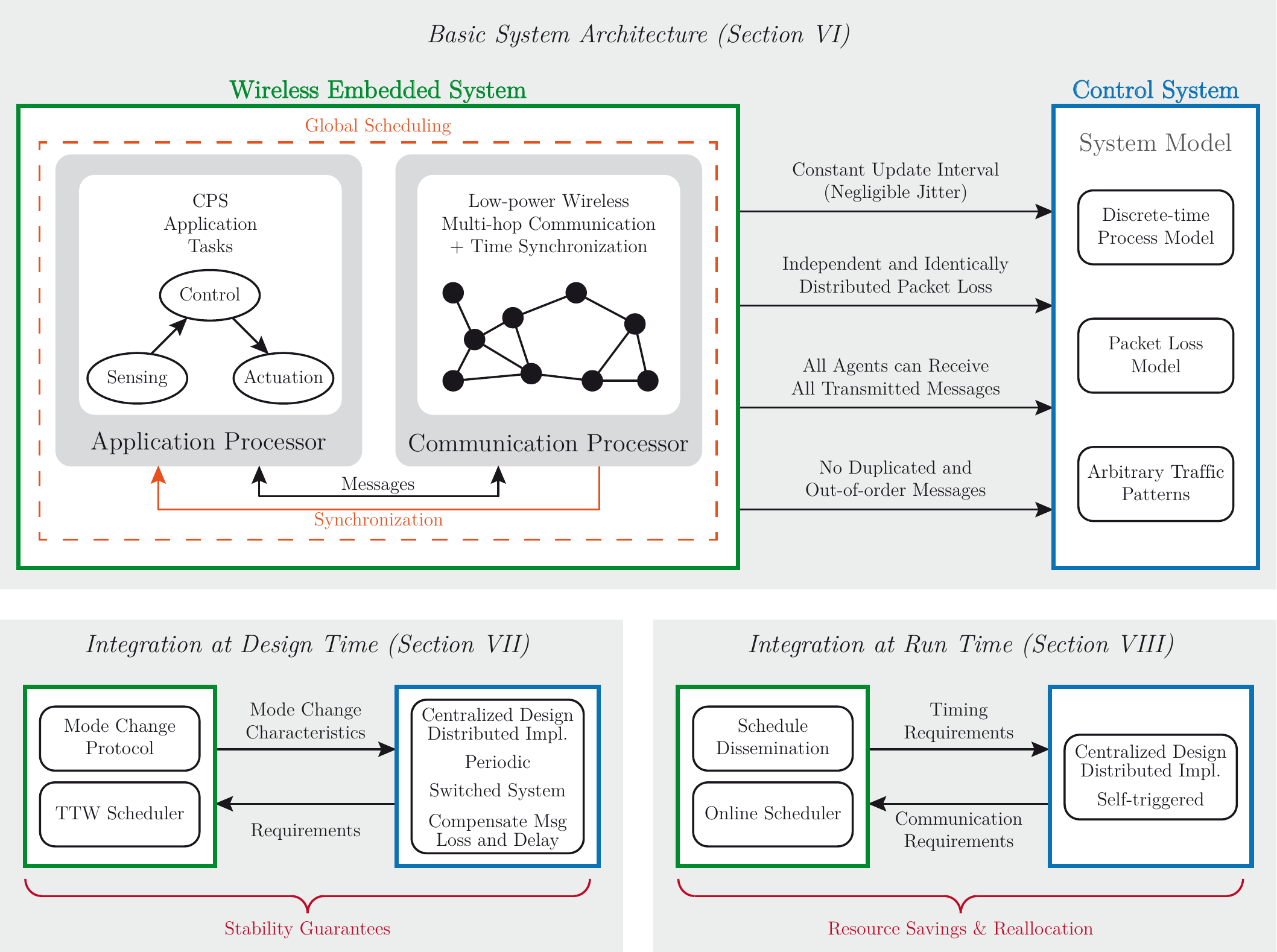}
 \vspace{-5mm}
 \caption{Overview of system architecture with co-design and integration approaches.}
 \label{fig:overview}
\end{figure*}

\section{Holistic Approach: Overview}
\label{sec:approach}
%\mz{Citations still need to be added.}

% To address the above-mentioned challenges and limitations of prior work, we argue for a \emph{holistic approach} that tightly integrates the control and wireless communication systems with all their hardware and software components both at design time and at run time.
In \secref{sec:challenges}, we identified three main challenges that need to be addressed to close the gap between vision and reality of smart manufacturing, while in \secref{sec:sota} we discussed to which extent those challenges are addressed by the current state of the art in research.
From this discussion, the approaches proposed in recent works~\cite{baumann2019fast,baumann2019control} resulted to be the only approaches that successfully address these challenges.
Those works propose a \emph{holistic approach} that tightly integrates the control and wireless communication systems with all their hardware and software components, both at design and at run time.
% To move towards closing the large gap between envisioned and currently realizable applications of smart manufacturing, we propose a \emph{holistic approach} that tightly integrates the control and wireless communication systems with all their hardware and software components, both at design and at run time.

The reasons why prior solutions
% Seb: at first, "solutions" was "limitations", which didn't make sense to me
 fail to address the above-mentioned challenges
%\db{check which word is used in earler sections}
lie in the imperfections of the wireless communication medium as compared to cable-based solutions:
\begin{itemize}
 \item[\textbf{C1}] \emph{Limited throughput.} Depending on the wireless technology used, the number of messages carrying sensor readings and control signals that can be exchanged among distributed agents per unit of time can be significantly lower compared with a wired communication system. Also, the throughput in a multi-hop network is generally lower than in a single-hop network where all agents can directly communicate with one another. Fundamental trade-offs between achievable throughput, communication range, and energy cost require non-trivial system design decisions to be able to meet the dependability and efficiency requirements.
 \item[\textbf{C2}] \emph{Unpredictable delays.} Communication delays hinder the system-wide coordination of distributed entities, can impair control performance, and can even make it impossible to stabilize a feedback system, especially when the delays vary unpredictably fashion (\ie distribution of the delays is unknown). Such unpredictable, time-varying delays are the norm rather than an exception in wireless networks, caused by, for example, retransmissions of lost messages, varying dwell times in message queues, and dynamic changes in the routing paths or communication schedules.
 \item[\textbf{C3}] \emph{Constrained communication patterns.} Although wireless communication is broadcast in nature, allowing all devices in communication range of a sender to receive a message if they have their radio turned on, the need for reliable and efficient communication has led to many protocols that support only specific communication patterns. For instance, in WirelessHART, all messages need to pass through the central gateway, which makes an effective coordination of all agents in a cell highly inefficient. Moreover, constrained communication patterns make it more challenging to provide certain dependability properties: Well-known fault-tolerance mechanisms~\cite{schneider90} rely on the ability to share information among a set of agents (\eg three agents, where one acts as the primary controller and the other two act as back-up controllers that seamlessly take over when the primary controller fails) and closed-loop control without visibility into the global system state can lead to poor performance and render stability infeasible~\cite{blondel00}.
 \item[\textbf{C4}] \emph{Correlated message loss.} It has been shown that communication in both wireless~\cite{srinivasan10} and wired~\cite{yajnik99} networks is often subject to correlated message loss, that is, multiple consecutive messages can be lost in a ``burst'' and the maximum length of these bursts is extremely difficult to predict. As correlated message loss interrupts a feedback loop in an unpredictable way, it can be very hard if not intractable to perform a valid closed-loop stability analysis.
 % \item[\textbf{C5}] \emph{Duplicated and out-of-order messages.} Retransmissions, dynamic (multi-path) routing, and opportunistic communication schemes are widely employed by current wireless systems, mainly to achieve high reliability.
 % However, these mechanisms are often the root cause of message duplicates and out-of-order message deliveries, which do not only impair efficiency but also hinder the design and analysis of feedback controllers and fault-tolerance mechanisms.
\end{itemize}

As indicated in \figref{fig:overview}, our holistic approach starts by taming those imperfections to the extent possible on the wireless embedded system side.
The resulting key characteristics are then embedded into a common system model that also captures the dynamics of the physical systems to be controlled.
% First of all, we need a common underlying system model capturing the key characteristics of the wireless embedded and the physical system to be controlled.
%\db{I think the holistic approach starts with a common system model. This is also important since many other approaches only care about either communication and control system and assume an idealistic model for the other part. 'First of all, we need a common underlying system model capturing the key characteristics of the wireless embedded and the physical system to be controlled.'}
Such a unified underlying model allows for the co-design of the control and communication systems based on relevant mutual properties. For instance, the control system needs to countervail inherent imperfections of wireless communication, such as message
% \fm{At the beginning of the paper we only talk about "messages" but from this point on, the term "packet" is also used often. I would not suggest changing all occurrences of one term with the other. Instead, I added a short footnote here to clarify this for the reader.}
loss and delays, while the communication system needs to account for the traffic demands of the control system.

Since all of this can be done prior to execution, we call this \emph{integration at design time}.
%\db{Then as a next point we need the integration at design time (I would use 'integration at design time' and 'integration at run time' since they are also used in the figure). Here I would try to be a bit more concrete. 'Next, control and communication systems must be co-designed based on relevant mutual properties. For instance, the control system needs to countervail inherent imperfections of wireless communication, such as message loss and delays, while the communication system needs to account for the traffic demands of the control system. Since all of this can be done prior to execution, we call it \emph{integration at design time}.'}
Nevertheless, at run time, the control and communication systems must adapt their functioning in a timely and safe manner to changes in the mutual properties to keep satisfying the requirements. Such changes can be due to varying application tasks, hardware or software updates, or environment dynamics, including benign and malicious ones. For example, often interference-free operation of a wireless \cps cannot be guaranteed. To keep the smart manufacturing system in a safe state, the control system needs to adapt and put more focus on robustness to reduce communication demands, while possibly sacrificing control performance.
%\db{I think this is a bit too abstract and it's unclear what 'reducing its functionality' could mean. I would here talk about sacrificing performance by changing to a more robust and less communication-hungry control strategy. 'cannot be guaranteed. To keep the smart manufacturing system in a safe state, the control system needs to adapt and put more focus on robustness to reduce communication demands, while possibly sacrificing performance.'}
In this way, dependability can be ensured without sacrificing adaptability and efficiency, yet it requires a tight integration of control and communication both at design time and \emph{at run time}.

Over the past three years, we have made significant progress toward such a holistic approach. \figref{fig:overview} illustrates the overall architecture of the wireless \cps along with our co-design and run time integration approaches. The basic system architecture, described in \secref{sec:architecture}, embodies the fundamental idea underlying our approach: Address challenges \textbf{C1}--\textbf{C4} to the extent possible in the design of the wireless embedded system such that the system model can account for them in an accurate and tractable way. The integration of the control and wireless embedded systems at design time and run time can then be achieved using straightforward methods, as detailed in Secs.~\ref{sec:design_time} and \ref{sec:run_time}.

As a result, we were able to demonstrate for the first time reliable and efficient control and coordination of multiple physical systems over low-power wireless multi-hop networks at update intervals of \SIrange[range-units=single, range-phrase=-]{20}{50}{\ms}~\cite{mager2019feedback}. Specifically, through integration at design time, we obtained formal closed-loop stability guarantees even while the wireless \cps adapts at run time to varying application needs and network dynamics~\cite{baumann2019fast}. Through integration at run time, we have shown significant bandwidth and energy savings as well as the dynamic reallocation of otherwise unused resources~\cite{baumann2019control}. With this, our holistic approach demonstrated the capability to meet certain dependability, adaptability, and efficiency requirements of future smart manufacturing that were impossible before.

\fakepar{Road map}
The following three sections describe our holistic approach in more detail. Afterward, \secref{sec:testbed} describes a full-fledged cyber-physical testbed we designed to systematically evaluate our approach on physical platforms and real wireless networks. \secref{sec:eval} presents experimental results that illustrate the capabilities of our approach in a range of scenarios that are representative of future smart manufacturing systems. While we believe that our work represents an important step toward realizing the smart manufacturing vision, several important challenges still remain to be solved. \secref{sec:outlook} discusses some of these challenges and opportunities for future research.
Essential parts of our work are available as open source (see \tabref{tab:open-source}).

\begin{table}[!tb]
\addtolength{\tabcolsep}{-1.3pt}
\caption{Availability of the different components of the presented approach and cyber-physical testbed}
\label{tab:open-source}
\begin{tabular}{p{0.175\textwidth}p{0.275\textwidth}}
\toprule
Component & Available \\
\midrule
Dual processor platform & \url{https://gitlab.ethz.ch/tec/public/dpp/-/wikis/home}\\
Inverted pendulum systems & Quanser systems (proprietary):\\
&\url{https://www.quanser.com/products/linear-servo-base-unit-inverted-pendulum/},\\
&Hardware details for the self-built pendulum systems and the interface between pendulums and the DPP are available on request.\\
Communication code & The communication code is based on Glossy, see \url{https://github.com/ETHZ-TEC/LWB}. \\
TTW scheduling framework & \url{https://github.com/romain-jacob/TTW-Artifacts}\\
Control code & The details of the control algorithms can be found in~\cite{mager2019feedback,baumann2019fast,baumann2019control}. Their implementation is highly specific to our hardware and physical systems and available on request.\\
\bottomrule
\end{tabular}
\addtolength{\tabcolsep}{1.3pt}
\end{table}

\section{Basic Architecture and System Model}
\label{sec:architecture}

% \db{In the following sections, the level of technical detail should be signficantly reduced. I will do this for the control related part, @Fabian: Could you please take care of the communication parts? The Wireless Embedded System part of this Section is currently two pages long, it would be great if this could be condensed to one page. In the other sections, the wireless part is already pretty short, but it would still be good if you could check whether we can leave out some details so simplify content}

The foundation of our approach toward building a dependable, adaptive, and efficient wireless control system is a wireless embedded system architecture that facilitates the co-design of communication and control at design time and at run time.
To realize this co-design, we proceeded in two steps:
\begin{enumerate}
  \item mitigate imperfections (\ie delay, jitter, packet loss) on the wireless embedded system side to the extent possible;
  \item transfer the resulting properties of the wireless system into a system model that is used for the control design.
\end{enumerate}

To this end, we designed a wireless embedded system consisting of a hardware platform, a wireless communication protocol, and a global scheduling policy at its heart.
We provide a brief overview of these components in the following; more detailed descriptions can be found in \cite{mager2019feedback,baumann2019fast}.

\subsection{Wireless Embedded System}
\label{sec:basicWirelessEmbeddedSystem}

\vspace{-1mm}
\fakepar{Hardware platform}
One of the key building blocks of our wireless embedded system is the hardware platform.
It combines all necessary components while providing the mechanisms needed to achieve the goals of dependability, efficiency as well as predictability (see challenge \textbf{C2}).
From a hardware perspective, we argue for low-power and low-cost commodity hardware as it can be deployed numerous and almost everywhere due to its small form factor. As discussed in \secref{sec:abstract_view}, we find multiple interconnected sensing-computing-actuation units in typical \cps applications. Thus, we have two kinds of tasks: the application task, which consists of sensing, computing, and/or actuation, and the communication task.
Both tasks are essentially independent of each other, except that the application generates the data to be communicated, and may execute concurrently with vastly different workloads.
% Although the data to be communicated are generated by the application (\eg by measuring the current system state through sensors), both tasks may execute concurrently with vastly different workloads.

We thus leverage a custom-built heterogeneous dual processor platform (\dpp).
This DPP features a communication processor~(\cp) and a more powerful application processor~(\ap).
%The latter is the MSP432P401R, a 32-bit low-power microcontroller running at 48 MHz and serving the more frequent and compute-intensive application tasks. The \cp is the CC430F5137, a 16-bit low-power microcontroller running at 13 MHz with an integrated low-power RF transceiver that operates at 250 kbit/s in the 868 MHz frequency band.
Because application and communication tasks execute on different processors, we avoid any resource interference between them, which could otherwise lead to unpredictable execution delays.
Furthermore, the data exchange between \ap and \cp is handled by \bolt\cite{Sutton2015a}, a processor interconnect with formally verified worst-case execution times, a critical property to enable real-time operation and to mitigate challenge \textbf{C2}.
% Because application and communication tasks execute on different processors, we avoid any resource competition between them, which could otherwise lead to unpredictable execution delays, see challenge \textbf{C2}.
% The data exchange between \ap and \cp is handled by \bolt\cite{Sutton2015a}, a processor interconnect that decouples both processors in terms of timing interference, power, and clock domains. This decoupling simplifies the software design and increases dependability and efficiency of operation while allowing for a high degree of adaptability.
% Even more important, \bolt provides formally verified worst-case execution times for the data exchange between \ap and \cp, a critical property to enable real-time operation and to mitigate challenge \textbf{C2}.

\fakepar{Wireless communication protocol}
%The requirements for the communication protocol largely depend on the considered control application. As described in \secref{sec:challenges}, they must satisfy the needs for dependability, adaptability, and efficiency.
In general, communication must bereliable enough to guarantee a safe execution of \cps, as captured by challenges \textbf{C2} (unpredictable delays) and \textbf{C4} (correlated message loss).
The available bandwidth should be sufficiently large
%and adaptive
in order to keep up with the dynamics of the physical processes to be controlled (see challenge \textbf{C1}).
Those requirements also need to be met when large distances are covered, that is, when messages must travel multiple hops.
%, \ie also under multi-hop communication or handover mechanisms, depending on the wireless technology that is used.
Finally, the protocol needs to ensure consistent functioning also when agents are moving in the network, under changes in the environment, or under dynamically changing communication patterns (see challenge \textbf{C3}).
In such cases, we are facing a dynamic network where the quality of communication links continuously changes, and links appear and disappear.

%\fm{Reviewer 3 asked for a quick discussion of communication protocols and a better introduction of the protocol that we use. I added this here, after we have named the communication requirements, because I think this does not really fit into the related work section where we discuss whole CPS solutions. I decided not to discuss specific protocols but rather refer to the benchmark competition as it becomes clear that Glossy based solutions are clearly the best performing protocols. Do you think this is sufficient to answer the reviewers comment?}
The research community and industry have developed numerous low-power wireless communication protocols over the past few years.
Comparing them is non-trivial as they were evaluated in different environments, on different hardware platforms, and using different experimental use cases.
Efforts have recently been made to benchmark some of these protocols~\cite{boano18}.
In particular, the results of the annual EWSN Dependability Competition~\cite{schuss17competition} clearly show that, under various considered benchmark scenarios, stateless flooding-based solutions have achieved the best performance in terms of latency, reliability, and energy efficiency, outperforming protocols based on stateful routing~\cite{boano17competitionResults}.
All top-three solutions in the competitions have in common that they are based on a variant of Glossy~\cite{Ferrari2011}.
% Since most existing protocols only support parts of these requirements, we present a wireless protocol based on Glossy floods~\cite{Ferrari2011}.

Glossy is a multi-hop communication primitive based on synchronous transmissions~\cite{zimmerling20survey}.
It inherently supports dynamic networks as it operates topology-agnostic, \ie independent of the underlying network state.
In a Glossy flood, one message is disseminated from one node to all other nodes in a (multi-hop) network with a reliability that has been found to be above 99.9\,\% on various real testbeds with more than 200 nodes~\cite{ferrari12lwb}.
The time required for this is close to the theoretical minimum for half-duplex radios.
In the rare event of message loss, it has been shown that these losses can be safely approximated by an \iid Bernoulli process~\cite{Zimmerling2013,Karschau2018}, which is an essential property for the control design.
Glossy can also time-synchronize all nodes in the network with an accuracy of a few microseconds.

% !TEX root = ../paper.tex
\begin{figure}[!tb]
	\centering
	\includegraphics[width=\linewidth]{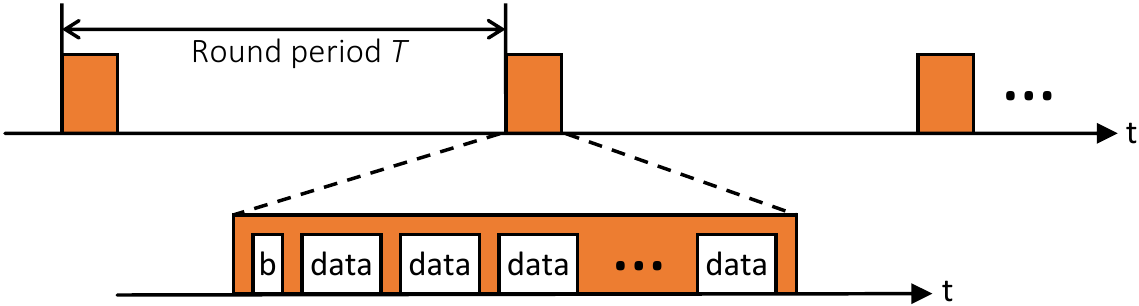}
	%\vspace{-3mm}
	\caption{Time-triggered operation of the low-power wireless protocol. \capt{Communication occurs in rounds that are scheduled with a given round period $T$. In the time interval between rounds, there is no communication between nodes, \ie communication sleeps for energy efficiency. Every beacon (b) and data slot corresponds to an one-to-all Glossy flood~\cite{Ferrari2011}.}} % \capt{Communication occurs in rounds with a given round period $T$. Every beacon (b) and data slot in a communication round is a network-wide Glossy flood.}}
	%\vspace{-2mm}
	\label{fig:overviewWirelessProtocol}
\end{figure}

Our wireless protocol is also based on Glossy.
As illustrated in \figref{fig:overviewWirelessProtocol}, nodes communicate in periodic rounds with period $T$ and sleep in between to save energy.
This is possible because of Glossy's time synchronization, which is used to schedule the start of the communication rounds at every node.
Each round consists of a series of \emph{slots}.
In each slot, data are sent by a single node using a Glossy flood.
The \emph{beacon slot} is always initiated by a dedicated node, called \emph{host}, serving as synchronization point for all other nodes.
All other communication slots after the beacon are \emph{data slots}, which contain the actual application messages,  one slot for each message.
As a result, one round realizes the data exchange required for one control cycle within a few tens of milliseconds, which makes our protocol applicable for many control applications and especially for fast feedback control.
Moreover, since every message can be received by all nodes, any communication pattern is inherently supported (see challenge \textbf{C2}). % by enabling nodes to simply keep only the relevant packets.
%Moreover, the presented wireless protocol can realize any traffic pattern because messages are always delivered to all nodes, which can then decide about keeping or dropping them.
%and can be filtered out, \eg based on node IDs, afterward.
This significantly simplifies the control design as it enables individual nodes to work based on global knowledge.
Although this kind of broadcast-only communication may seem wasteful, many solutions based on synchronous transmissions that have been developed over the past few years demonstrate that this approach can outperform traditional routing-based approaches by a significant margin in terms of efficiency and reliability across a wide range of real-world scenarios~\cite{zimmerling20survey}. 
% \mz{Revised the text and added a reference.}

% \input{img/schedulingFramework}

\fakepar{Global scheduling}
In order to provide end-to-end guarantees for the overall \cps, a global scheduling of the different tasks executing on the two processors of the DPP platform and a time-predictable wireless data exchange is needed to cope with delays and jitter (see challenge \textbf{C2}).
The time needed from sensing the state of the system until applying the corresponding control command is the end-to-end delay \Tdelay.
Intervals between consecutive sensing and actuation tasks define the update interval \Tupdate.
An appropriate task scheduling achieves two goals:
\begin{enumerate}
  \item the most recent data is available for communication;
  \item variations of \Tupdate and \Tdelay are minimized to the extent that both can be bounded by a small jitter, which considerably simplifies the control design and the further analysis of the overall \cps (\eg in terms of closed-loop stability).
\end{enumerate}
The basis of our scheduling is an accurate global time reference provided by the communication protocol, at every node.
An exemplary schedule and further details are explained in \cite{mager2019feedback}.

\fakepar{Essential properties}
With the presented wireless embedded system, the control design can build on the following properties:
\begin{itemize}
  \item[\textbf{P1}] The theoretical worst-case jitter on \Tupdate and \Tdelay is bounded by \SI{\pm50}{\us} for update intervals of up to \SI{100}{\ms}~\cite{mager2019feedback}.
  \item[\textbf{P2}] Message losses across the wireless network can be safely approximated by an \iid Bernoulli process~\cite{Zimmerling2013, Karschau2018}.
  \item[\textbf{P3}] The application or control logic can make use of arbitrary communication patterns over a multi-hop network.
  \item[\textbf{P4}] Message duplicates and out-of-order message deliveries are impossible by design.
\end{itemize}

\subsection{Control System}

\begin{figure}
\centering
\subfloat[remote control\label{sfig:ctrl}]{%
       \includegraphics[height=0.15\textheight]{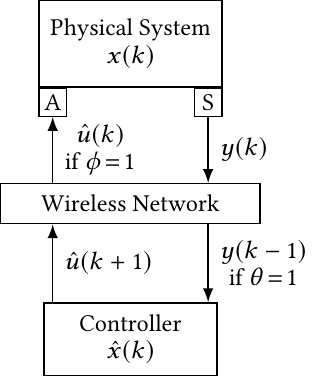}}
    \hfill
  \subfloat[distributed control\label{sfig:sync}]{%
        \includegraphics[height=0.15\textheight]{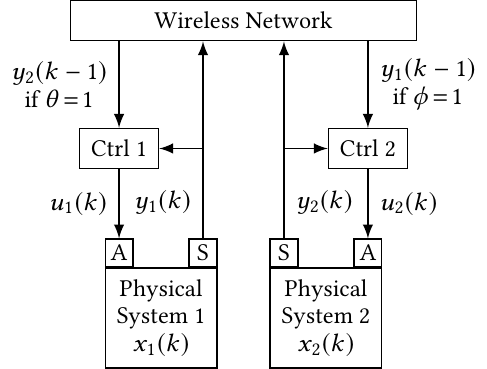}}
  \caption{Block diagram of the overall control system. We consider two scenarios, namely remote control (\figref{sfig:ctrl}) and distributed control (\figref{sfig:sync}).}
  \label{fig:WirelessSystemModel}
\end{figure}

In this section, we present a mathematical model of the overall \cps capturing the key properties that result from the wireless embedded system design. We consider two settings, depicted in \figref{fig:WirelessSystemModel}: remote control and distributed control.
We first discuss the common underlying system model and then comment on differences between both settings.

The mathematical model shall describe the physical system and the key properties of the wireless embedded system. 
As stated in property \textbf{P1}, the jitter is orders of magnitude lower than the update intervals of tens of milliseconds we target with mechanical systems.
Thus, the jitter can be safely neglected for control design~\cite[p.~48]{cervin2003integrated}.
This allows us to model the system in discrete time, where one discrete time step corresponds to one communication round. For each controlled physical system $i$, we assume linear and time-invariant (LTI) dynamics,
%\ie we have
\begin{subequations}
\begin{align}
x_i(k+1) = A_ix_i(k) + B_iu_i(k) + v_i(k),
\end{align}
with $x_i(k)\in\R^n$ the state of the system, $u_i(k)\in\R^m$ the control input, $v_i(k)\in\R^n$ a normally distributed random variable (capturing, for instance, model uncertainty), $k\in\mathbb{N}$ the discrete time index, and $A_i$ and $B_i$ matrices of appropriate dimensions.
Further, we assume that we can measure the full state, but measurements are corrupted by noise,
\begin{align}
y_i(k) = x_i(k) + w_i(k),
\end{align}
\end{subequations}
where $w_i(k)$ is again a normally distributed random variable.

Message loss is modeled through variables $\theta(k)$ and $\phi(k)$ (\cf \figref{fig:WirelessSystemModel}), which are two independent Bernoulli random variables indicating successfully received ($\theta(k)=1$, $\phi(k)=1$) respectively lost ($\theta(k)=0$, $\phi(k)=0$) messages. While the Bernoulli assumption is oftentimes not satisfied for traditional wireless systems based on routing~\cite{Zimmerling2013},
%\db{refer back to discussion in earlier section or add ref}
it is indeed valid for our wireless embedded system design as per property \textbf{P2}. 
To ease the presentation and since both $\theta$ and $\phi$ are \iid, we omit the time dependence of the variables in the following.

For remote control (see \figref{sfig:ctrl}), we consider a remote computing unit, the controller, connected to the physical system over a wireless network. In this case, sensor measurements and control commands need to be sent over the wireless network. In this scenario, all computations are executed at a remote location, and no computational power is required directly at the physical system.
Thus, in case a data packet containing the next control input is lost, we perform a zero-order hold (ZOH), \ie we keep applying the previous control input:
\begin{align}
u(k) = \phi\hat{u}(k) + (1-\phi)u(k-1).
\end{align}
Here, $\hat{u}(k)$ denotes the control input computed by the controller, to be made precise in the following section. We present the control strategy and analysis for remote control for the single-loop case, \ie with only one system. Therefore, we drop the index $i$ whenever discussing the remote control scenario.

In the distributed control scenario (see \figref{sfig:sync}), we assume systems to be collocated with a local controller. Communication over the wireless multi-hop network is then needed to solve the distributed control task. 
\figref{sfig:sync} illustrates a scenario where agents communicate their sensor measurements over the network. However, the general methodology developed in the next sections extends straightforwardly to other scenarios, and we will also discuss such examples throughout the text.

% !TEX root = ../paper.tex

\section{Integration at Design Time}
\label{sec:design_time}

% We are now extending the basic architecture from the previous section at \emph{design time}, which means ... \fm{Insert small explanation what integration at design time essentially means.}
%We are now extending the basic architecture from the previous section with the goal to adapt the system's behavior at run time within a set of well-defined modes. A mode describes the operation of the overall \cps, including the scheduling of tasks, the control objectives, and the content and period of communication rounds. Mode changes happens in response to internal or external events. Because the modes are calculated at \emph{design time}, \ie prior to execution, we can verify the closed-loop stability for each mode and also for switches between any pair of modes.

%As will become apparent in the remainder of this section, the nature of the wireless embedded system allows us to use fairly straightforward methods from control theory to design a suitable control strategy and prove stability. The stability proof results in a condition that is computationally cheap to evaluate and can be used to check stability for any LTI system using our wireless embedded system design and control strategy. Moreover, we show how the support for arbitrary traffic patterns greatly simplifies solving distributed control tasks.

%\db{I wouldn't tailor this so much towards mode changes. From my point of view the mode changes are only one part of the overall integration at design time. A new proposal from my side:}Alternative section intro:

Building on the basic architecture from the previous section, we now present an integration of the wireless embedded system and the control system at \emph{design time}. Prior to execution, we derive a global schedule that serves the requirements of the control system (see \figref{fig:overview}). On the control side, we take the key properties of the wireless embedded system design, come up with a suitable control strategy, and formally prove closed-loop stability of the overall \cps. As will become apparent in the remainder of this section, the nature of the wireless embedded system allows us to use fairly straightforward methods from control theory to design a suitable control strategy and prove stability. The stability proof yields a condition that is computationally cheap to evaluate and can be used to check stability for any LTI system using our wireless embedded system design and control strategy. Moreover, we show in two examples how the support for arbitrary communication patterns greatly simplifies solving distributed control tasks.

To additionally be able to influence the behavior during execution, we further extend the basic architecture and enable the \cps to change between a well-defined set of modes. A mode describes the operation of the overall \cps, including the scheduling of tasks, the control objectives, and the content and period of communication rounds. Since we calculate the modes a priori, we can verify closed-loop stability for each mode and also for switches between any pair of modes.

\subsection{Wireless Embedded System}
We extend the basic wireless embedded system with an offline and an online component.
We leverage Time-Triggered Wireless (TTW)~\cite{jacob2020ttw} to compute different operating modes prior to execution (offline), which includes the order and time offsets of application and communication tasks.
To switch between modes and thus adapt to changing application demands during execution (online), we design a mode change protocol.

\fakepar{Time-Triggered Wireless}
% The scheduling problem in \figref{fig:schedulingFramework} is rather easy to solve as just two nodes are involved in a static scenario without mode changes.
With an increasing number of nodes and possibly multiple control loops being closed over the same wireless network, the scheduling problem, as described in \secref{sec:basicWirelessEmbeddedSystem}, quickly gets more complex, which calls for an automated solution.
To this end, we use TTW, a framework tailored to solve this type of scheduling problem~\cite{jacob2020ttw}.
TTW co-schedules task executions and message transmissions together with the communication rounds.
% This scheduling problem is a complex optimization that cannot be solved online.
It statically synthesizes the schedule of all tasks, messages, and rounds offline by formulating the corresponding mixed-integer linear program.
This way, the application's real-time constraints can be guaranteed while minimizing the computational demand at runtime.

\fakepar{Mode change protocol}
The computed modes are distributed to all nodes prior to execution.
This allows us to realize a mode change at runtime by exchanging unique mode identifiers rather than all information that is relevant for a mode.
As a result, the communication bandwidth required to realize a mode change becomes extremely small, which reduces end-to-end latencies and energy costs on the communication side.

Our mode change protocol ensures a timely and safe transition from one mode to another.
The beacon slot (see \figref{fig:overviewWirelessProtocol}), originally only used to signal the start of a communication round, is extended with information about the current mode ID, the next mode ID, and a counter that describes the number of rounds until the mode change becomes active.
Nodes in the network can learn about the mode change until the counter reaches 0.
In the unlikely event that a node misses all beacons during this time, it enters a resynchronization state where it does not participate in any communication until it receives again a beacon with the new mode ID.
This prevents nodes from executing the wrong mode and disturbing the network, which ensures a safe operation.
See~\cite{baumann2019fast} for more details.

\subsection{Control System}
\label{sec:ctrl}
% \begin{itemize}
% \item Describe the model that results from the wireless embedded system (constant delay, discrete time, bernoulli drops), i.e. the model of the overall \cps
% \item state that one can do a principled feedback control design (eg stabilization) based on this; that is, present remote stabilization as \emph{one} example of what can do with this including sound theory (recap the proof; but refer to other papers for all details)
% \item Derive a similar model for the distributed/multi-agent case as we have for the single agent/stabilization; could have the same key features (delay, drops...), but also abstract information to be available everywhere
% \item Explain that this is an excellent basis for distributed control; cases to be discussed in following sections
% \end{itemize}
Next, we outline the control strategy  for both remote control and distributed control tasks, which builds on the essential properties of the preceding wireless embedded system design.

\fakepar{Control design}
We first discuss the remote control scenario illustrated in \figref{sfig:ctrl}. The controller aims to achieve desired (\eg stable) closed-loop behavior of the physical system by sending appropriate input signals. To this end, we start by designing a controller for the nominal system, that is, without delays and message loss. This can be done using well-known techniques from control theory, such as the linear quadratic regulator (LQR)~\cite{Anderson2007}. We then enhance the control design to cope with constant delays and rare \iid message loss.

We know that messages that are sent over the wireless channel arrive with a delay of one time step (see \figref{sfig:ctrl}). Further, we know when a data packet should be sent in each communication round, and thus we also know when a packet has been lost. The controller can now compensate for the delay and message loss by performing state predictions based on a mathematical model of the physical system,
\begin{align}
\begin{split}
\hat{x}(k) &= \theta Ay(k-1)+(1\!-\!\theta)A\hat{x}(k-1)+B\hat{u}(k-1) \label{eqn:prediction} \\
&= \theta Ax(k-1) + (1-\theta)A\hat{x}(k-1)+B\hat{u}(k-1)\\
&+\theta Aw(k-1).
\end{split}
\end{align}
Here, $\hat{x}(k)$ denotes the predicted state and $\hat{u}(k)$ the control input computed by the controller.

With $\hat{x}(k)$, the controller has an estimate of the current state of the system. However, the control input that is sent over the network will be delayed by another time step. The controller anticipates this delay by making a further prediction and computing the next control input based on this,
\begin{align}
\label{eqn:pred_inp}
\hat{u}(k+1)&=F\left(A\hat{x}(k)+B\hat{u}(k)\right),
\end{align}
where $F$ has been designed to render the nominal system stable (\eg an LQR design). The input $\hat{u}(k+1)$ is then sent over the wireless network.

\fakepar{Stability analysis}
The nature of the wireless embedded system and the control design presented above allow for providing theoretical stability guarantees. 
% Stability is here understood in the mean-squared sense. This implies, for instance, that $x(k)\to 0$ almost surely as time goes to infinity~\cite[p.~131]{boyd1994linear}. 
In the following, we will give an intuition of how these stability results can be obtained and refer to~\cite{mager2019feedback,baumann2019fast} for detailed descriptions and the full proofs.

Two main ingredients for the stability analysis are properties \textbf{P1} and \textbf{P2} of the wireless embedded system.
Since jitter is negligible, we can accurately describe the dynamics with the discrete-time model developed so far, assuming a fixed sampling interval.
The \iid property of message loss further makes it possible to apply well-known tools from the theory of linear matrix inequalities (LMIs)~\cite[Ch.~9]{boyd1994linear}.
In essence, the system is stable as long as the message loss probability is \enquote{low enough} and the update interval is \enquote{short enough.}
For a certain message loss probability and update interval, stability of the overall \cps can be assessed by evaluating a single LMI.
The concrete LMI and stability test are given in~\cite[Thm.~1]{baumann2019fast}. 

This analysis is valid as long as we stay in a given mode. In the face of mode changes, the analysis needs to be enhanced. A mode change might, for instance, indicate a change of the length of a discrete time step and, thus, different $A$ and $B$ matrices. In control theory, this is referred to as a \emph{switched system}. It is well known that a switched system can become unstable even if all subsystems are stable~\cite{lin2009stability}. However, if we ensure that the system, on average, stays in each mode for a sufficient amount of time, called the \emph{average dwell time}, the stability of the subsystems is still sufficient. Thus, as long as the mode changes respect the average dwell time, we can guarantee stability of the overall system. In~\cite{baumann2019fast}, we discuss how to calculate the dwell times and show in examples that respecting those on average is not overly restrictive.

\fakepar{Distributed control}
We now turn our attention to distributed control, as depicted in \figref{sfig:sync}.
We reintroduce agent index $i$ as we are dealing with multiple systems.
%Since in distributed control, we are naturally dealing with multiple systems at the same time, we reintroduce the index $i$.
While we here assume a local controller to be collocated with each system, information of other systems is needed to fulfill the distributed task. That is, the control input $u_i(k)$ of each system now not only depends on its local state, but it also needs to take the states of other systems into account.
Considering again  static linear feedback, the control law can be written as
\begin{align}
\label{eqn:inp_distr}
u_i(k) = F_{ii}x_i(k) + \sum\limits_{j\in\Omega_i}F_{ij}\hat{x}_j(k-1),
\end{align}
where $\Omega_i$ denotes the set of all agents whose state is relevant for agent $i$. In general, if we want to find the optimal solution for the distributed control problem, $\Omega_i$ may comprise all other agents.
Per property \textbf{P3} the wireless embedded systems supports arbitrary communication patterns, and thus we can also cater for the most extreme case where every agent needs information from all other agents (\ie all-to-all communication).
%Since property \textbf{P3} allows us to make use of arbitrary traffic patterns, also the most extreme case, where all agents need information of all others (\ie all-to-all communication), can be supported.

Similar to the remote control case, the state $\hat{x}_j(k-1)$ in \eqref{eqn:inp_distr} denotes the current estimate that agent $i$ has of agent $j$'s state. As can be seen from~\eqref{eqn:inp_distr}, the delay is already incorporated since at time $k$ the value $\hat{x}_j(k-1)$ is used. In case of message loss, we use ZOH, \ie for agent 2 from \figref{sfig:sync} we have
\begin{align}
\hat{x}_2(k-1) &= \theta y_2(k-1) + (1-\theta)\hat{x}_2(k-2).
\end{align}

In the following, we give two concrete examples of distributed control tasks.
%\lt{I think this section is overly detailed and lengthy. One may consider skipping one of the examples as a simple solution.}
% -> Sebastian:  We kind of need both examples as we discuss both in the experiments.  Also, IMO, it is important to illustrate the versatility of control solutions we can implement without any extra overhead.
In both cases, we present the two-agent setting for illustration as per \figref{sfig:sync}. The general framework straightforwardly extends to a larger number of agents, as we demonstrate through testbed experiments in \secref{sec:eval}.
\begin{example}[Consensus]
\label{exp:consensus}
We consider a variant of the well-known average consensus problem~\cite{olfati2007consensus}. Therein, the agents aim to reach the desired state $x_\mathrm{des}=\tfrac{1}{N}\sum_ix_i$, which is the average of the individual states $x_i$ of  $N$ agents.
Here, we divide the problem into two parts. First, the agents should agree on a common $x_\mathrm{des}$ via average consensus. Then, this state should be tracked. Thus, communication between agents is needed only in the first part.

A straightforward way to achieve average consensus is to initialize the local $x_{i,\mathrm{des}}(0)$ of each agent with its current state. Then, in each round, all agents receive the $x_{j,\mathrm{des}}(k-1)$ from the agents they can communicate with and update their $x_{i,\mathrm{des}}(k)$ following a standard consensus update law~\cite{olfati2007consensus}
\begin{align}
\label{eqn:avg_cons}
x_{i,\mathrm{des}}(k) = p_{ii}x_{i,\mathrm{des}}(k-1) + \sum_{j\in\Omega_i^j}p_{ij}\hat{x}_{j,\mathrm{des}}(k-1),
\end{align}
where $\Omega_i^j$ denotes the set of all agents $j$ from which agent $i$ receives information, and $p_{ii}$ and $p_{ij}$ are nonnegative real numbers designed such that $p_{ii}+\sum_{j\in\Omega_i^j} p_{ij} = 1$.
Consensus can be trivially achieved if all agents can communicate their states to all others.
Then, choosing $p_{ii}=p_{ij}=\tfrac{1}{N}$ for all $i,j$ yields consensus after one round.
In \secref{sec:eval_consensus}, we make this advantage of many-to-all communication concrete by contrasting to an example with nearest-neighbor communication.
%Clearly, the easiest way to achieve consensus in this way is if all agents can communicate their states to all others. Then, we can chose $p_{ii}=p_{ij}=\sfrac{1}{N}\,\forall i,j$ and have consensus after one round.
%Since property \textbf{P3} allows for arbitrary traffic patterns, also all-to-all communication is supported by our wireless embedded system.
As a second step, after consensus has been reached at some time $k^*$, all agents can implement the control law (analogous to~\eqref{eqn:inp_distr})
\begin{align}
\label{eqn:consensus_adapted}
u_i(k) &= F_i (x_i(k)-x_{i,\mathrm{des}}(k^*))
%\begin{split}
%u_1(k) &= F_1 (x_1(k)-x_{1,\mathrm{des}}(k^*))\\
%& =F_1\frac{x_1(k)}{2} + F_1 \frac{\hat{x}_2(k^*)}{2},
%\end{split}
\end{align}
in order to track the agreed-upon goal $x_{i,\mathrm{des}}(k^*) = x_\mathrm{des}$.
%For the two agents in \figref{sfig:sync}, the control law~\eqref{eqn:inp_distr} for agent 1 reduces to
%\begin{align}
%\label{eqn:consensus_adapted}
%\begin{split}
%u_1(k) &= F_1 (x_1(k)-x_{1,\mathrm{des}}(k^*))\\
%& =F_1\frac{x_1(k)}{2} + F_1 \frac{\hat{x}_2(k^*)}{2},
%\end{split}
%\end{align}
%where $k^*$ is the time at which consensus was reached.

% While this straightforward design is enabled in our approach through property \textbf{P3}, reaching consensus can be more involved and require special designs for general networks, see \cite{olfati2007consensus}.
\end{example}

The consensus algorithm presented in Example~\ref{exp:consensus} is an intuitive and basic example of distributed control.
It solely focuses on agreeing on a common target state and tracking that target state.
In real applications, however, systems typically need to find a trade-off between the distributed control task and, for instance, stabilizing themselves locally. One possibility to also account for this is the use of optimal control methods, which we present in the second example.

\begin{example}[Optimal Distributed Control: Synchronization]
\label{exp:sync}
As in Example~\ref{exp:consensus}, we seek to have $x_1(k)$ and $x_2(k)$ evolve as closely as possible, while now trading this goal off with further objectives.
%We con  But we assume that they need to find a trade-off between satisfying the distributed control task and further objectives.
To distinguish it from Example~\ref{exp:consensus}, we will refer to this problem as \emph{synchronization}.

As commonly done in optimal control \cite{Anderson2007}, we formulate the different objectives as a quadratic cost function
%The multiple objectives of each system can be expressed in terms of a quadratic cost function, \eg
\begin{align}
\label{eqn:lqr_cost}
J &= \lim_{K\to\infty}\frac{1}{K}\E\left[\sum_{k=0}^{K-1}\sum_{i=0}^2\left(x_i(k)^\transp Q_ix_i(k) + u_i(k)^\transp R_i u_i(k)\right)\right.\nonumber\\
&\left.+(x_1(k)-x_2(k))^\transp Q_\mathrm{sync}(x_1(k)-x_2(k))\vphantom{\sum_{k=0}^{K-1}}\right].
\end{align}
The positive definite matrices $Q_\mathrm{sync}$, $Q_i$, and $R_i$ indicate our objectives of satisfying the synchronization objective (\ie keeping the difference between states small by choosing large penalty $Q_\mathrm{sync}$), being stable (\ie keeping the states close to the equilibrium state at $x_i(k)=0$), and keeping the inputs small, respectively. 
Using augmented states, this can be brought into standard LQR form and be solved using readily available tools~\cite{Anderson2007}.
% By introducing augmented states $\tilde{x}(k)=(x_1(k),x_2(k))^\transp$ and input $\tilde{u}(k)=(u_1(k),u_2(k))^\transp$, we can rewrite~\eqref{eqn:lqr_cost} as
% \begin{align}
% \label{eqn:lqr_cost_aug}
% J &= \lim_{K\to\infty}\frac{1}{K}\E\left[\sum_{k=0}^{K-1}\left(\tilde{x}^\transp\begin{pmatrix}Q_1+Q_\mathrm{sync}&-Q_\mathrm{sync}\\-Q_\mathrm{sync}&Q_1+Q_\mathrm{sync}\end{pmatrix}\tilde{x}(k)\right.\right.\nonumber\\
% &\left.\left.+\tilde{u}^\transp(k)\begin{pmatrix}R_1&0\\ 0&R_2\end{pmatrix}\tilde{u}(k)\right)\right].
% \end{align}
% Now, \eqref{eqn:lqr_cost_aug} is in standard LQR form, and the feedback controller that minimizes~\eqref{eqn:lqr_cost_aug} can readily be found using standard tools~\cite{Anderson2007}.
As a solution, we obtain a static feedback matrix $F$ and the control input~\eqref{eqn:inp_distr} for agent 1 can be written as
\begin{align}
u_1(k) &= F_1\hat{x}(k)
= F_{11}x_1(k) + F_{12}\hat{x}_2(k).
\end{align}
In general, the feedback matrix $F$ is dense; thus, for optimal control, every agent needs information from all other agents. Using our wireless embedded design, this can be straightforwardly supported since information sent by one agent can be received by all others in the wireless network (property \textbf{P3}).
\end{example}

% !TEX root = ../paper.tex

\section{Integration at Run Time}
\label{sec:run_time}

The presented integration of communication and control at design time already fulfills many of the requirements and provides means to solve the associated challenges of future \cps, as outlined in \secref{sec:challenges}.
%\db{must be written way more precise once the earlier sections are finalized}, enabling, for instance, feedback and distributed control over wireless.
% Until here, we presented a tight integration of wireless embedded and control system at \emph{design time}. This fulfills many of the requirements for future \cps outlined in\db{must be written way more precise once the earlier sections are finalized}, enabling, for instance, feedback and distributed control over wireless.
%
However, decisions about the usage of the wireless medium are taken in an uninformed way, \ie sensor and control messages are transmitted at the highest rate supported by the wireless embedded system, regardless of whether or not those messages contain relevant information. In wireless \cps, this is undesirable for two main reasons. First, the bandwidth in a wireless network is severely limited, and with that also the number of systems that can participate in a communication round (see challenge \textbf{C1}, limited throughput).
% Thus, also the number of systems that can participate in a communication round is limited.
% If all systems are scheduled to communicate in every round, no additional traffic can be served.\fm{I see the point of this but it is not correct to formulate it this way. I think the previous sentence already captues that.}
Second, the RF transceiver consumes power, which is a significant factor in the overall energy consumption of the hardware platform. For embedded sensors and mobile devices,
%\db{as for instance in the application examples\ldots}
which are untethered and usually powered by batteries, this is a key concern (see efficiency challenge described in \secref{sec:challenges}).

To mitigate both problems, the presented integration at design time needs to be complemented by an integration at \emph{run time}.
As illustrated in \figref{fig:overview}, during operation, the control system reasons about its future communication demands and informs the communication system accordingly.  To achieve this, recent event-triggered and self-triggered control approaches \cite{heemels2012introduction,miskowicz2018event} can be used, where in contrast to traditional periodic or discrete-time control \cite{aastrom2013computer} sensor and controller updates occur only when needed.
The communication system then uses the information about whether an update is necessary or not to allocate freed resources to lower priority traffic (\eg status information or additional sensor measurements) or to shut down resources completely for saving energy. We call this integration at run time \emph{control-guided communication}. Below, we outline the main ideas behind control-guided communication and refer the reader to~\cite{baumann2019control} for further details.
%In contrast to traditional periodic or digital control \cite{aastrom2013computer}, where sensor and control updates occur periodically communicates at fixed rates, recent event-triggered and self-triggered control

% The network manager takes the information about communication demands received from all agents and allocates slots according to a scheduling policy.\db{not sure what to write here}
% The concrete scheduling policy can be adjusted based on the requirements of the application at hand.
% As an example, assume we need to assign slots to all control systems that signal communication needs to ensure stability.
% If after allocating those slots further slots are available, some of them can be allocated to lower priority traffic (\eg status information or additional sensor measurements).
% Any further free slots is left empty.
% Since radios are only turned on during allocated slots, free slots are then used to save energy.

\subsection{Control System}
\label{sec:run_time_ctrl}
While the model of the wireless \cps derived in \secref{sec:architecture} is still valid, we need to adapt the control strategy if we envision a control-guided communication. Apart from stabilizing a system or solving a distributed task, the new control strategy has two additional goals: 1) only use the wireless channel if necessary, and 2) decide about communication demands in advance. To achieve both goals, we employ a self-triggered control strategy. In self-triggered control, the control algorithm decides at each communication instant when to communicate the next time~\cite{anta2010sample}. This information can be included in the data packet that would have been sent anyways in the current communication round.
Various self-triggered control algorithms have been proposed in the literature.
Here, we present the approach from~\cite{baumann2019control}.

We start by defining an error $e(k)$, which can, for instance, be the deviation from the equilibrium for a remote control scenario or the deviation between system states for a synchronization task. Next, we set a triggering threshold $\delta$. The threshold $\delta$ is basically the deviation from the control objective that we are willing to tolerate. A straightforward triggering design is then to trigger communication whenever $e^\transp(k)e(k)>\delta$. However, to enable the communication system to adapt to this demand, we need to decide about the next triggering instant in advance. Thus, at time $k$, we reason about when we expect the error to exceed the threshold the next time based on all data collected so far. Formally, we find the smallest $M>1$ such that
\begin{align}
\label{eqn:self-trigger}
\E[e^\transp(k+M)e(k+M)\vert\mathcal{D}(k)] > \delta,
\end{align}
where $\mathcal{D}(k)$ is the data available at time $k$. The value $M$ is included in the data packet that is sent over the network.

In the following, we make the triggering design precise for the distributed control scenario from Example~\ref{exp:sync}.
\begin{example}[Optimal Distributed Control: Synchronization (cont.)]
\label{exp:ST}
In Example~\ref{exp:sync}, we assumed that agent 2 directly sends its sensor measurements to agent 1.
Since each agent is aware of the full feedback matrix $F$, agent 2 can also, instead of sending the sensor measurements, compute and communicate the control input $u_{12}(k) = F_{12}x_2(k)$.
Then, the control law for agent 1 reads
\begin{align}
\label{eqn:distr_ctrl_st}
u_1(k) = F_{11}x_1(k) + u_{12}(k).
\end{align}
For this case, we define the error as $e_{12}(k) \coloneqq u_{12}(k)-u_{12}(k_\ell)$, where $k_\ell$ denotes the last time instant agent 2 transmitted $u_{12}$ to agent 1. 
As shown in~\cite{baumann2019control}, we can find an analytical expression for~\eqref{eqn:self-trigger}, which an agent checks at each communication instant.
That is, at each communication instant an agent estimates when it needs to communicate the next time and includes this information in the data packet.
% Checking $\E[e^\transp_{12}(k+M)e_{12}(k+M)]>\delta$ then amounts to finding the smallest $M>1$ such that
% \begin{align}
% \label{eqn:st_trigger_condition}
% &\lVert F_{12}(\tilde{A}_2^{M}x_2(k) + \sum\limits_{i=0}^\mathrm{M} \tilde{A}_2^{M-i}B_2u_{21}(k))-u_{12}(k)\rVert^2\nonumber \\
% &+ \Tr(F_{12}^\transp (\tilde{A}_2^\transp P_2(k+M|k)\tilde{A}_2+\Sigma_2)F_{12})>\delta,
% \end{align}
% with $\Tr$ the trace of a matrix, $\tilde{A}_2 = A_2+B_2F_{22}$, $P_2(k+1|k) = F_{12}^\transp (\tilde{A}_2^\transp P_2(k|k)\tilde{A}_2+\Sigma_2)F_{12}$, and $\Sigma_2 = \E[v_2(k)v_2^\transp(k)]$. All information needed to evaluate~\eqref{eqn:st_trigger_condition} are available to agent 2 at time $k$. Thus, it can, at every communication instant, compute $M$ and signal this information to the network manager.
% \fm{This is the first time we use the term network manager in context with our approach. We explain it in the following Wireless Embedded System section. Suggestion: "Thus, it can, at every communication instant, compute M and signal this information to the network."}
\end{example}

\subsection{Wireless Embedded System}
With this approach, the number of agents participating in the next communication round depends on the current state of the control system.
An offline derivation and distribution of communication schedules is not sufficient anymore.
Instead, we extend the wireless embedded system design with an online scheduler that disseminates the schedule at the beginning of each communication round, as shown in \figref{fig:overview}.
This allows us to adapt to the demands signaled by the control system dynamically and to either reallocate resources or to save energy in case the control system does not need all slots in a round.

During operation each agent computes the number of rounds $M$ until it needs to transmit again control information.
The computed $M$ is piggybacked onto the messages sent in the data slots (see \figref{fig:overviewWirelessProtocol}).
% This information is passed from the application processors (\ap) to the respective communication processors (\cp), which in turn piggyback it onto the messages that are exchanged in the network.
The owner of the beacon slot acts as the \emph{network manager}: It collects all communication demands and computes an appropriate communication schedule according to an exchangeable policy.
Depending on the application, an appropriate policy can be implemented; for example, an energy-saving policy may shut down all slots that are not needed to meet the communication demands.
% The actual policy depends on the application, \eg an energy-saving policy could shut down all slots that are not needed to meet the resource demands.
The communication schedule is sent by the network manager in the beacon slot, containing information about the number of data slots and the mapping of nodes to slots in the current communication round.
Because the network manager knows the communication demands of all nodes, it can infer when messages are lost.
Since messages contain the future communication demand, the network manager allocates a slot in the next communication round for every lost message.
This potentially leads to allocated but unused resources.
However, it does ensure that enough resources are allocated to fulfill the communication demands.

%\input{content/problem}
%\input{content/emb_sys_design}
%\input{content/control}
% !TEX root = ../paper.tex

\section{Cyber-physical Testbed}
\label{sec:testbed}
% \st{I would use one dedicated section for the testbed only, not yet describing the experiments/use cases}
% \begin{itemize}
%     \item Stress importance of real-world evaluation
%     \item Explain our choices for why we build this testbed (can probably draw from CPSBench paper)
%     \begin{itemize}
%         \item Fast and well studied physical system
%         \item Large area for multi-hop network
%         \item Not in a clean lab environment to also have some possible disturbances (people walking around, other communication networks, \ldots)
%     \end{itemize}
%     \item Discuss the different options that we have (e.g., sim + real, thus scaling up)
%     \begin{itemize}
%         \item Different tasks: Stabilization, synchronization
%         \item Stable and unstable mode
%         \item Combine simulated systems with real systems to scale up
%         \item Any node can be a controller, thus, can have one controller per system, one controller for all systems, \ldots
%     \end{itemize}
%     \item Describe extended testbed: Cover larger space, more cart-pole systems, different type of cart-pole systems (Quanser vs. self built) (in general, distributed control of inhomogeneous systems is more challenging, so that should be stressed)
% \end{itemize}

Next to a systematic co-design and a thorough theoretical analysis, it is essential to validate wireless control solutions on realistic cyber-physical testbeds~\cite{lu2015real}.
We presented such a testbed in~\cite{baumann2018evaluating}.
In this section, we elaborate on our design choices, the capabilities of the testbed, and the testbed extensions used in this paper compared to the original version.

\begin{figure*}
\centering
\includegraphics[width=\textwidth]{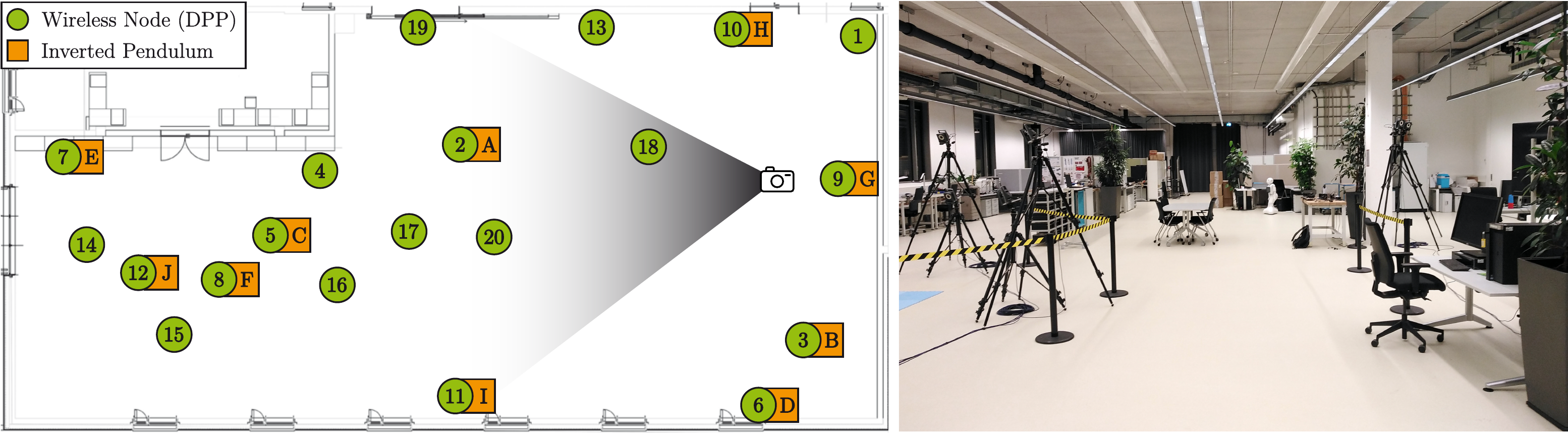}
\vspace{-5mm}
\caption{Layout of the cyber-physical testbed with 20 dual processor platform (\dpp) wireless nodes, 6 real physical systems (A--F), and 4 simulated physical systems (G--J), deployed in a robotic laboratory of approximately \SI{20}{\meter} by \SI{30}{\meter}. The viewing angle of the photograph is indicated in the schematic on the left.}
\label{fig:testbed}
\end{figure*}

\figref{fig:testbed} shows the layout of the testbed used for the experiments in \secref{sec:eval}.
It consists of 20 \dpp wireless nodes, 6 real and heterogeneous physical systems (A--F), and 4 simulated physical systems (G--J) deployed in a robotic laboratory environment of approximately \SI{20}{\meter} by \SI{30}{\meter}.
To analyze the network diameter during testbed deployment, we ran multiple link tests.
These tests indicate that for the chosen radio transmit power of -12\dBm we obtain a 3-hop network. %, where none of the nodes can reach all others within a single transmission.
% In the end, we chose a radio transmit power of -12\dBm, which forms a stable 3-hop network, with none of the nodes being able to reach all others with a single transmission.
% The signal strength is chosen such that the 20 nodes form a 3-hop network.
%The robotic laboratory environment is different from the office environment in which the original testbed was deployed.
Communication is subject to interference from various electronic equipment and other experiments in the same and adjacent rooms.
% In the laboratory, the testbed is subject to various kinds of interference, what would also be the case in any real-world scenario.
% Apart from radiation of electric components, which may interfere with wireless signals, the nodes also need to deal with additional wireless communication in different frequency bands stemming from other kinds of experiments conducted in the same room.
% The original testbed from~\cite{baumann2018evaluating} was set up in an office environment and, thus, did not have to deal with these kinds of disturbances.
% Further, it only featured two physical systems.

\begin{figure}[!tb]
\centering
\subfloat{%
       \includegraphics[height=0.15\textheight]{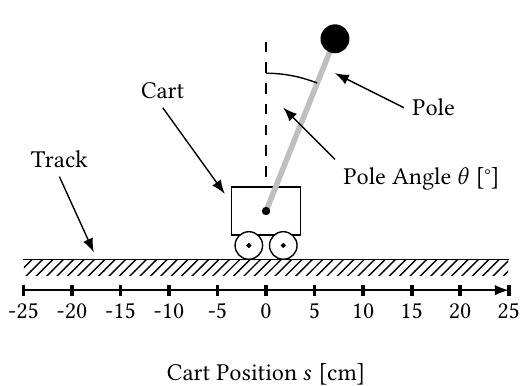}}
    \hfill
  \subfloat{%
        \includegraphics[height=0.15\textheight]{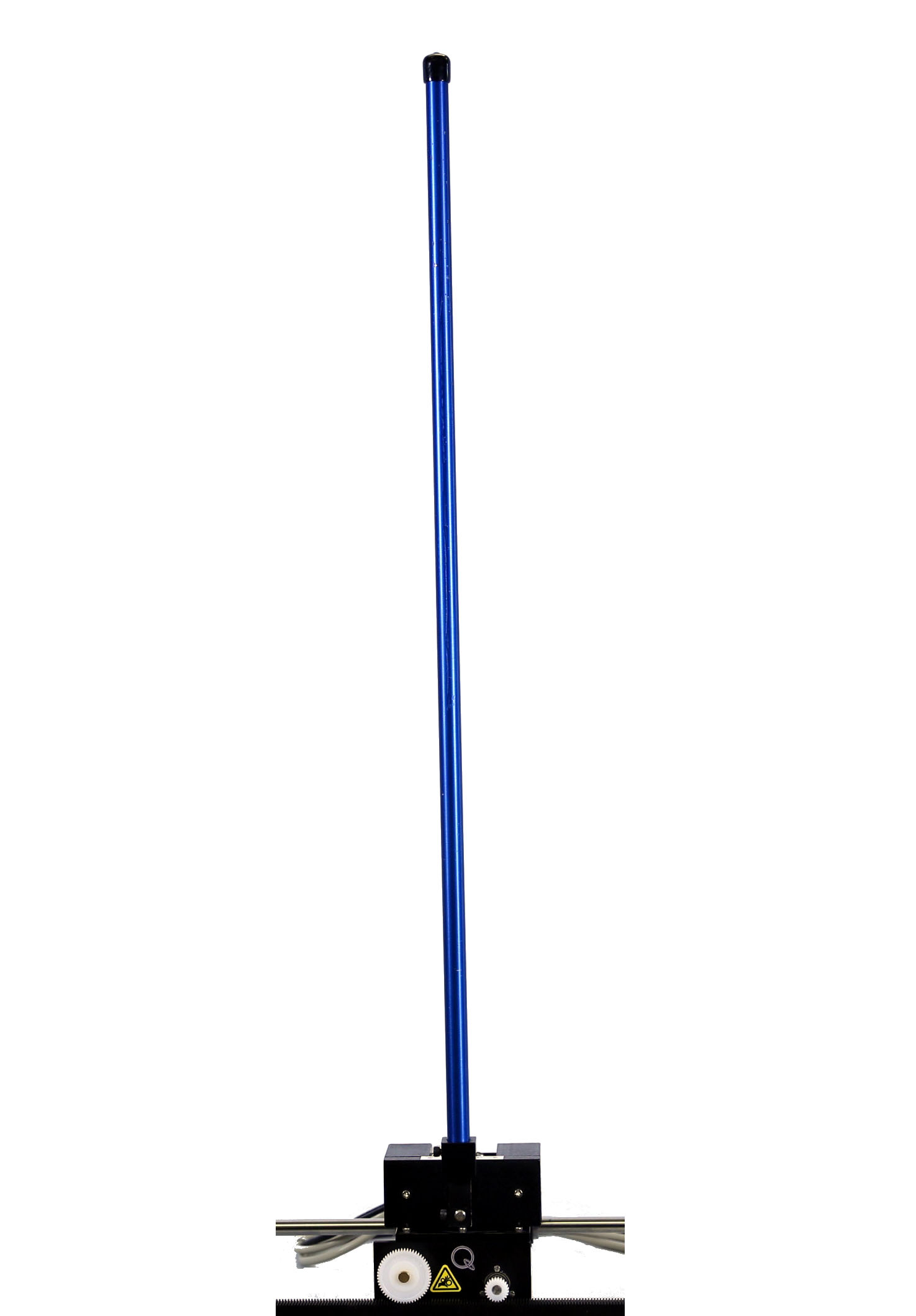}}
  \caption{We use multiple heterogeneous cart-pole systems as physical systems in the cyber-physical testbed. In addition to two types of real cart-pole systems with different physical characteristics, the testbed also supports the deployment of simulation models to scale up the number of agents in a cost-efficient way.}
  \label{fig:cartPoles}
\end{figure}

A crucial design point for a cyber-physical testbed is the choice of a physical system.
The physical system should be well understood and have fast dynamics that challenge the control, computing, and networking elements of \cps.
% Ideally, a physical system for a cyber-physical testbed should, on the one hand, be a well-known and -studied system and, on the other hand, have fast dynamics that match timescales of real applications and challenge the control, computing, and network elements.
For these reasons, we use cart-pole systems in our testbed, as depicted in \figref{fig:cartPoles}.
Cart-pole systems have been widely studied in control~\cite{boubaker2012inverted}, and their dynamics match the timescales of (mechanical) systems envisioned for smart manufacturing, such as drones.

Moreover, the testbed should be versatile in the sense that it accommodates a variety of tasks to evaluate different aspects of the wireless control solution.
The most widely studied control task for the cart-pole system is probably the \emph{inverted pendulum}, stabilizing the pole in an upright position through appropriate motions of the cart.
Due to the fast dynamics of the cart-pole system, this requires fast and reliable feedback.
% Due to the fast dynamics of the system, stabilizing the pole in an upright position requires fast and reliable feedback.

For wireless control, scenarios with remote controllers are of particular interest.
For example, sensor measurements need to be communicated over multiple hops to a remote controller, which computes and sends back the actuator commands.
Our testbed supports such scenarios, as one or multiple pendulums may be stabilized via one or multiple remote controllers. 

%To realize such a scenario, we need short update intervals (\ie communication and control computations need to be fast), high reliability, a suitable control strategy, and a stability analysis to guarantee safe operation.
% Because the controller can be implemented on any node in the network, we can have one controller controlling all systems, one dedicated controller per system, \ie multiple control loops running in parallel, or anything in between.
% Further, the controller can be implemented on any node in the wireless network.
% Thus, we can have one controller controlling all systems, one dedicated controller per system, \ie multiple control loops running in parallel, or anything in between.
% \db{could discuss some more things like set-point tracking, stable and unstable equilibrium}

Besides remote control, our testbed also supports distributed control tasks.
For instance, we may attempt to synchronize the positions of the carts to let them move in concert.
%This allows us to, for instance, compare the effectiveness of different communication structures that may be realized with different wireless protocols.
% This scenario is useful to compare the effectiveness of different communication structures that may be realized with different wireless protocols.
% Since, in the most general case, all agents in the network may need information of all others, all-to-all communication is desirable.
The difficulty of this distributed control task can also be varied.
% For distributed control, the difficulty of the task can be controlled through different choices.
In the most straightforward case, we may try to synchronize the cart positions with unmounted poles and thus neglect the stabilization problem.
% The difficulty can be increased by adding the poles and using an additional controller locally at each pendulum for stabilization, while communication is still needed for synchronization.
Making it more difficult, we may add the poles but equip each pendulum with a local controller, so stabilization happens locally at faster update intervals than synchronization over the network.
In this case, communication is only needed for synchronization and not critical for closed-loop stability.
In the most challenging case, both synchronization and stabilization happen over the (same) wireless network.
% The most challenging case covers both, synchronization and stabilization, over the network.

With this set of supported control tasks, we can mimic various use cases from \figref{fig:smartManufacturingAbstraction}.
% The remote control task maps to the situation where an agent has only sensing and actuation units and requires another agent with a computing unit to carry out the control computation.
% Remote control represents an agent with only sensing and actuation units, who needs to be connected to an edge computing unit or a cloud service, which carries out computations and sends actuation commands.
A remotely controlled cart-pole system may represent an agent with only sensing and actuation units that is connected directly or over the wireless network with an edge computing unit or a cloud-based service that carries out the control computations.
The distributed control scenarios relate to a fleet of agents that need to coordinate their actions, either using local or remote computing units.

Most of the envisioned applications in \secref{sec:vision} require a large number of agents.
It should therefore be possible to scale up the testbed at low cost to challenge the scalability of \cps designs.
We accommodate for this by enabling an easy integration of simulated physical systems.
A mathematical model of a cart-pole system can be deployed on any of the \dpp nodes in our testbed.
One advantage of simulated systems is a better reproducibility because, unlike real physical systems, they always react in the same way given the same control inputs.

In general, distributed control with heterogeneous agents is more challenging than with homogeneous agents~\cite{lunze2012synchronization}.
However, this is what is expected in smart manufacturing, for example, when different types of mobile robots need to work together on a given task.
To this end, our testbed includes physical systems that can be bought off-the-shelf\footnote{We use Quanser's \emph{Linear Servo Base Unit with Inverted Pendulum}.}
% \footnote{\url{https://www.quanser.com/products/linear-servo-base-unit-inverted-pendulum/}}
as well as self-built cart-pole systems; 
only the former are used in the original testbed~\cite{baumann2018evaluating}.
% Due to different hardware configurations, the dynamics of these systems are different.
Due to different hardware configurations, also the dynamics of these systems are different.
% Finally, the behavior of the simulated cart-pole systems can be influenced by changing parameters in their mathematical model.
Instead of affording different types of real cart-pole systems, distributed control with heterogeneous agents can also be studied by altering parameters in the mathematical model of the simulated cart-pole systems.

% !TEX root = ../paper.tex

\section{Application Case Studies}
\label{sec:eval}

In this section, we use our wireless cyber-physical testbed to illustrate the performance and capabilities of the proposed co-design methodology under realistic conditions.
To this end, we consider a series of experimental scenarios that resemble envisioned application use cases of smart manufacturing:
\begin{itemize}
 \item The AGV example from \secref{sec:usecase:agv} includes multiple agents that need to coordinate their actions with each other but are stable on their own.
% However, all individual agents are stable, thus, we have a purely distributed control task.
Accordingly, in \secref{sec:eval_consensus}, we show how to reliably coordinate multiple stable agents.
In particular, we demonstrate the benefits of arbitrary communication patterns supported by our wireless protocol.
 \item Distributed control becomes more difficult if we consider unstable agents, for example, a team of drones that jointly inspect machines in a factory.
% In manufacturing applications, this is, for instance, the case if teams of drones inspect the machines in the factory.
Each drone needs to stabilize its flight, while also coordinating its actions with the other drones in the team.
% We need coordination among the agents in the team, but each drone additionally needs to stabilize itself.
Moreover, drones may join or leave the team, or change between different tasks at run time.
We address this use case in \secref{sec:eval_sync}, where we show distributed control of unstable agents with mode changes.
 \item Using the same distributed control scenario, we showcase in \secref{sec:eval_cgc} how the integration at run time allows for scheduling additional traffic or saving energy based on the communication resources requested by the control system.
 \item Besides distributed control, remote control is also of great importance for future smart manufacturing, for instance, when an edge devices controls a mobile robot. \secref{sec:eval_remote} demonstrates remote control of multiple unstable agents.
 \item Because wireless communication is, in general, more susceptible to message loss than wired communication, we investigate in \secref{sec:eval_drops} the impact of message loss on the control performance of a remotely stabilized agent.
\end{itemize}

% In \secref{sec:eval_cgc} we furthermore showcase the integration at run time for the same scenario, \ie we solve the distributed task while simultaneously scheduling additional traffic, such as data from additional remote sensors, and saving energy without sacrificing control performance too much.

%in the remote control scenario.
% In all scenarios, robustness to failures is another important design goal.
% For wireless control, especially the higher probability of loosing messages as compared to cable-based solutions is of interest.
% Since the remote control scenario is the most vulnerable, we validate the robustness of our co-design by artificially introducing message loss in such a scenario in \secref{sec:eval_drops}.

Most of the control strategies in the following experiments rely on a mathematical model of the physical system to be controlled.
% The physical systems used for the evaluation are cart-pole systems as introduced in \secref{sec:testbed}.
For the cart-pole systems D and E in \figref{fig:testbed} and all simulated systems, we use the same model as in~\cite{mager2019feedback,baumann2019fast}.
For the self-built systems A, B, C, and F, we identify the mathematical model using standard methods from system identification~\cite{ljung1999system}.
Specifically, we apply a chirp signal and estimate the $A$ and $B$ matrices using the method of least squares. This yields
\begin{align*}
A &\approx \left(\begin{smallmatrix}1&\num{-3e-3}&\num{8e-3}&\num{5e-4}\\
                \num{-2e-3}&\num{1.02}&\num{-6e-3}&\num{1e-2}\\
                \num{1e-2}&\num{-0.11}&\num{0.94}&\num{2e-2}\\
                \num{1e-2}&\num{-2e-2}&\num{-8e-2}&\num{1.03}
            \end{smallmatrix}\right)\\
B &\approx \left(\begin{smallmatrix}
                \num{3e-4}\\
                \num{9e-4}\\
                \num{8e-3}\\
                \num{1e-2}
            \end{smallmatrix}\right)
\end{align*}
for a sampling interval of \SI{10}{\milli\second}.
In the following experiments, we always let all 20 \dpp nodes participate in the wireless communication, but we use the physical system only for a selected subset of the nodes, referred to as agents.

\subsection{Average Consensus}
\label{sec:eval_consensus}

We begin with the average consensus scenario introduced in \secref{sec:ctrl} (Example~\ref{exp:consensus}).
Our experiments show that:
% Reaching consensus under different communication architectures is a major concern in the consensus literature, and various algorithms have been developed.

\fakepar{Finding}
% In the consensus literature, guaranteeing to reach consensus under different communication architecture is a major concern, and various algorithms have been developed.
\emph{Thanks to the many-to-all communication support, our wireless control system reaches consensus within one time step using a straightforward and computationally cheap algorithm.}
% Through the many-to-all communication primitive supported by our wireless embedded system design, we can reach consensus already after one time step with a straightforward and computationally cheap algorithm.
% Other communication primitives need significantly longer to reach consensus when using the same algorithm.

\fakepar{Setup}
We use 5 agents with real cart-pole systems (A--E in \figref{fig:testbed}) and different initial cart positions.
% We consider the 5 carts, \ie the systems in \figref{fig:cartPoles} with unmounted poles, A -- E in \figref{fig:testbed}.
%Their carts start at different initial positions.
Agents communicate their state information across the 3-hop wireless network to solve the average consensus problem explained in \secref{sec:ctrl} (Example~\ref{exp:consensus}).
After agreeing on a common desired position, the agents apply control commands to track this position.
% Agents send around their state information and apply the consensus algorithm until they agreed on a common desired position and only then start tracking this desired position.

%Because of property \textbf{P3}, we can realize arbitrary communication patterns. 
We compare two approaches.
In the first one, every agent shares its state information with all other agents, referred to as \emph{all-to-all communication}.
As discussed in Example~\ref{exp:consensus}, we choose $p_{ii}=p_{ij}=\sfrac{1}{N}$ for all $i,j$ with $N=5$.
% We can, thus, as also discussed in Example~\ref{exp:consensus}, choose $p_{ii}=p_{ij}=\sfrac{1}{N}\,\forall i,j$ with $N=5$.
For the second approach, we implement \emph{nearest neighbor communication}; that is, agent A can only send information to agent B, which can only send information to agent C, and so on.
We set all weights $p$ to $\sfrac{1}{2}$.
% For this case, we choose all weights $p$ equal to $\sfrac{1}{2}$.
Using this approach, agents only have partial information, which makes it more difficult to determine when global consensus is reached.
% If each agent only receives information of one neighbor, it is harder to extract when global consensus is reached.
Therefore, we require that the desired position of neighbors stays the same for 3 consecutive communication rounds before agents start to track this position.

% We let the agents send around information and apply the consensus algorithm until they agreed on a common desired position and only then start tracking this desired position.
We set $F_i=50$ (see Example~\ref{exp:consensus}) and introduce an additional integrator with a gain of 5 to ensure that the desired position is reached.
The update interval of the control loop for exchanging state information over the wireless network is \SI{100}{\milli\second}.
A second control loop for tracking the desired position becomes effective once consensus is reached.
This control loop is independent of the first one and runs at a shorter update interval of \SI{10}{\milli\second}.

\begin{figure}[!tb]
    \centering
    \subfloat[All-to-all communication.]
    {
        \includegraphics[width=\linewidth]{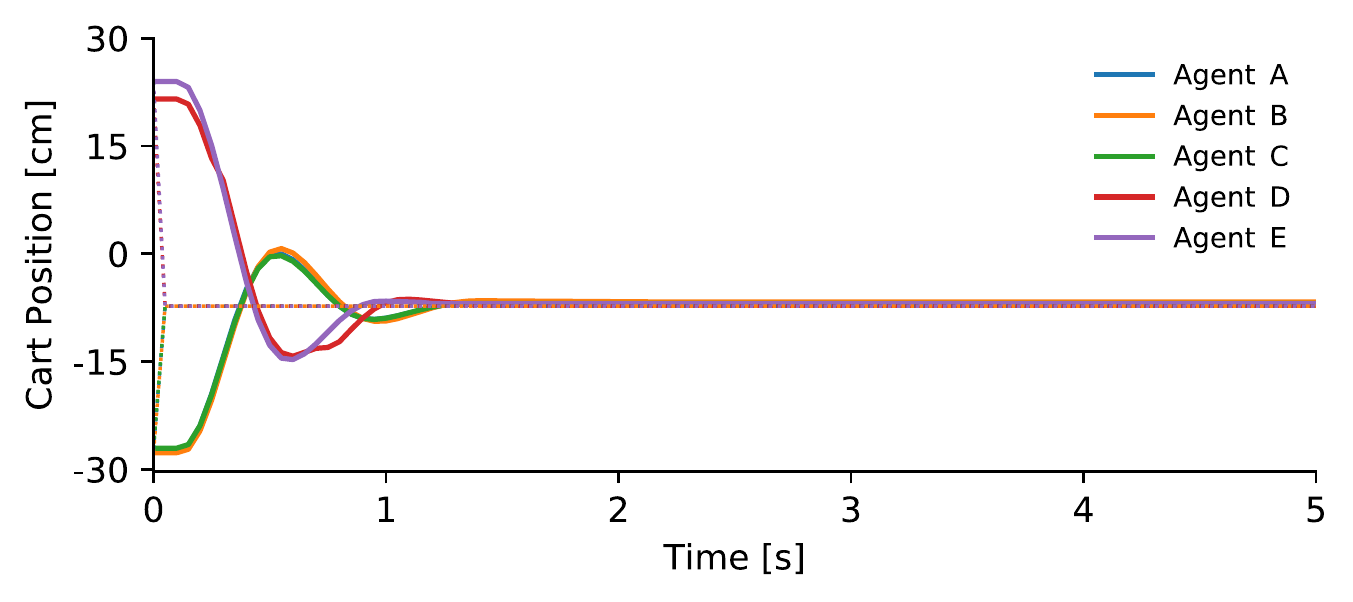}
        \label{fig:one_step_consensus}
    }
    %\hfil
    \newline
    \subfloat[Nearest neighbor communication.]
    {
        \includegraphics[width=\linewidth]{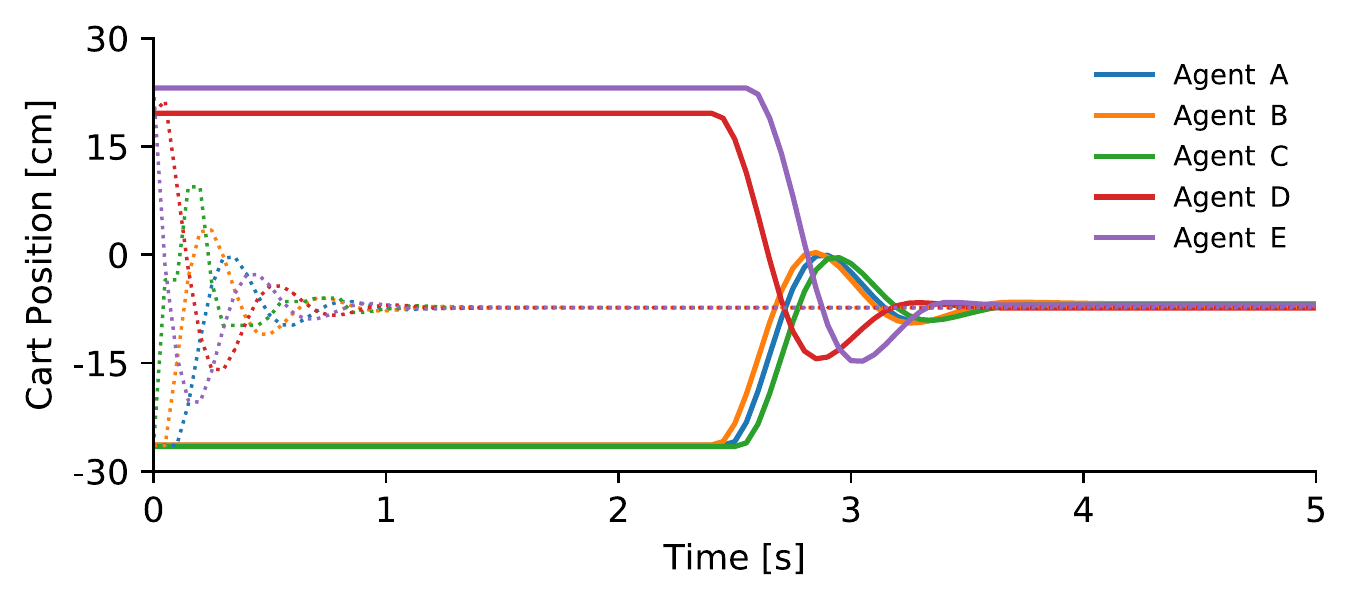}
        \label{fig:nearest_neighbor_consensus}
    }
    \caption{Five agents with real cart-pole systems first aim to reach consensus on their common desired position (dotted lines) by exchanging state information over the wireless network, and then track this cart position (solid lines) using a local control loop. \capt{It takes only \SI{100}{\milli\second} to reach consensus using all-to-all communication compared to \SI{2.5}{\second} steps with nearest neighbor communication.}}
    % \caption{Consensus of five carts (nodes 2, 3, 5, 6, 7 in \figref{fig:testbed}) using all-to-all communication. Top: desired position of the carts; bottom: actual position.}
    \label{fig:consensus}
\end{figure}

\fakepar{Results}
\figref{fig:consensus} shows the results for all-to-all and nearest neighbor communication; solid lines are the cart positions and dotted lines are the calculated desired positions of the 5 agents over time.
We see from the dotted lines in \figref{fig:one_step_consensus} that with all-to-all communication agents need only \SI{100}{\milli\second} (\ie 1 time step) to agree on a common desired position.
Using nearest neighbor communication, shown in \figref{fig:nearest_neighbor_consensus}, this takes significantly longer: \SI{2.5}{\second} (\ie 25 time steps).
Because the tracking of the desired position is independent of the communication approach, we see a similar performance after consensus is reached.
These results clearly showcase the benefits of the support for arbitrary communication patterns of our wireless control system. 

\subsection{Optimal Distributed Control}
\label{sec:eval_sync}
% Many of the works on consensus target integrator-type systems, such as the carts in the previous section.
% Because our wireless embedded system design provides full information at every node, we can tackle more complex problems in a simple way.
% Since through our wireless embedded system design we have full information available on all nodes, we can also tackle more complex problems.
Increasing the difficulty of the distributed control task, we now turn to a scenario where agents have two control objectives: stabilizing their mounted poles while also synchronizing their cart positions.
Because of the self-stabilization task, it is not sufficient for the agents to initially agree on a common desired position.
% As each agent still needs to stabilize itself, initially agreeing on a fixed target will not be enough.
Instead, the agents need to continuously exchange their states and adapt their control input based on their own state and the states of the other agents.
To showcase the adaptability of our co-design solution, we additionally let agents switch between different operating modes at run time. We find that:
% The scenario becomes even more complex as we are dealing with a heterogeneous multi-agent system, as described in \secref{sec:testbed}, which is more difficult compared to homogeneous systems~\cite{lunze2012synchronization}.

\fakepar{Finding}
\emph{The support for many-to-all communication allows for solving distributed control tasks in a straightforward way by designing a centralized optimal controller and implementing this controller in a distributed fashion locally on each agent.
We thus can successfully synchronize heterogeneous agents and safely switch between different operating modes at run time.}
% This allows us to successfully synchronize a team of heterogeneous agents.
% Further, the mode change protocol allows agents to leave and join the team without sacrificing performance.

\fakepar{Setup}
We use 10 agents (A--F with real and G--J with simulated physical systems), and adopt the optimal control design from \secref{sec:ctrl} (Example~\ref{exp:sync}) with parameter settings from~\cite{mager2019feedback,baumann2019fast}.
Agents exchange state updates over the network every \SI{100}{\milli\second} for synchronization and run local control loops with an update interval of \SI{10}{\milli\second} for stabilization.
% To also showcase the adaptivity of the proposed co-design solution, we additionally consider mode changes in this scenario.
All agents start in a local stabilization mode without synchronization.
% In the beginning, all agents locally stabilize themselves.
After \SI{60}{\second} they switch to a synchronization mode, while still stabilizing their poles locally.
After \SI{180}{\second} agent A leaves the team (\ie switches back to local stabilization), corresponding to a second mode change.
A third and final mode change is after \SI{300}{\second} when agent B also leaves the team.
All other agents keep synchronizing their cart positions until the end of the experiment.

\begin{figure*}[!tb]
\centering
\includegraphics[width=\linewidth]{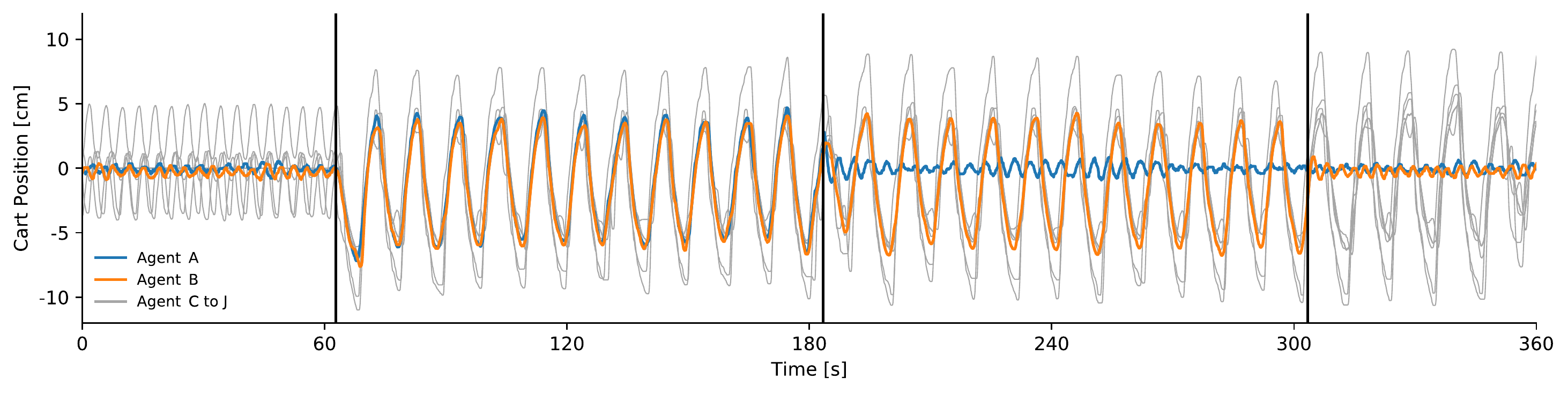}
\caption{Cart positions over time as agents stabilize their poles locally. Black lines indicate mode changes. After the first mode change, all 10 agents start to synchronize their cart positions over a 3-hop wireless network. At the second and third mode change, one agent each drops out of the synchronization task, while the other agents keep synchronizing their cart positions. \capt{Agents successfully synchronize and smoothly switch from one operating mode to another.}}
\label{fig:sync_mode_change}
\end{figure*}

\fakepar{Results}
\figref{fig:sync_mode_change} shows the cart positions of all 10 agents over time.
During the initial local stabilization mode, the 10 agents move independently of each other with different frequencies and amplitudes (\ie varying usage of track length).
After the first mode change, the agents synchronize their cart positions and start to move in concert with the same frequency and similar amplitudes.
We further observe that agents A and B leave the team at the second and third mode change, respectively, without causing any disturbance to the other agents remaining in the team, which continue to move in concert.
All 10 agents successfully stabilize their poles across all mode changes.

\subsection{Online Resource Allocation}
\label{sec:eval_cgc}

In the previous experiment, agents transmit state information in every communication round.
% Dynamic scheduling of additional traffic, \eg battery status information, or shutting down unneeded resources was not possible.
However, the actual communication demand varies during execution.
Indeed, depending on an agent's current state and the communication period, it is possible to transmit less often, allowing to dynamically reuse freed slots for other traffic or to shut slots down completely to save energy.
% We call this integration at run time, which was discussed in \secref{sec:eval}.
% In the previous example, we considered periodic communication, \ie agents transmitted information in every communication round and scheduling additional traffic was not possible.
% To also allow for scheduling of additional traffic and shutting down unneeded resources, we now showcase the integration at run time discussed in \secref{sec:eval}.
% We implemented this control-guided communication with a self-triggered control design, where each agent includes the information about the next time it wants to transmit in addition to the actual state information.
Such dynamic resource allocation can be realized through the control-guided communication approach presented in \secref{sec:run_time}: When an agent communicates, it already decides about the next time it needs a communication slot and piggybacks this demand information onto its data packet.
The network manager collects these communication demands and ensures that all demands are served. Our results show that:

\fakepar{Finding}
\emph{Through integration at run time, we can reliably synchronize 5 agents while serving additional traffic and saving energy.
Using less bandwidth for control comes at the cost of reduced control performance.
However, as long as about \SI{25}{\percent} of the bandwidth is available for control traffic, the control performance is comparable to the periodic control baseline.}

\fakepar{Setup}
% The following results were obtained using the testbed from~\cite{baumann2019control}\footnote{For this initial submission, the results are the same as the ones presented in~\cite{baumann2019control}. We will repeat these experiments on the new testbed. However, the key findings will be the same as the ones shown in this section.}.
We use 5 agents (D and E with real as well as G--I with simulated physical systems), whose cart positions are to be synchronized.
%In this experiment we use 2 real (D,E) and 3 simulated (G,H,I) cart-pole systems, whose cart positions we want to synchronize.
% It features 2 real and 3 simulated cart-pole systems, whose cart positions we want to synchronize.
The control design follows the architecture outlined in \secref{sec:run_time_ctrl} (Example~\ref{exp:ST}) with the same parameterization as in~\cite{baumann2019control}.
Each communication round consists of 5 data slots, which is sufficient to exchange control data among all 5 agents in every round.
The communication period is set to \SI{50}{\milli\second}.
As before, the local control loops for stabilization run at a shorter update interval of \SI{10}{\milli\second}.
We use an example scheduling policy where the network manager always assigns, if possible, one of the free slots to other traffic while the remaining free slots, if any, are shut down to save energy.

 \begin{figure*}[!tb]
 \centering
 \includegraphics[width=0.9\linewidth]{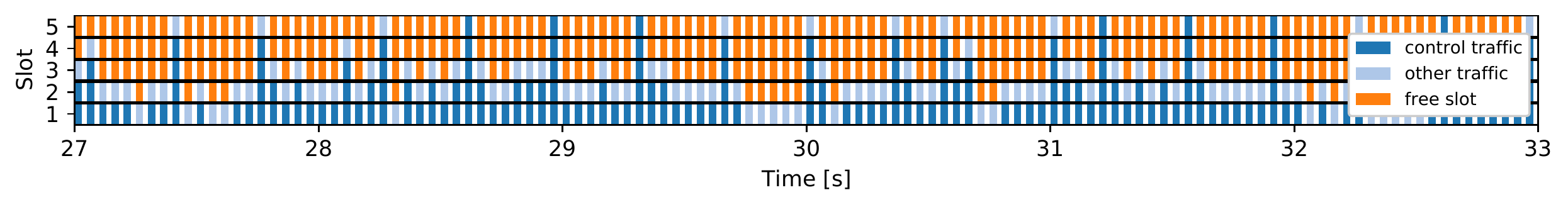}
 \caption{Bandwidth utilization with control-guided communication over time, recorded during an experiment with 5 agents that synchronize their cart positions over a 3-hop wireless network. \capt{Because of the self-triggered control design, the bandwidth used for control traffic in each communication round varies between 0 and 5 slots. In almost every communication round one slot is used for other traffic, while the remaining free slots are shut down to save energy.}}
%  \caption{Synchronization of 5 pendulums over a 3-hop network using a self-triggered control strategy. The top plot shows the root mean squared error of the cart position, which exhibits a sine movement due to the sine disturbance added to one of the pendulums. The bottom plot displays for a time window of \SI{6}{\second}, which slots are used for control traffic, additional traffic, and which slots are left free to save energy. Recreated from~\cite{baumann2019control}.}
 \label{fig:slot_usage}
 \end{figure*}

\begin{figure*}[!tb]
\centering
\subfloat[Control performance. \label{sfig:st_ctrl}]{
    \includegraphics[width=0.30\linewidth]{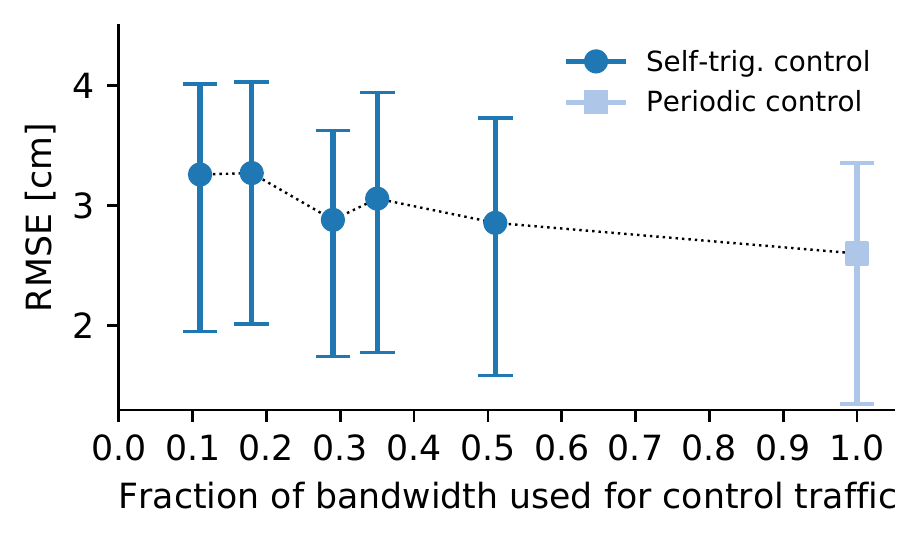}
}
\subfloat[Communication energy costs. \label{sfig:st_radio_duty_cycle}]{
    \includegraphics[width=0.30\linewidth]{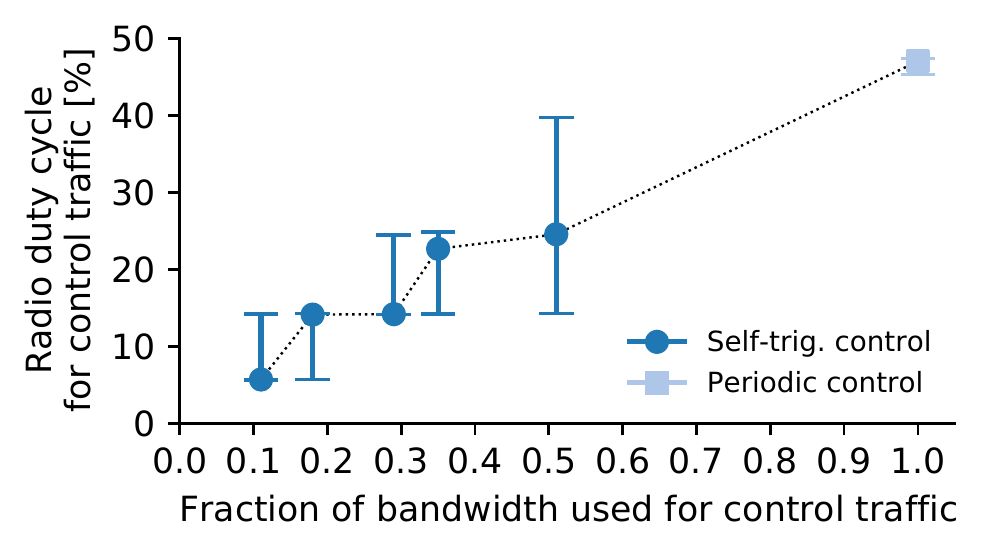}
}
\subfloat[Bandwidth available for other traffic \label{sfig:st_other_traffic}]{
    \includegraphics[width=0.30\linewidth]{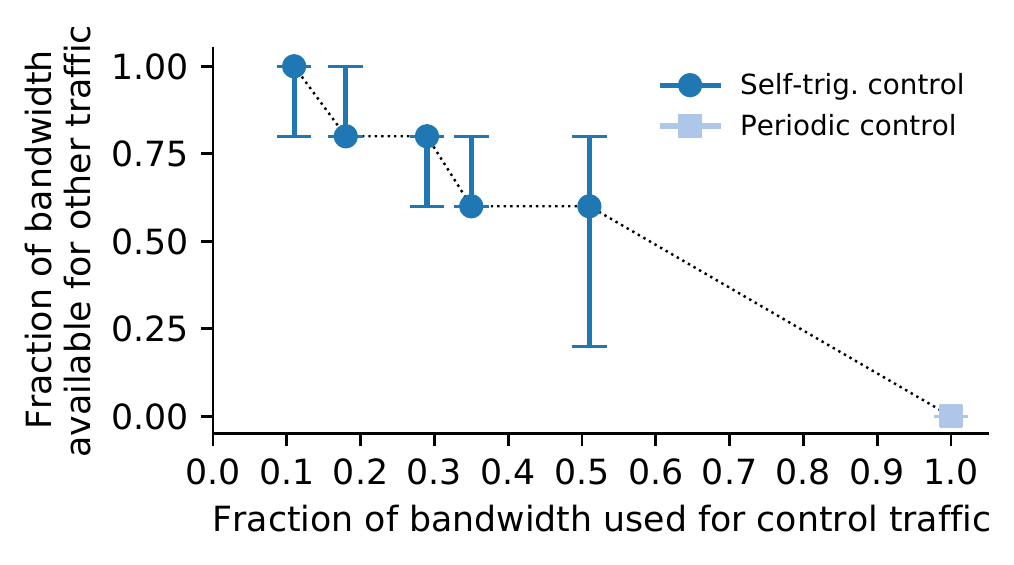}
}
\caption{Trade-off between control performance, the associated communication energy costs, and the ability to serve other traffic for different fractions of bandwidth available for control traffic. Graphs show the median and the 25th and 75th percentiles. \capt{Control performance decreases when less bandwidth is used for control traffic. On the other hand, bandwidth that is not used for control can be used to save significant communication energy or to serve other traffic.}}
\label{fig:st_tradeoffs}
\end{figure*}

\fakepar{Results}
\figref{fig:slot_usage} shows for a representative time interval from our experiments how the 5 data slots are used in each communication round, for a self-trigger threshold of $\delta=0.03$.
% An example policy for a fixed threshold $\delta=0.03$ is shown in \figref{fig:slot_usage}.
We see that only about one third of the bandwidth is needed for control traffic.
The remaining bandwidth is used to serve other traffic or shut down to save energy.
\figref{fig:st_tradeoffs} shows the trade-off among control performance (in terms of the root mean squared synchronization error), associated energy costs (in terms of radio duty cycle for control traffic), and fraction of bandwidth available for other traffic.
We vary the fraction of bandwidth used for control traffic through different choices of the self-trigger threshold $\delta$.
% We did multiple experiments and changed the self-trigger threshold $\delta$ each time, which directly affects the fraction of the communication bandwidth that is used for control traffic, as shown on the abscissa.
For $\delta=0$ all agents communicate control data in every round, which we use as a periodic control baseline.
We see that using less bandwidth for control leads to a drop in control performance.
% Saving energy and serving additional traffic comes at the cost of reduced control performance.
However, when using only about \SI{25}{\percent} of the bandwidth for control, we can still obtain reasonable control performance while being able to serve around \SI{75}{\percent} of additional traffic and to reduce energy costs by about~\SI{80}{\percent} compared with the periodic control baseline.

\subsection{Reliable Remote Control in the Face of Mobile Agents}
\label{sec:eval_remote}

Complementing the distributed control scenarios above, we now look at remote control with mobile agents, a highly relevant scenario for future smart manufacturing. We find that:

%as an addition to the already presented distributed scenarios.
%We have two cart-pole systems sending their state information to a controller that runs on another node in the network.
%The controller computes control inputs for both agents and communicates them back to the physical systems.
% After having presented several variants of distributed control, we now investigate remote control scenarios over multi-hop networks.

\fakepar{Finding}
\emph{We can reliably stabilize two agents with fast dynamics over a 3-hop wireless network despite agent mobility.
% The setup remains stable also in the face of mobile nodes, \ie we can move the controller node without any degradation of control performance.
Apart from the practical demonstration, our co-design methodology also allows us to provide theoretical stability guarantees.}

\fakepar{Setup}
We use two agents with real cart-pole systems (A and B) and stabilize their poles via a remote controller running on agent 1.
Thus, we have two agents that send their state information over the 3-hop wireless network to a remote controller.
The controller computes control inputs and sends them back across the network to both the two agents.
% In the first scenario, we consider stabilizing of cart-pole systems A and B by a remote controller running on node 1 (\cf \figref{fig:testbed}).
% The cart-pole systems send their sensor information to the remote controller over the wireless network, which in turn computes and sends the control input.
The control design follows the description in \secref{sec:ctrl}, concrete parameters are chosen as in~\cite{mager2019feedback,baumann2019fast}.
Communication over the wireless network happens at an update interval of \SI{40}{\milli\second}.
% State information are sent with a communication period of \SI{40}{\milli\second}.
Using a conservative estimate of \SI{0.1}{\percent} for the message loss probability, we obtain a theoretical stability guarantee.
During the experiment, we pick up the agent running the controller and move around the testbed area.
% To demonstrate support of mobile nodes, we move the controller node in the laboratory during the experiment.

\begin{figure*}
\centering
\includegraphics[width=0.8\linewidth]{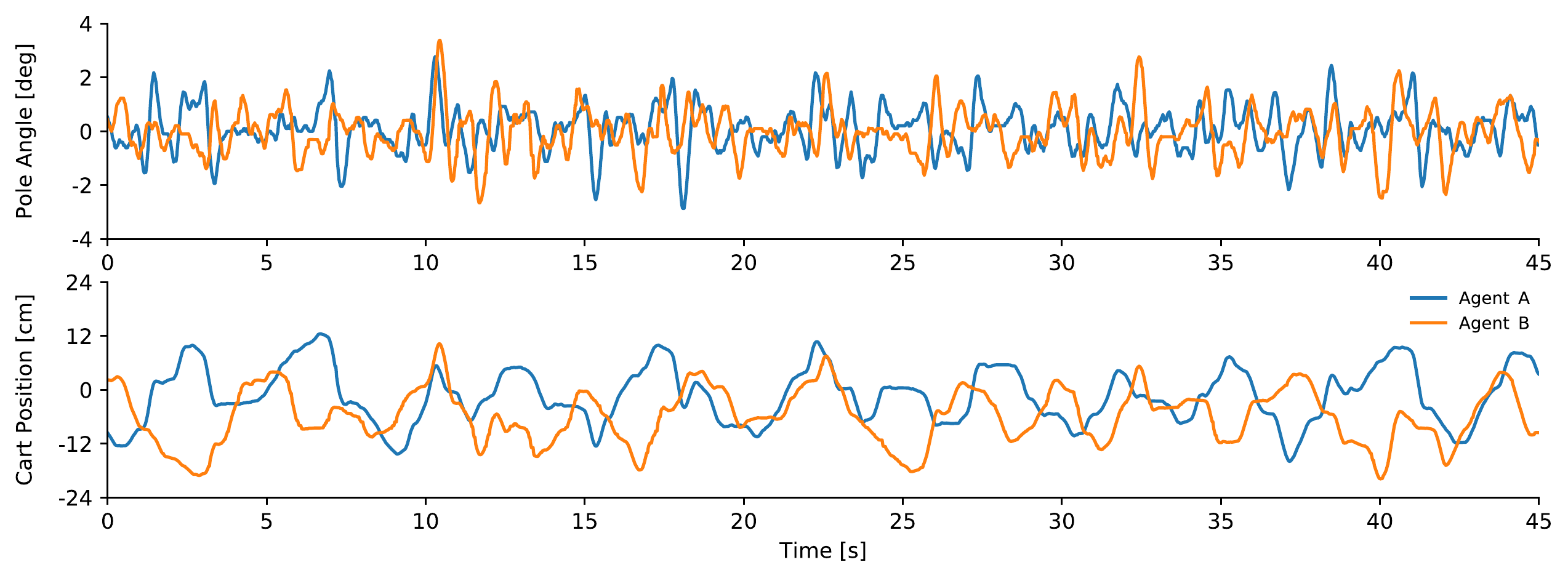}
\caption{Remote stabilization of two agents with real cart-pole systems over a 3-hop wireless network. During the experiment, the agent running the remote controller is moved around the testbed area. \capt{The poles of both agents are safely stabilized, and the control performance is not affected by agent mobility.}}
\label{fig:remote_stab}
\end{figure*}

\fakepar{Results}
\figref{fig:remote_stab} shows the evolution of the pole angle and the cart position of both agents throughout the experiment.
We see that both agents stay in safe regimes over the entire duration of the experiment and never come close to reaching the limits of their tracks.
In other words, we can safely stabilize them via the wireless multi-hop network.
% We can safely stabilize the two systems via the remote controller, both systems stay in safe regimes for the entire duration of the experiment and never come close to reaching the track limits at around $s=\SI{30}{\centi\meter}$.
Moving the controller node during the experiment has no impact on control performance\footnote{A video of such an experiment can be found at \url{https://youtu.be/19xPHjnobkY}.} as the operation of our wireless protocol is not affected by changes in the network as long as agents remain connected~\cite{Ferrari2011,ferrari12lwb}.

\subsection{Reliable Wireless Control Despite Node Failures}
\label{sec:eval_drops}
Great efforts are being made toward \enquote{ultra-reliable} wireless communication, targeting message loss rates of $10^{-9}$ or even lower.
% \fm{A reference would be nice.}
%\db{relate to introductory sections}
% -> Seb: didn't know where to relate this to, just removed for the time being.
We demonstrate that, given a proper co-design of the communication and control systems, the resulting \cps can deal with much higher loss rates. In particular, we find that:

\fakepar{Finding}
\emph{We can stabilize two agents via a remote controller over a multi-hop wireless network despite \SI{10}{\percent} message loss.}

\fakepar{Setup}
We consider the setup from the previous experiment, but artificially introduce message loss.
That is, in addition to the possible message loss over the wireless network, we let agent B drop \SI{10}{\percent} of control commands according to an \iid Bernoulli process.
Even at such high message loss we expect the agent to remain stable according to our theoretical analysis.

\begin{figure}[!t]
    \centering
    \subfloat[No artificial message loss.]
    {
        \includegraphics[width=0.47\linewidth]{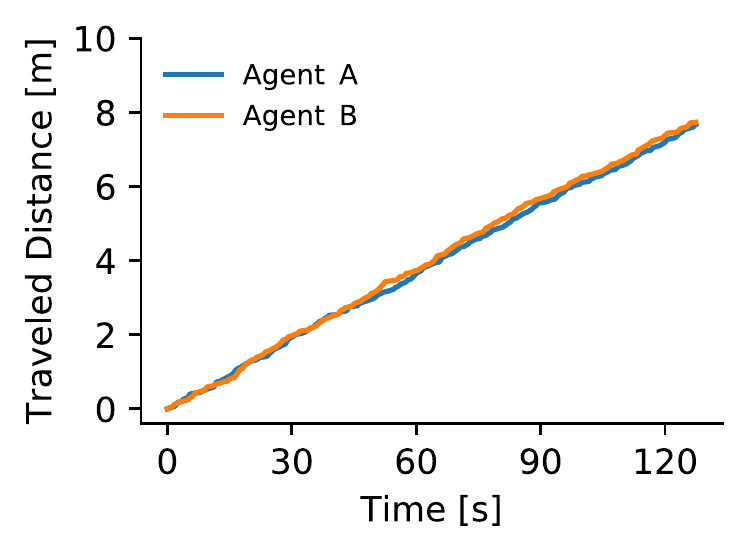}
        \label{fig:remote_stab_distance}
    }
    % \hfil
    % \newline
    \subfloat[Agent B drops 10\% of messages.]
    {
        \includegraphics[width=0.47\linewidth]{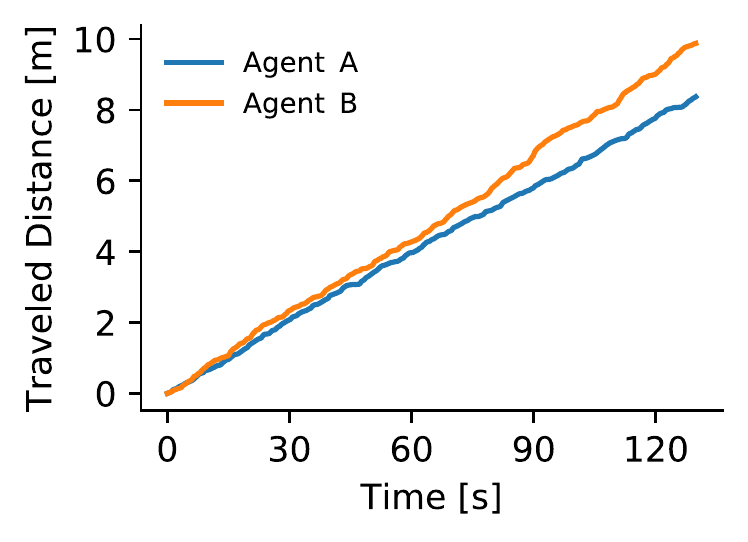}
        \label{fig:remote_stab_drops_distance}
    }
    \caption{Traveled distance of two argents while stabilizing their poles in remote control experiments. \capt{Agent B travels slightly farther when it artificially drops messages, which indicates a lower control performance. However, the drop in control performance is not severe, and stability is still guaranteed.}}
    \label{fig:distance}
\end{figure}

\fakepar{Results}
The resulting position and angle trajectories are nearly indistinguishable from the ones shown in \figref{fig:remote_stab} without artificial message loss.
Thus, we do not plot them here, and conclude that we can safely stabilize the agent despite \SI{10}{\percent} of message loss.
Differences in the control performance can be seen, for example, by looking at the distances traveled by the carts during the experiment.
\figref{fig:remote_stab_distance} shows the traveled distances from the previous experiment without artificial message loss.
The results in \figref{fig:remote_stab_drops_distance} with artificial message loss show that agent B compensates the message loss by moving slightly more.
% \figref{fig:remote_stab_drops_distance} shows the distance the carts traveled during the experiment.
% The cart for which we introduced artificial message loss had to move further to stabilize the system, but the differences are rather small.
We draw three conclusions from these results: (\emph{i}) If the control and communication systems are carefully designed in tandem, message loss rates well above typically targeted rates can be tolerated. (\emph{ii}) Instead of focusing on ``ultra-reliable"  wireless communication, focus should be put on designing wireless solutions for which the assumption of \iid message loss is a valid approximation. (\emph{iii}) Message loss has an impact on control performance; however, even for loss rates of up to \SI{10}{\percent}, the decrease in control performance is not severe, as long as the losses are \iid
The \iid assumption is highly valid for our wireless solution, as per property \textbf{P2}.

% \begin{figure}
% \centering
% \includegraphics[width=0.4\textwidth]{img/pendulums_pos_ang_remote_stab_drops_10}
% \caption{Traveled distance of the cart during a remote control experiment. For the second pendulum we artificially introduced \SI{10}{\percent} message loss.}
% \label{fig:remote_stab_drops_distance}
% \end{figure}

% !TEX root = ../paper.tex

\section{Open Challenges and Opportunities}
\label{sec:outlook}

%\mz{Start with a oerspective on what prior and our work already provides for future manufacturing.}
%\mz{The below is still a very rough collection of ideas.}

Starting from a perspective on future smart manufacturing, we derived several requirements and associated challenges, such as predictability of operation in terms of functional correctness and timeliness, as well as the efficiency of operation.
Constructing systems that satisfy these requirements is particularly difficult in systems that have low and non-deterministically changing resource availabilities and require a highly dynamic operation.
We presented a design methodology based on a holistic approach that integrates distributed feedback control and multi-hop wireless communication, both at design and at run time (\cf \figref{fig:overview}).
%Building on a unified architecture of wireless embedded system and control system (\cf \figref{fig:overview}),
In particular, through integration at design time,
%Through a tight integration of wireless embedded and control system \emph{at design time},
we were able to achieve fast distributed feedback control on the order of tens of milliseconds over multi-hop networks with stability guarantees also in the face of mode changes.
Further savings and flexible reallocation of resources were achieved by
%deepening the integration toward \emph{run time} and
having controllers inform the communication system at run time about their current need for communication.  Taken together, integration at design time and run time allows for high-performance distributed control while making efficient use of the available resources, as well as for adaptive behavior such as switching seamlessly between modes of operation.

Through this holistic approach, we were able to
%address important aspects as pertaining to the core requirements of dependability, adaptability, and efficiency (\cf \secref{sec:challenges}).
%and thus
address many challenges that currently prevent wireless technology from being adopted in the manufacturing industry.
However, to fully realize the smart manufacturing vision, there are many more challenges to be overcome.
Below, we outline some important directions for future work,
%to open new perspectives for next-generation smart manufacturing systems,
structured along the main challenges of dependability, adaptability, and efficiency (see~\secref{sec:challenges}).

%\db{I would restate the three main points again in the beginning and then from there start the discussion, which parts of these three main challenges did we not address? Then we can have maybe three fakepars discussing those three points. The way it is written now it is hard to see to which points the fakepars connect}

\subsection{Dependability}
The presented solution allows for reliable wireless communication (fast updates, highly uncorrelated packet drops, negligible jitter).  Further, through the integration of control and communication at design time, we can provide a formal proof of stability, which is a base requirement for any feedback control system.  The theoretical guarantee encompasses all components of the \cps (computation, communication, control, LTI physical process), mode changes, and dynamic network topologies.
%In this way, we were able to address essential dependability challenges.

\fakepar{Formal guarantees}
The type of theoretical guarantee given herein concerns the dynamical stability of the closed-loop control system with all main components of the \cps.
%While this is a base requirement for any control system,
It is desirable to complement this with further guarantees and certification.
For example, \emph{formal verification} of protocols, algorithms, or entire \cps for certain safety and liveness properties is highly relevant.  While approaches like model checking \cite{baier2008principles} hinge on the derivation of an accurate model and can be challenging to apply for complex systems, the automatic design of software components (\eg scheduler, controller, protocols) that are \emph{correct by construction} is an orthogonal approach.
Also in the realm of feedback control, some aspects have not been addressed yet.  For instance, extending our current stability guarantees to incorporate further uncertain components, such as (partly) unknown or data-driven physical system models and learning-based controllers, is highly relevant, especially in the context of the envisioned adaptability in smart manufacturing.

\fakepar{Fault detection and identification}
The larger scale of future smart manufacturing systems in terms of number of components and their interconnection complexity makes the prediction, early detection, and identification of faults highly demanding. It is necessary to have means to observe the operation of the various components by collecting key performance indicators and combine the data toward information and knowledge concerning the overall system state, causes of failures, as well as predictive maintenance. Whereas many results are available using advanced methods for data analysis and machine learning, it is unclear how to employ these methods and the associated data collection in low-resource environments.  As the occurrence of failures in a complex system is inevitable, the careful design of contingency and fault handling strategies is another core task.  Here, the flexibility and redundancy envisioned in smart manufacturing (\eg in terms of agents) can be of help.

%\db{Also would include emergency modes in this discussion, maybe redundant controllers. Also dependability}

\fakepar{Distributed architectures}
With the help of smart manufacturing, production processes are becoming significantly more flexible and versatile while minimizing its operators' management effort. In traditional manufacturing systems, the collaboration between machines and other \cps devices is mainly controlled by a central management system. Such a concept has a high risk of failure since the overall system's availability and reliability largely depend on the underlying communication network and the centralized management unit's availability. At the same time, this centralized approach is no longer manageable with the increasing number of collaborating \cps devices and the vast complexity managing a large number of interconnected devices. The same observation holds for centralized scheduling and coordination to synchronize and schedule the computational as well as communication tasks. Better scalability and fault tolerance concepts are needed toward fully distributed approaches without a central scheduler or controller while still being able to provide formal guarantees for the correct operation.  As an attractive middle ground, such distributed operation may also be temporary or limited to certain parts of the hierarchy (\eg a specific automation cell).

% Comment Marco:
%We may want to mention that, as a middle point between always centralized and always fully distributed, such distributed operation may also be temporary and/or limited to certain parts of the hierarchy (e.g., a specific automation cell). Some high-level automation of the entire plant using a centralized "brain" seems almost inevitable to me.

%\db{This I would classify as dependability, I would maybe not have it as a first example since it's not that obvious. Maybe start with fault detection.}

\fakepar{Security and privacy}
Deliberately, the discussion did not touch on any security- or privacy-related challenges in the design of future smart manufacturing systems. One may argue that well-known security concepts from the conventional computing domain can be simply applied and are sufficient. Obviously, this is not the case when considering problems like side-channel attacks via the physical embedding of sensors and actuators, key-management over the whole lifetime of the large number of distributed components, novel attack scenarios exploring the low-power wireless connectivity, as well as attacking the overall system functionality by manipulating sensor readings as available to the control algorithms.  Likewise, when sharing data and methods across users, factories, and possibly companies, important privacy concerns must be answered.
Novel security and privacy methods are thus needed to address the unique challenges of future smart manufacturing systems.

\subsection{Adaptability}
Because of the underlying synchronous transmissions, the presented wireless control solution is ignorant of any graph structure of the network and requires no routing.  Hence, the solution is inherently adaptive to dynamic nodes, which is key to provide the mobility needed in many smart manufacturing scenarios.  Furthermore, the current design and stability guarantees provide for some adaptability when switching between different operation modes.  The set of possible modes, however, must be known in advance.
%can accommodate for mode changes, which, however, must be known in advance.

\fakepar{Enhanced adaptability at run time}
While in some scenarios such as highly safety-critical ones, it is plausible that the set of possible application requirements and thus operation modes is fixed and known beforehand, it is vital for the vision of smart manufacturing to enhance adaptability also to situations that are previously unknown at design time.  For example, when robots with new manufacturing capabilities are added or for highly individualized tasks, the smart manufacturing system must adapt at run time.  The goals of instantaneous adaptation to unknowns while giving guarantees throughout operation are obviously conflicting in general.  Providing meaningful solutions and trade-offs tailored to domain and problem classes is thus a formidable task for basic research.  A promising direction for enhanced adaptability is the combination of recent advances in machine learning and data science with control and communication, as discussed next.

\fakepar{Learning-based control}
In the complete vision of smart manufacturing, properties of the computation and communication environment, as well as of the controlled physical processes, are constantly changing. In addition, when scaling the complexity and size of smart manufacturing systems, it is increasingly unfeasible to manually derive appropriate plant and control algorithms that integrate computation and (wireless) communication, even if they are parameterizable. Therefore, recent approaches attempt to combine classical control-theoretic approaches with data-driven concepts, \eg machine learning. Again, it is unclear how these novel design approaches that are partially based on massive data collection can be viably implemented in low-resource environments, and how dependability requirements can be met alongside.

%\db{That's adaptability. Here I would take up that we were able to adapt through mode changes, but everything needs to be known in advance.
%Learning might enable reacting to previously unseen situations.}

\subsection{Efficiency}
The presented solution builds on low-power wireless communication technology, which is in contrast, \eg to envisioned 5G communication.  While there are potential and huge expectations associated with future 5G solutions, the industry is also facing significant issues such as high running costs, network infrastructure not being owned by the manufacturer, insufficient data protection, and current inability to meet certain requirements on latency, jitter, and reliability.  Here, we presented an alternative solution that addresses similar use cases as being discussed in the context of the 5G vision, yet is based on existing commodity hardware.
%
% Comment Marco (incorporated):
%I'd briefly restate the issues industry currently has with 4G/5G solutions: high running costs, unable to meet certain requirements on latency, jitter, reliability, etc., insufficient data protection, long service deployment times as the network infrastructure is owned by the network operator and not the manufacturing company. ...

In addition to making use of low-power hardware, we further improved efficiency utilizing resource-aware control algorithms and the integration of control and communication at run time.   Through the concept of control-guided communication, communication bandwidth can be flexibly allocated at run time to the control processes that are in need or be saved otherwise.  While constituting only a first approach toward such integration at run time,
% first approach based on self-triggered control,
we were already able to demonstrate significant resource savings and flexible reallocation.

\fakepar{Truly event-triggered wireless control}
The presented solution for control-guided communication is based on a self-triggered control approach.
With self-triggered control, the control algorithm decides at the time of the current control computation, when the next control update needs to happen.  Including this information in the current data packet, the network manager can plan resource allocation in advance and, in this way, ensure efficient use of communication bandwidth.  The downside of the self-triggered approach, however, is that the need for communication must be planned in advance.  For better efficiency and also adaptability, it would be desirable to instead decide \emph{instantaneously} about the need for communication.  While in the control community, many so-called event-triggered estimation and control approaches have been developed in the last two decades \cite{heemels2012introduction,miskowicz2018event}, it remains largely unclear whether and how these can be \emph{integrated} with the communication system and indeed result in demonstrable resource reallocation, savings, or other advantages for wireless systems in practice.

% Comment Marco:
%So, to summarize, it is still unclear whether the event-triggered estimation and control concepts developed by the control community indeed result in demonstrable resource savings (or other advantages) for wireless systems in practice, without impairing stability.

\fakepar{Advances in wireless communication}
%\fakepar{Truly event-based for better efficiency and adaptability}
Scaling up the network diameter, number of agents, or supporting higher update rates may require faster low-power wireless physical layers (\eg Bluetooth Low Energy (BLE 5), ultra-wide band communication).  Further, protocol innovations such as network coding have the potential of increasing the effective capacity of the network and thus enhancing efficiency. These and other advances on the physical and protocol layers will directly benefit wireless control approaches as the one presented herein.

%\fakepar{Energy-harvesting devices}
%\st{@Marco:  If you have time to write this, please go ahead.  Otherwise, we can also remove this fakepar.}

%\db{I would reinclude the event-triggered part, since here we can nicely discuss again what we're already able to do and what is still missing}

%\fakepar{End-to-end designs}
%Control and coordination across wireless \emph{and} wired networks, leveraging the programmability of the latter.

%\fakepar{Open testbeds} mobile robots, real-time localization, etc.
\subsection{Real-world Deployment}
All core aspects of our presented wireless control solution have been deployed and demonstrated on the \cps testbed (\secref{sec:testbed}), which we developed for this research.  We deliberately chose the components of the testbed to be representative of many smart manufacturing use cases, as well as to pose a real challenge for the state of the art in wireless control.  For instance, we opted for inverted pendulums as physical processes because they represent prototypical control tasks and \emph{require} fast feedback for their operation at their unstable equilibrium. In addition to many hours of testing in the lab, we have also deployed our \cps testbed on multiple occasions outside the laboratory such as during a demonstration at the \emph{2019 Cyber-Physical Systems and Internet-of-Things Week} in Montreal, Canada.\footnote{This video gives an impression: \url{https://youtu.be/AtULmfGkVCE}.}    Despite being in a room with hundreds of conference attendees and much other wireless equipment operating in parallel, we were able to run the coordination and remote stabilization demos similar to those presented in \secref{sec:eval_sync} and \secref{sec:eval_remote} for about three hours without a single failure.

At the same time, we are convinced that more real-world demonstrators and deployments will be essential on the way of making wireless control a reliable and accepted technology in smart manufacturing.
In addition to theoretical proofs based on rigorous analysis, it is key to systematically evaluate the end-to-end system with all hardware and software components on physical platforms and real-world wireless networks in scenarios that resemble as close as possible the targeted smart manufacturing use cases.  We as a community thus need more of such testbeds with robots, flying drones, conveyor belts, etc. Ideally, these testbeds are open and extensible, and researchers around the world can use them. Besides practical validation, such testbeds are also needed to enable comparability and reproducibility of research claims and results, and they may serve as a common benchmarking platform.
To this end, efforts on testbeds in control and communication should no longer run separately, but be united and work toward a common goal.
% -> Seb: This previous sentence could also be left out IMO; either way is fine with me.
Ultimately, the final step is then, of course, real-world trials in actual manufacturing and production plants. But without prior demonstrators, the companies will be unlikely to allow this.
% (private conversation with Kalle).

%end-to-end system evaluation/validation on relevant scenarios and environments as close as possible to the target ones

% Comment Marco:
% In addition to theoretical proofs based on rigorous analysis etc. it is key to systematically evaluate the end-to-end system with all HW/SW components on physical platforms and real-world wireless networks in scenarios that resemble as close as possible the targeted smart manufacturing use cases. So we need more of such testbeds with robots, flying drones, conveyor belts, etc. Ideally, these testbeds are open and extensible, and researchers around the world can use them. Besides practical validation, such testbeds are also needed to enable comparability and reproducibility of research claims and results and may serve as a common benchmarking platform. To this end, major efforts on testbeds in control and communication should no longer run separately but be united and work toward a common goal. Final step is then, of course, real-world trials in the actual manufacturies. But without prior demonstrators, the companies will not allow this (private conversation with Kalle).

% !TEX root = ../paper.tex

\section*{Acknowldegements}

We would like to thank Romain Jacob for his contributions to the research detailed in this paper.
We would also like to thank Harsoveet Singh, Joel Bessekon Akpo, and Felix Grimminger for their
help with the testbed, and the TEC group at ETH Zurich
for making the design of the DPP platform available to the
public.

% \fi
\IEEEpeerreviewmaketitle

\ifCLASSOPTIONcaptionsoff
  \newpage
\fi



\bibliographystyle{IEEEtran}
\bibliography{IEEEabrv,ref}


\begin{IEEEbiography}[{\includegraphics[width=1in,height=1.25in,clip,keepaspectratio]{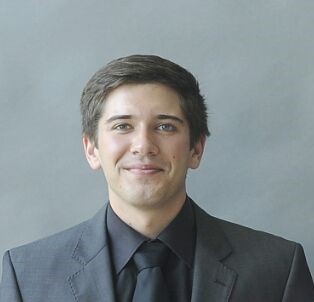}}]{Dominik
Baumann} received the Dipl.-Ing. degree in electrical engineering from TU Dresden,
Germany, in 2016. He is currently a joint PhD student with the Intelligent Control Systems Group at the Max Planck Institute for Intelligent Systems, T\"ubingen,
Germany and the Division of Decision and Control Systems, KTH Stockholm, Sweden.

His research interests include the control of cyber-physical systems and the interconnection of control theory and machine learning.
More info at \url{https://is.mpg.de/~dbaumann}.
\end{IEEEbiography}

\begin{IEEEbiography}[{\includegraphics[width=1in,height=1.25in,clip,keepaspectratio]{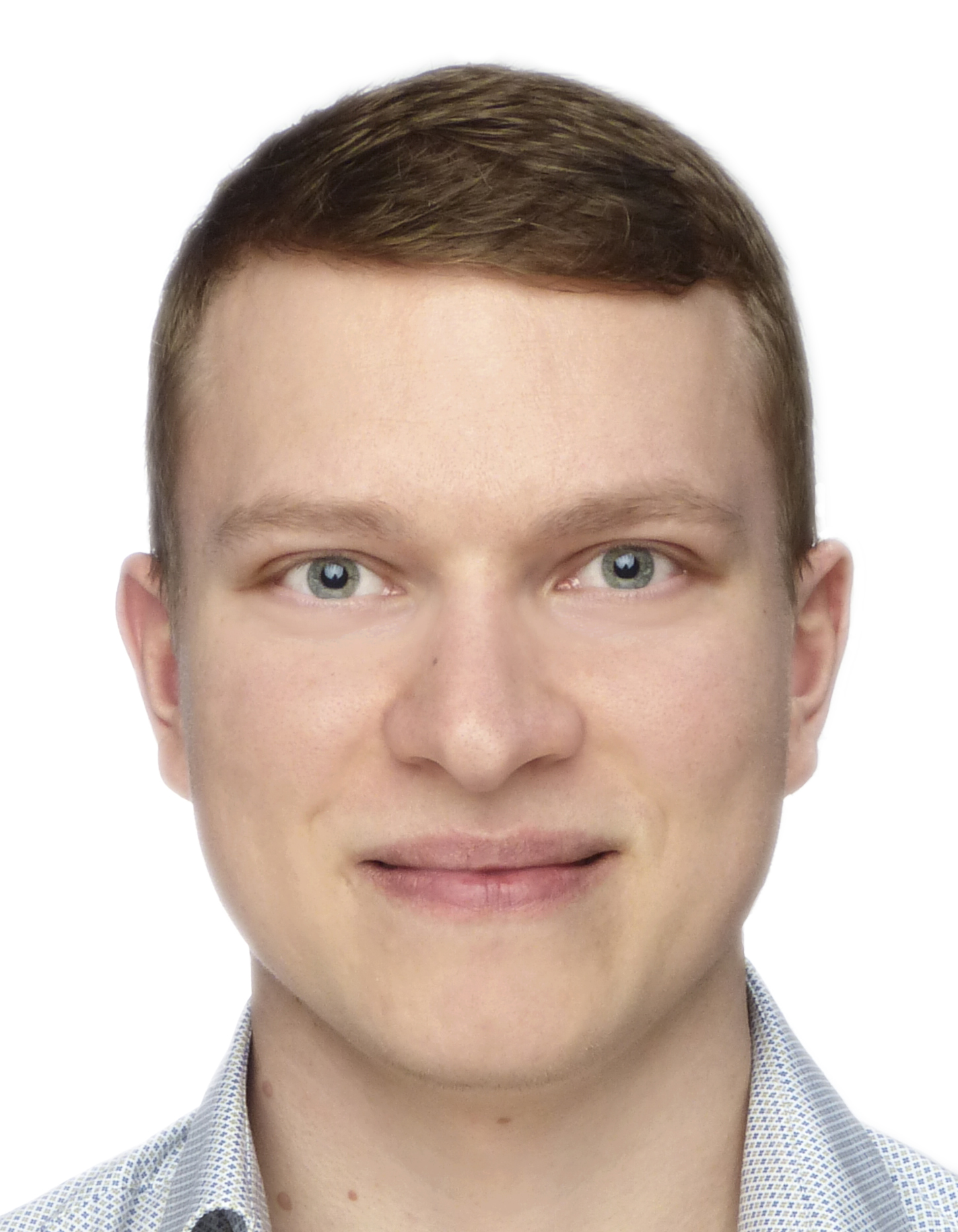}}]{Fabian Mager} received the Dipl.-Medieninf. degree in media computer science from TU Dresden, Germany, in 2015.
He is currently a PhD student in the Networked Embedded Systems Lab within the Center for Advancing Electronics at TU Dresden.

His research interests include cyber-physical systems and wireless communication, especially in the embedded low-power domain.
More info at \url{https://nes-lab.org/wordpress/?team_member=fabian-mager}.
\end{IEEEbiography}

\begin{IEEEbiography}[{\includegraphics[width=1in,height=1.25in,clip,keepaspectratio]{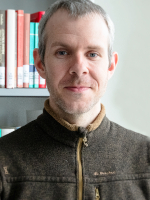}}]{Ulf Wetzker}
holds a Diploma degree in computer science and systems engineering (2008) from \mbox{TU~Ilmenau}, Germany.
Since October 2008 he works as a research scientist at the Fraunhofer Institute for Integrated Circuits IIS, Division Engineering of Adaptive Systems EAS where he joined the Industrial Wireless Communication group in 2012.

His research interests are in the area of wireless communication systems, data analytics and machine learning, with a special focus on anomaly detection and root cause analysis in wireless networks. 
More info at \url{https://www.eas.iis.fraunhofer.de/en/research_topics/wireless-networked_automation.html}.
\end{IEEEbiography}

\begin{IEEEbiography}[{\includegraphics[width=1in,height=1.25in,clip,keepaspectratio]{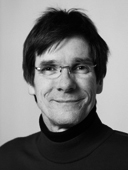}}]{Lothar Thiele}
is professor for Computer Engineering at ETH Zurich, Switzerland.
His research interests include models, methods, and software tools for the design of embedded systems, Internet of Things, cyber-physical systems, sensor networks, embedded software, and bioinspired optimization techniques.
Lothar Thiele is associate editor of INTEGRATION -- the VLSI Journal, Journal of Signal Processing Systems, IEEE Transaction on Industrial Informatics, Journal of Systems Architecture, IEEE Transactions on Evolutionary Computation, Journal of Real-Time Systems, ACM Transactions on Sensor Networks, ACM Transactions on Cyber-Physical Systems, and ACM Transaction on Internet of Things.

In 1986, he received the Dissertation Award of the Technical University of Munich, in 1987, the Outstanding Young Author Award of the IEEE Circuits and Systems Society, in 1988, the Browder J. Thompson Memorial Award of the IEEE, and in 2000-2001, the IBM Faculty Partnership Award.
In 2004, he joined the German Academy of Sciences Leopoldina.
In 2005, he was the recipient of the Honorary Blaise Pascal Chair of University Leiden, The Netherlands.
Since 2009, he is a member of the Foundation Board of Hasler Foundation, Switzerland.
Since 2010, he is a member of the Academia Europaea.
In 2013, he joined the National Research Council of the Swiss National Science Foundation (SNF).
Lothar Thiele received the EDAA Lifetime Achievement Award in 2015.
Since 2017, he is Associate Vice President of ETH for Digital Transformation.
In 2019, he joined the Foundation Board of Technopark Zurich as president.
More info at \url{https://people.ee.ethz.ch/~thiele/}.
\end{IEEEbiography}

\begin{IEEEbiography}[{\includegraphics[width=1in,height=1.25in,clip,keepaspectratio]{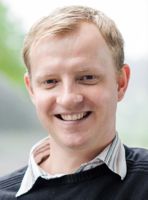}}]{Marco Zimmerling}
holds a Diploma degree in computer science (2009) from TU Dresden, Germany, and a Ph.D.~degree in computer engineering (2015) from ETH Zurich, Switzerland.
Since November 2015 he leads the Networked Embedded Systems Lab within the Center for Advancing Electronics at TU Dresden.
His research interests are in the general area of cyber-physical systems, with a special focus on dependable and sustainable wireless embedded systems. 
Dr.~Zimmerling received the 2015 ACM SIGBED Paul Caspi Memorial Dissertation Award, the 2016 EDAA Outstanding Dissertation Award, and Best Paper Awards at ACM/IEEE ICCPS 2019, ACM SenSys 2013, and ACM/IEEE IPSN 2011. More info at \url{https://wwwpub.zih.tu-dresden.de/~mzimmerl/}.
\end{IEEEbiography}

\begin{IEEEbiography}[{\includegraphics[width=1in,height=1.25in,clip,keepaspectratio]{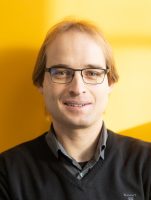}}]{Sebastian Trimpe}
(M'12) received the B. Sc. degree in general engineering science and the M.Sc. degree (Dipl.-Ing.) in electrical engineering from Hamburg
University of Technology, Hamburg, Germany, in 2005 and 2007, respectively, and the Ph.D. degree (Dr. sc.) in mechanical engineering from ETH Zurich, Zurich, Switzerland, in 2013. 
Since 2020, he is a full professor at RWTH Aachen University, Germany, where he heads the Institute for Data Science in Mechanical Engineering.  Before, he was an independent Research Group Leader at the Max Planck Institute for Intelligent Systems in Stuttgart and Tübingen, Germany. His main research interests are in systems and control theory, machine learning, networked and autonomous systems. 

Dr. Trimpe is recipient of several awards, among others, the triennial IFAC World Congress Interactive Paper Prize (2011), the Klaus Tschira Award for achievements in public understanding of science (2014), the Best Demo Award of the International Conference on Information Processing in Sensor Networks (2019), and the Best Paper Award of the International
Conference on Cyber-Physical Systems (2019).  More info at \url{https://is.mpg.de/~strimpe}.
\end{IEEEbiography}


%\newpage



%\vfill


\end{document}